\documentclass[11pt,a4paper]{article}
\pdfoutput=1
\usepackage[pdftex]{graphics}
\usepackage{jheppub}
\usepackage{amsmath,amssymb,amsfonts}
\usepackage{enumitem}
\usepackage{multirow}
\usepackage{array,booktabs}
\usepackage{slashed}
\usepackage{url}

\usepackage{accents}
\usepackage{mathrsfs}
\usepackage{mathtools}
\usepackage{bm}

\usepackage[usenames,dvipsnames]{xcolor}
\definecolor{labelcolor}{RGB}{194, 175, 116}
\definecolor{rmkcolor}{RGB}{15,120,255}
\definecolor{bred}{RGB}{240,8,8}
\definecolor{linkcolor}{RGB}{63,47,40}

\usepackage[export]{adjustbox}
\setlist[itemize]{
    label=\adjustbox{scale=0.7}{$\bullet$}, itemsep=0pt,topsep=4pt
}

\hypersetup{
    citecolor=linkcolor,
    linkcolor=linkcolor,
    urlcolor=linkcolor
}

\usepackage[final]{showlabels} % use "inline" or "final"

\usepackage{tikz}
\usetikzlibrary{calc} % to use relative coordinates
\usetikzlibrary{shapes.geometric} % to draw regular polygons
\usetikzlibrary{positioning} % to use right=of 
\usetikzlibrary{fit} % for fit size
\usepackage[a]{esvect} % arrow styling %f
\tikzset{empty/.style = {inner sep = 0pt, outer sep = 0, minimum size = 0}}
\tikzset{b/.style = {inner sep = 2pt, outer sep = 4pt, minimum size = 12pt}}
\tikzset{c/.style = {inner sep = 2pt, outer sep = 4pt, minimum size = 12pt}}
\tikzset{w/.style = {inner sep = 1pt, outer sep = 2pt, minimum size = 12pt, anchor = west}}
\tikzset{s/.style = {inner sep = 2.5pt, outer sep =2.5pt, minimum size = 1pt, font = \small}}
\tikzset{lin/.style = {draw, line width = 0.5pt}}
\usepackage[export]{adjustbox}

\renewcommand{\eqref}[1]{Eq.\,(\ref{#1})}
\newcommand{\eqrefs}[2]{Eqs.\,(\ref{#1}) and (\ref{#2})}
\newcommand{\eqrefss}[3]{Eqs.\,(\ref{#1}), (\ref{#2}), and (\ref{#3})}

\newcommand{\Sec}[1]{Sec.\,\ref{#1}}
\newcommand{\Secs}[2]{Secs.\,\ref{#1} and \ref{#2}}
\newcommand{\Secss}[3]{Secs.\,\ref{#1}, \ref{#2}, and \ref{#3}}

\newcommand{\Secsto}[2]{Secs.\,\ref{#1} to \ref{#2}}
\newcommand{\App}[1]{App.\,\ref{#1}}
\newcommand{\rcite}[1]{Ref.\,\cite{#1}}
\newcommand{\rrcite}[1]{Refs.\,\cite{#1}}

\newcommand{\transition}[1]{\qquad\adjustbox{scale=0.95}{\text{#1}}\qquad}
\newcommand{\transit}[1]{\quad\adjustbox{scale=0.95}{\text{#1}}\quad}

\let\oldc\c
\let\oldi\i

\DeclareMathOperator{\diag}{diag}

\def\mem{\hspace{0.1em}}
\def\hem{\hspace{0.05em}}
\def\nem{\hspace{-0.1em}}
\def\hnem{\hspace{-0.05em}}
\def\hhem{\hspace{0.025em}}
\def\hhnem{\hspace{-0.025em}}
\def\hhhem{\hspace{0.0125em}}

\def\iq{{{\implies}\quad}}
\def\qiq{{\quad\implies\quad}}
\def\qfq{{\quad\iff\quad}}

\def\a{\alpha}
\def\b{\beta}
\def\c{\gamma}
\def\d{\delta}
\def\e{\epsilon}
\def\ve{\varepsilon}
\def\m{\mu}
\def\n{\nu}
\def\r{\rho}
\def\s{\sigma}
\def\k{\kappa}
\def\l{\lambda}
\def\t{\tau}
\def\L{\Lambda}
\def\S{\Sigma}

\def\bpsi{{\smash{\bar{\psi}}\kern0.02em\vphantom{\psi}}}

\def\te{\tilde{\epsilon}}
\def\mathe{{\scalebox{1.025}[1]{$\mathrm{e}$}}}
\def\pt{{\hnem\nem\mathscr{P}}}

\def\rmA{{\mathrm{A}}}
\def\rmB{{\mathrm{B}}}

\def\mwedge{{\mem\wedge\mem\hhem}}
\def\swedge{{\mem{\wedge}\,}}

\def\motimes{{\mem\otimes\mem}}
\def\modot{{\mem\odot\mem}}

\def\mplus{{\mem+\mem}}
\def\mminus{{\mem-\mem}}
\def\mtimes{{\mem\times\mem}}
\def\mdot{{\mem\cdot\mem}}

\def\da{{\dot{\a}}}
\def\db{{\dot{\b}}}
\def\dc{{\dot{\c}}}
\def\dd{{\dot{\d}}}

\newcommand{\wrap}[1]{{\smash{#1}\vphantom{\b}}}

\def\tdo{{\tilde{o}}}

\renewcommand{\o}{o}
\def\i{\iota}

\def\lsq{{
    \kern-0.037em
    \adjustbox{scale=0.99,valign=c}{$
        {\lfloor \llap{\reflectbox{\rotatebox[origin=c]{180}{$\lfloor$}}}}
    $}
    \kern-0.04em
}}
\def\rsq{{
    \kern-0.04em
    \adjustbox{scale=0.99,valign=c}{$
        {\rlap{\reflectbox{\rotatebox[origin=c]{180}{$\rfloor$}}} \rfloor}
    $}
    \kern-0.037em
}}

\newcommand{\lp}[1]{{
    \langle #1 \rangle
}}
\newcommand{\rp}[1]{{
    \lsq #1 \rsq
}}

\def\rdr{\rp{\pi\hem d\pi}}
\def\ldl{\lp{\lambda\hem d\lambda}}

\def\id{{\rlap{1} \hskip 1.6pt \adjustbox{scale=1.1}{1}}}

\newcommand{\dbar}{
    d\kern-.20em\makebox[0pt][l]{$\bar{}$}\kern.20em
}
\newcommand{\deltabar}{
    \delta\kern-.20em\makebox[0pt][l]{$\bar{}$}\kern.20em
}

\newcommand{\lrp}[1]{\left(\,{#1}\,\right)}
\newcommand{\bigbig}[1]{\big(\mem{#1}\mem\big)}
\newcommand{\BB}[1]{\Big(\,{#1}\,\Big)}
\newcommand{\bb}[1]{\bigg(\,{#1}\,\bigg)}

\newcommand{\bbsq}[1]{\bigg[\,{#1}\,\bigg]}

\newcommand{\act}[1]{[\,{#1}\,]}

\newcommand{\bbn}[1]{\bb{\nem\hnem{#1}\hnem\nem}}

\def\kerr{{\smash{\text{$\kern-0.075em\sqrt{\text{Kerr\hem}}$}}}}

\def\PD{{{Pleba\'nski-Demia\'nski}}}

\def\Pleb{{{Pleba\'nski}}}

\let\oldexp\exp
\renewcommand{\exp}{\oldexp\nem}
\def\Pexp{\mathrm{P}\kern-0.1em\exp}

\def\Pexp{\mathrm{P}\kern-0.1em\exp}

\def\M{{\mathcal{M}}}
\def\N{{\mathcal{N}}}

\def\P{{\mathbb{P}}}
\def\mflat{\mathbb{M}}
\def\F{{\mathcal{F}}}
\def\PT{\mathbb{P}\mathcal{T}}
\def\pt{\mathbb{PT}}
\def\CP{\mathbb{CP}}

\def\O{\mathcal{O}}

\def\tk{\tilde{\kappa}}
\def\tpi{\tilde{\pi}}

\def\Diff{\mathrm{Diff}}

\def\npad{\kern0.49em}
\def\mpad{\kern0.525em}
\def\hpad{\kern1.36em}
\def\wpad{\kern1.75em}

\def\Cinfty{C^\infty\hnem}

\def\C{\mathbb{C}}
\def\Z{\mathbb{Z}}
\def\NN{\mathbb{N}}

\newcommand{\TwoByTwoMatrixMed}[4]{
    \lrp{\begin{array}{cc}
        \hphantom{\hpad}\mathclap{
            #1
        }\hphantom{\hpad}&\hphantom{\hpad}\mathclap{
            #2
        }\hphantom{\hpad}
        \\
        \hphantom{\hpad}\mathclap{
            #3
        }\hphantom{\hpad}&\hphantom{\hpad}\mathclap{
            #4
        }\hphantom{\hpad}
    \end{array}}
}

\newcommand{\TwoByTwoMatrixCustom}[5]{
    \lrp{\begin{array}{cc}
        \hphantom{\kern#1}\mathclap{
            #2
        }\hphantom{\kern#1}&\hphantom{\kern#1}\mathclap{
            #3
        }\hphantom{\kern#1}
        \\
        \hphantom{\kern#1}\mathclap{
            #4
        }\hphantom{\kern#1}&\hphantom{\kern#1}\mathclap{
            #5
        }\hphantom{\kern#1}
    \end{array}}
}

\def\Rec{{\mathscr{R}\hhem}}

\def\ss{{\kern0.05em/\kern-0.25em/\kern0.10em}}

\def\inn{{\:\in\:}}
\def\eqq{{\:=\:}}
\def\too{{\:\to\:}}

\def\Cinfty{C^\infty\hnem}

\def\vex{\vec{x}}
\def\vea{\vec{a}}

\def\doz{{\dot{0}}}
\def\diz{{\dot{1}}}

\def\Cinfty{C^\infty\hnem}

\newcommand{\cont}[2]{\big\langle\hem{#1},\hem{#2}\hem\big\rangle}
\def\Tstar{T^*\nem}

\def\lb{\{\kern-0.15em\{}
\def\rb{\}\kern-0.15em\}}

\newcommand{\comm}[2]{[\hem{#1},{#2}\hem]}

\def\Sym{\mathrm{Sym}}

\def\ta{\widetilde{\a}}
\def\tb{\widetilde{\b}}

\def\LL{\mathscr{L}}

\def\QQ{\mathscr{Q}}

\def\SL{\mathrm{SL}}

\newif\ifToggleMacros
\ToggleMacrostrue  %% View comments
\ifToggleMacros
    \newcommand{\jh}[1]{{\color[RGB]{240,25,25} {JH:#1}}}
\else
   \newcommand{\jh}[1]{}
\fi

\def\next{\,\,\,\mapsto\,\,\,}

\def\CD{\mathscr{D}}

\def\Pdeg{\hnem\mathscr{P}\hem}

\def\pref{\rp{p\eta}}
\def\refq{\rp{\eta q}}
\def\pq{\rp{pq}}

\newcommand{\ratio}[1]{
    \bb{\nem\hnem{
        \frac{\pref}{\refq}
    }\hnem\nem}^{\hnem\nem\nem#1}
}

\newcommand{\zet}[1]{
    \bb{\nem\hnem{
        \frac{\l_0}{\l_1}
    }\hnem\nem}^{\hnem\nem\nem#1}
}

\def\Mm{{M^\star}}
\def\Qm{{Q^\star}}

\def\trho{\widetilde{\rho}}

\def\sfa{\mathsf{a}}
\def\sfb{\mathsf{b}}
\def\sfc{\mathsf{c}}

\def\K{\mathcal{K}}

\def\vK{\vec{\K}}

\def\AK{
    \big|\hem{
        \vec{\K}
    }\hem\big|
}

\def\tphi{\widetilde{\phi}}

\begin{document}

\begin{titlepage}

\begin{flushright}
    \footnotesize\scshape
    CALT-TH 2026-028
\end{flushright}
\vskip 150pt
{\LARGE\bfseries
    Incidence Relations for
    \\[0.6\baselineskip]
    Self-Dual Black Holes
}\\[1.5\baselineskip]
\parbox{0.7\linewidth}{
    {\scshape
        Joon-Hwi Kim
    }\\[0.25\baselineskip]
    {\footnotesize\itshape
        Walter Burke Institute for Theoretical Physics,
        \\[-0.2\baselineskip]
        California Institute of Technology, Pasadena, CA 91125
    }\\[0.25\baselineskip]
    {\footnotesize
        \textit{e-mail:} \url{joonhwi@caltech.edu}
    }
}\\[1.6\baselineskip]

\noindent\parbox{0.76\linewidth}{
    \noindent\textsc{Abstract}:\,
    In Penrose’s nonlinear graviton construction,
    a self-dual spacetime
    arises as the moduli space of holomorphic twistor lines in curved twistor space,
    whose explicit parametrization is concretely given by
    the incidence relation.
    We demonstrate a concrete local construction of
    the incidence relations for self-dual black hole spacetimes in Kerr-Schild coordinates:
    Eguchi-Hanson, self-dual Taub-NUT, and self-dual Pleba\'nski-Demia\'nski.
    This is facilitated by implementing
    the Dunajski-Mason recursion in the framework of Pleba\'nski’s second heavenly equation,
    for which we explicitly find the Pleba\'nski scalar.
    Our results describe closed-form formulae.
    Notably, for the self-dual Taub-NUT solution,
    the exact incidence relation 
    exhibits linear dependence on the gravitational coupling.
    The construction of zero-rest-mass fields
    on self-dual black hole backgrounds
    is briefly sketched as an application.
    % 
    % Brief remarks or observations are made on
    % a multivaluedness reflecting the Misner string geometry,
    % the coincidence loci for real black holes,
    % and
    % the coexistence of two single copy correspondences
    % via the heavenly equation or the Kerr-Schild metric.
}

\end{titlepage}

\bibliographystyle{utphys-modified}
\renewcommand*{\bibfont}{\footnotesize}

\setcounter{tocdepth}{2}

\beforetochook
\hrule
\begingroup
    \renewcommand{\baselinestretch}{0.9}\normalsize
    \tableofcontents
\endgroup
\afterTocSpace
\hrule
\afterTocRuleSpace
\setcounter{footnote}{0}
\pagestyle{myplain}\pagenumbering{arabic}
% \newpage

\newpage
\section{Introduction}

Twistor theory is a unique approach to
four-dimensional physics
where objects in spacetime
% arise from
% originate from
are reformulated in
a complex projective space,
known as twistor space
\cite{penrose:maccallum,Atiyah:2017erd}.
The relationship between flat spacetime and the twistor space
is given by the incidence relation,
\begin{align}
    \label{inc0}
    \mu^\da
    \,=\,
        x^{\da\a}\hem \lambda_\a
    \,.
\end{align}
Here, $x^{\da\a}$ are null Cartesian coordinates
of flat complexified spacetime,
while $(\lambda_\a,\mu^\da)$ give projective coordinates on the twistor space.
The incidence relation in \eqref{inc0}
can be geometrically interpreted as
representing self-dual (SD) null planes,
known as $\a$-surfaces.
Each point $[(\lambda,\mu)]$ on twistor space
corresponds to an $\a$-surface
$x^{\da\a} \sim x^{\da\a} + y^\da \l^\a$
in spacetime.
Conversely, each spacetime point $x$
corresponds to a Riemann sphere $\CP^1$ embedded in twistor space $\CP^3$,
by taking \eqref{inc0} as a parametrization of the $\mu$-coordinates
in terms of the $\lambda$-coordinates on $\CP^1$
for fixed $x^{\da\a}$.
This $\CP^1$ holomorphically embedded in $\CP^3$ is called a twistor line.
% Via the incidence relation,
% geometric data in twistor space
% are transcribed to spacetime
% and vice versa.
The incidence relation facilitates
transcribing geometric data in twistor space to spacetime,
or conversely,
lifting spacetime objects to twistor space.

The celebrated nonlinear graviton construction of Penrose \cite{Penrose:1976js}
shows that
SD, Ricci-flat Riemannian four-manifolds
arise by
deforming the complex structure of twistor space.
In the resulting curved twistor space,
the holomorphic lines corresponding to spacetime points no longer appear ``straight.''
Reproducing the terminology of  Mason and Skinner \cite{GravityMHVTwistors},
such curved twistor lines are
described by a ``deformed'' incidence relation
of the form
\begin{align}
    \label{inc}
    \mu^\da
    \,=\,
        F^\da(x,\lambda)
    \,.
\end{align}
The deformed incidence relation in \eqref{inc}
gives an explicit parametrization of the twistor line associated with the spacetime point $x$,
and thereby encodes the relation between curved spacetime and its twistor space.
Equivalently,
\eqref{inc} represents
$\a$-surfaces in curved spacetime.
This traces back to
the Mason-Newman \cite{MasonNewman:1989} characterization 
of SD spacetimes,
establishing
integrability of translations along SD null directions
via a choice of tetrad.

It is desirable to comprehend
the explicit content of
% the deformed incidence relation.
the deformation entailed in \eqref{inc},
when compared to \eqref{inc0}.
This is viable
in the framework of
Pleba\'nski's second heavenly equation
\cite{plebanski1975some},
for instance.
The self-duality of spacetime
can be characterized as
the closure of the Pleba\'nski anti-self-dual (ASD) two-forms $\Sigma^{\a\b}$ \cite{Plebanski:1977zz,Capovilla:1991qb},
$d\Sigma^{\a\b} = 0$.
On account of the Darboux theorem \cite{darboux1882probleme},
two of these ASD two-forms can be represented as
$\Sigma^{01} = dq^\da \swedge dp_\da$ and $\Sigma^{11} = \te_\wrap{\da\db}\mem dq^\da \swedge dq^\db$
by a choice of coordinates
$x^{\da1} = q^\da$, $x^{\da0} = p^\da$.
This essentially implements
a lightcone gauge for the metric field,
shows that
a single scalar field $\Phi$ 
called the second {\Pleb} scalar
dictates the SD spacetime geometry,
and
derives the celebrated
second heavenly equation of Pleba\'nski \cite{plebanski1975some}
as the equation of motion for $\Phi$;
see \rcite{Adamo:2021bej}, for instance.
It is known that,
once the second heavenly coordinates 
$x^{\da1} = q^\da$, $x^{\da0} = p^\da$
are established,
the deformed incidence relation
can be recursively constructed
in the affine patch $\lambda_1 \neq 0$ of $\CP^1$
as
\cite{dunajski2000hyper,Adamo:2021bej}
\begin{align}
    \label{incpleb}
    F^{\da}\hnem(x,\lambda)
    \,=\,
        x^{\da\a}\hem \lambda_\a
        \mem+\mem
        \k\,
            \frac{\partial \Phi}{\partial p_\da}
            \mem\frac{{\l_0}^2}{{\l_1}}
        \mem+\mem
        \k\,
            \frac{\partial \Phi}{\partial q_\da}
            \mem\frac{{\lambda_0}^3}{{\lambda_1}^2}
        \hem
        \mem+\mem
        \cdots
    \,,
\end{align}
where $\k$ is an auxiliary parameter counting the power of $\Phi$.

\newpage

To see the physical intuition behind \eqref{incpleb},
it may be instructive to assume a perturbative context
in which $\kappa$ encodes a small gravitational coupling.
In this case,
it could be argued that
\eqref{incpleb}
implements
the deformed incidence relation
in \eqref{inc} as 
a ``general-relativistic correction''
on
% the special-relativistic incidence relation in 
\eqref{inc0}
due to gravitation.

More precisely,
\eqref{incpleb} describes a series in $\lambda_0/\lambda_1$,
while $x^{\da\a}\hem \lambda_\a = q^\da\mem \lambda_1 + p^\da\mem \lambda_0$.
It is known that
\eqref{incpleb}
can be derived by
implementing the framework known as
Dunajski-Mason (DM) recursion 
\cite{dunajski2000hyper,dunajski1996heavenly,dunajski1997recursion}.
The DM recursion is a tool that enables
a Laurent series construction of
functions of $x$ and $\lambda$
that descend to the twistor space
by integrability along the $\a$-surfaces.
The function $F^\da(x,\lambda)$ in \eqref{inc}
is certainly one such function,
since it parametrizes the very $\a$-surfaces.
That is, \eqref{inc}
is read as
% the function 
$F^\da(x,\lambda)$
taking a constant value $\m^\da$
along each SD null surface
in the curved spacetime.
% For a given harmonic function $\psi(x)$ on spacetime,
% the DM recursion
% constructs 
% \begin{align}
%     \hat{\psi}
%     \,= 
%         \sum_{n=-\infty}^{\infty} 
%         \bb{\nem
%             \frac{{\l_0}}{{\l_1}}
%         \nem}^{\nem\nem\hnem n}\,
%             \Rec^n\act{
%                 \psi(x)
%             }
%     \,,
% \end{align}
% by repeatedly acting on
% the DM recursion operator $\Rec$.
The DM recursion constructs the Laurent series
in terms of a recursion operator $\Rec$.
In the second heavenly formalism,
the repeated action of 
this recursion operator $\Rec$
on the coordinates $x^{\da1} = q^\da$
is given by
\begin{align}
    q^\da
    \,\,\,\mapsto\,\,\,
        p^\da
    \,\,\,\mapsto\,\,\,
        \k\,
            \frac{\partial \Phi}{\partial p_\da}
    \,\,\,\mapsto\,\,\,
        \k\,
            \frac{\partial \Phi}{\partial q_\da}
    \,\,\,\mapsto\,\,\,
        \cdots
    \,,
\end{align}
from which one derives the explicit formula in \eqref{incpleb}.

\medskip
Recently, a certain class of SD vacuum spacetimes has attracted considerable attention under the name
``SD black holes''
\cite{Adamo:2023fbj,Adamo:2025fqt,Crawley:2021auj,Crawley:2023brz,Guevara:2023wlr,Guevara:2024edh,Guevara:2025psg,probe-nj,note-sdtn,nja,AAS,Skvortsov:2025ohi,Doran:2026bng}.
Justifications for this terminology
can be found,
e.g., in \rrcite{Adamo:2023fbj,Adamo:2025fqt,AAS}.
In particular,
\rcite{AAS} gave a precise mathematical definition of SD black holes
in terms of complexifications of
Plebański–Demiański \cite{PD} metrics
and also provided
their exhaustive classification:
\begin{enumerate}[label=\textsc{(\alph*)},itemsep=1pt,topsep=4pt]
    \item 
        The Eguchi-Hanson (EH) instanton
        \cite{Eguchi:1978xp,Eguchi:1978gw,Eguchi:1979yx},
    \item
        The SD Taub-NUT (SDTN) solution
        \cite{hawking1977gravitational,Gibbons:1978tef},
    \item
        The SD {\PD} (SDPD) solution
        \cite{Casteill:2001zk,Nozawa:2015qea,Araneda:2023xnv}.
\end{enumerate}
Furthermore,
\rcite{AAS} explicated a beautiful organization behind
this classification,
based on dual twistor quadrics.
In turn,
\rcite{AAS} established
a remarkable fact that
all SD black hole solutions
admit Kerr-Schild (KS) metrics,
by building upon
a theorem of Tod \cite{Tod:1982mmp}
that concerns cases in which
the second heavenly equation
gets ``linearized'' in a certain sense.

The fact that
the
EH solution
admits a KS metric
% is attributed to
% Sparling and Tod
has been known 
for a while
\cite{Tod:1982mmp,BurnettStuart:1979sparlingtod,Berman:2018hwd}.
The fact that
the
SDTN solution
admits a KS metric
is a recent finding due to \rcite{note-sdtn},
according to the literature.
The fact that
the
SDPD solution
admits a KS metric is
a new discovery of \rcite{AAS}.

The construction of
the curved twistor spaces
for SD black holes
has been an interest
in the literature
and has been investigated through,
e.g.,
\rrcite{sparling1976nonlinear,hitchin1979polygons,lebrun1991complete,Adamo:2025fqt,AASS}.
In particular,
see \rcite{Adamo:2025fqt} 
for an explicit recent account on the deformed incidence relation
of SDTN.
The recent work \cite{AASS}
% which appeared 
% % several days ago
% earlier this month
also
identified the deformed complex structure
of the SDTN solution
in the KS description,
yet without providing the deformed incidence relation
explicitly.

% \newpage
% \medskip
The purpose of this article
is to demonstrate
an explicit construction of 
the deformed incidence relations
for SD black holes,
in local patches.
This is viable
by implementing the DM recursion
in the {\Pleb} second heavenly framework.
\begin{itemize}
    \item 
        We explicitly construct 
        a second {\Pleb} scalar $\Phi$
        for each SD black hole solution:
        EH, SDTN, and SDPD.
    \item 
        It is immediate that
        plugging this $\Phi$ into \eqref{incpleb}
        yields
        an explicit formula for the deformed incidence relation
        up to 
        % the leading order in $\kappa$.
        a certain order in $\l_0/\l_1$.
    \item 
        Going further,
        we carry out the DM recursion fully
        and obtain exact, closed-form formulae 
        by resummation.
        We work in the affine patch of $\CP^1$ where $\lambda_1 \neq 0$.
\end{itemize}
The higher-order terms omitted in \eqref{incpleb}
entail choosing a primitive in general,
due to a nonlocality inherent in the DM operator $\Rec$.
By specializing in the specific cases of EH, SDTN, and SDPD,
however,
we were able to
deduce general formulae to all orders.

Crucially,
we explicate that
the KS coordinates
for SD black holes
are also {\Pleb}'s second heavenly coordinates
at the same time.
For the EH solution,
this observation has been available since
Tod \cite{Tod:1982mmp}.
For SDTN,
this observation
has been reported in \rcite{note-sdtn},
which provided an explicit {\Pleb} scalar as well.
For SDPD,
our explicit construction of 
a second {\Pleb} scalar
in KS coordinates
is a new result,
to our understanding.

We summarize our results below.
For EH, SDTN, and SDPD, respectively, we find
\medskip\vspace{1.5pt}
\hrule\vspace{0.5pt}
\hrule\vspace{-5pt}
\begin{subequations}
\label{summary}
\begin{align}
\label{summary:EH}
    F^\diz
    \,&=\,
        \sqrt{
            \bigbig{x^{\diz\a}\l_\a}^{\nem2}
            - \frac{16\k
                \smash{
                    \bigbig{x^{\diz0}\l_0}^{\nem2}
                }\vphantom{p^\doz}
            }{(x^2)^2}
        \mem}
    \,,\,\,\,\mem
    F^\doz F^\diz
    \mem=\mem
        \bigbig{x^{\doz\a}\l_\a}\nem
        \bigbig{x^{\diz\a}\l_\a}
        - \frac{16\k\mem
            x^{\doz0} x^{\diz0} {\l_0}^2
        }{(x^2)^2}
    \,,\\[0.15\baselineskip]
\label{summary:SDTN}
    F^\da
    \,&=\,
        x^{\da\a} \l_\a
        + \sqrt{2}\hem\k\mem
            \delta^{\da\a}\mem
            \bbsq{
                \l_\a
                \hem
                \log\nem\bb{\nem{
                    1-
                    \frac{x \mplus iy}{|\vex| \mplus z}
                    \mem \frac{\l_0}{\l_1}
                }\nem}
                % + \o_\a\mem
                + \delta_\a{}^0\hem
                    \l_0\mem
                    % \frac{\l_0}{\l_1}\mem 
                    \frac{x \mplus iy}{|\vex| \mplus z}
            }
    \,,\\[0.1\baselineskip]
\label{summary:SDPD}
\begin{split}
    % \mathllap{\smash{%
    %     \adjustbox{raise=-0.5\baselineskip}{$\left\{\vphantom{\displaystyle\frac{\Big|}{\Big|}}\right.$}%
    % }}
    F^\diz
    \,&=\,
        Y_+^{\frac{\Delta+1}{2\Delta}}
        \hem
        Y_-^{\frac{\Delta-1}{2\Delta}}
    \,,\quad
    F^\doz F^\diz
    \mem=\mem
        % \frac{\ta^\doz}{\ta^\diz}\mem Y_+ Y_-
        % + \frac{b}{\sqrt{2}\hem \ta^\diz}\mem x^{\diz\a} \l_\a\mem \l_1
        % (x^{\doz\a}\l_\a)
        % (x^{\diz\a}\l_\a)
        \bigbig{x^{\doz\a}\l_\a}\nem
        \bigbig{x^{\diz\a}\l_\a}
        + \k\hem b\mem \ta^\doz \ta^\diz {\l_0}^2
    \,.
\end{split}
\end{align}
\end{subequations}
\vspace{-8pt}
\hrule\vspace{0.5pt}
\hrule\vspace{-1pt}
\medskip
\noindent
Notably, the SDTN case in \eqref{summary:SDTN}
exhibits an exact linearization
in the coupling parameter $\k$.
For \eqref{summary:SDPD},
see \eqref{SDPD:notations} for
the definition of $\ta^\da$, $Y_\pm$, and $\Delta$.
\eqref{summary:SDTN} can be reproduced from \eqref{summary:SDPD}
via a proper limiting procedure that 
involves setting the acceleration parameter to zero:
$1/b \to 0$.
All of \eqrefss{summary:EH}{summary:SDTN}{summary:SDPD}
nicely reduce to the flat incidence relation in \eqref{inc0}
in the $\k\to0$ limit.

\medskip
Some remarks are in order.
First of all,
we emphasize that
there is no claim of originality
on the idea that
the deformed incidence relation
can be concretely constructed by
implementing DM recursion in the second heavenly framework;
in particular,
see 
Sec.\,3.7 of \rcite{dunajski2000hyper}.
Still,
at least
for the SDTN and SDPD cases,
it seems that
our closed-form formulae
% for the deformed incidence relations
in \eqrefs{summary:SDTN}{summary:SDPD}
had not been reported in the previous literature.
For the EH case,
our results are consistent with
\rrcite{dunajski2000hyper,Bittleston:2023bzp,hitchin1979polygons},
for instance.
Secondly,
we clarify that
possible subtleties and complications
% due to 
regarding
global topology
will be largely swept under the rug
in our discussion.
The role of this article
is to simply provide a demonstration,
concretely by working locally
(both in spacetime and in the Riemann sphere).
% Further investigations
% can follow
% in more comprehensive forms,
% which are beyond the scope of this work.
% Lastly,
% we choose to work in the holomorphic, complexified category
% for simplicity.
% \newpage

We begin by a review of
{\Pleb}'s second heavenly coordinates,
curved twistor space,
deformed incidence relation,
and the DM recursion
in \Sec{REVIEW},
% which is
based on
\rrcite{%
    dunajski2000hyper,dunajski1996heavenly,dunajski1997recursion,%
    plebanski1975some,Adamo:2021bej,boyer1983geometry,hull1991geometry,%
    MasonNewman:1989,chakravarty1991canonical,chacon2019self,gindikin1986construction,GravityMHVTwistors%
}.
\Sec{TYPEN} discusses a preliminary example 
due to \rcite{dunajski2000hyper}
as a check.
\Secss{EH}{SDTN}{SDPD}
then give the core part of this paper,
% then construct
where
the explicit deformed incidence relation
is constructed
in KS coordinates
for each SD black hole solution:
EH, SDTN, and SDPD.
Finally, \Sec{STATES}
gives some physical use of our exact deformed incidence relations:
constructions of massless fields on SD black hole backgrounds
by Penrose transform,
such as plane-wave states
or perturbations
representing insertions of ``mini'' ASD black holes.

\section{Review of Self-Dual Spacetimes and Curved Twistor Theory}
\label{REVIEW}

This section 
aims to provide a friendly review of
some essential concepts regarding
SD spacetimes and
curved twistor theory.
Statements in this section are brought from
\rrcite{%
    dunajski2000hyper,dunajski1996heavenly,dunajski1997recursion,%
    plebanski1975some,Adamo:2021bej,boyer1983geometry,hull1991geometry,%
    MasonNewman:1989,chakravarty1991canonical,chacon2019self,gindikin1986construction,GravityMHVTwistors%
}.
Readers familiar with these basic ideas
can skip this section entirely.
Also, \Secs{REVIEW>RESID}{REVIEW>DM>CAN}
may be skipped for all readers.

For simplicity,
we work in the holomorphic, complexified category.
Spacetime is
an oriented complex four-manifold $\M$
equipped with a holomorphic metric $g$.
Its tangent bundle decomposes into SD and anti-self-dual (ASD) spinor bundles:
\begin{align}
    T\M 
    \,\cong\,
    {S^+\nem\M}\hem {\,\otimes\,} {S^-\nem\M}
    \,.
\end{align}
The Levi-Civita connection $\nabla$ of $g$
splits into SD and ASD parts,
defining connections on the two spinor bundles.

In this paper,
a spacetime $(\M,g)$ will be said to be SD
iff its ASD spinor bundle $S^-\nem\M$ is flat.
This implies that
the ASD connection coefficients can be set to zero
in a local patch,
so
the structure group of $\M$ is $\mathrm{SL}(2,\mathbb{C}) \cong \mathrm{Sp}(2,\mathbb{C})$.
That is, $\M$ is hyperk\"ahler.
Note that,
in this definition,
SD spacetimes are necessarily Ricci-flat
and thus are vacuum
for vanishing cosmological constant.

\subsection{{\Pleb} Coordinates}
\label{REVIEW>M}

In this subsection, we review 
the description of SD spacetimes
in the framework of
Pleba\'nski's second heavenly equation
\cite{plebanski1975some,dunajski2000hyper,Adamo:2021bej,boyer1983geometry,hull1991geometry}.
To expedite our exposition,
we present a quick pathway
rather than deriving the entire construction from general assumptions.

\subsubsection{Self-Dual Spacetime as a Cotangent Bundle}
\label{REVIEW>PLEB}

Suppose a two-dimensional symplectic manifold $\N$ with local coordinates $q^\da$.
Consider its cotangent bundle
$\M = \Tstar\N$
and let $p_\da$ denote the fiber coordinates.
Suppose a scalar field $\Phi \in \Cinfty(\M)$
and denote its derivatives along the fiber direction as
\begin{align}
    \label{pderivs}
    \Phi^\da
    \mem:=\mem
        \frac{\partial \Phi}{\partial p_\da}
    \,,\,\,\,
    \Phi^{\da\db}
    \mem:=\mem
        \frac{\partial^2 \Phi}{\partial p_\da \partial p_\db}
    \,,\,\,\,
    \Phi^{\da\db\dc}
    \mem:=\mem
        \frac{\partial^3 \Phi}{\partial p_\da \partial p_\db \partial p_\dc}
    \,,\,\,\,
    \Phi^{\da\db\dc\dd}
    \mem:=\mem
        \frac{\partial^4 \Phi}{\partial p_\da \partial p_\db \partial p_\dc \partial p_\dd}
    \,,
\end{align}
and so on.
If $\Phi$ satisfies the equation
\begin{align}
    \label{heavenly}
    \Box\mem \Phi
    \,=\,
        \k\,
            \Phi_\wrap{\da\db}\mem \Phi^{\da\db}
    \,,
\end{align}
then the symmetric rank-$2$ tensor field
$h \in \Gamma(\Sym^2\Tstar\M)$,
\begin{align}
    \label{h-in-qp}
    h
    \,=\,
        \Phi_\wrap{\da\db}\, dq^\da \modot dq^\db
    \,,
\end{align}
\newpage\noindent
defines a SD metric $g$ on the four-manifold $\M$
via $g = \eta + 2\k\hem h$.
Here, 
we have introduced a customary constant parameter $\k$
and defined
\begin{align}
    \label{flats}
    \eta \,:=\, 2\mem dp_\da \modot dq^\da
    \,,\quad
    \Box
    \,:=\,
        2\mem \frac{\partial^2}{\partial p_\da \partial q^\da}
    % \,=\,
    %     2\mem \te^{\da\db} \frac{\partial^2}{\partial p^\da \partial q^\db}
    \,.
\end{align}

\eqref{heavenly}
is known as Pleba\'nski's second heavenly equation
\cite{plebanski1975some},
where $\Phi$ is called the {\Pleb} second heavenly scalar.
The set of four functions $x^{\da\a}$,
where
\begin{align}
    \label{xcoords}
    x^{\da0} \,:=\, p^\da
    \,,\quad
    x^{\da1} \,:=\, q^\da
    \,,
\end{align}
provides a local coordinate chart on $\M$.
In this paper,
we wish to refer to these coordinates as 
{\Pleb}'s second heavenly coordinates
or ``{\Pleb} coordinates'' in short.

\smallskip
The above proposition can be shown in at least two ways.
The first method works with differential forms.
The observation is that
a coframe $e^{\da\a}$ can be chosen such that
the metric $g = \eta + 2\k\hem h$
arises as
\begin{align}
    \label{metg}
    g
    \,&=\,
        \e_\wrap{\a\b}\mem \te_\wrap{\da\db}\mem
            e^{\da\a} \modot e^{\db\b}
    \,=\,
        2\mem dp_\da {\:\odot\:} dq^\da
        + 2\hem\k\mem \Phi_\wrap{\da\db}\mem dq^\da \modot dq^\db
    \,.
\end{align}
In particular, we take
\begin{align}
    \label{tet-e}
    e^{\da0} 
    \,=\, 
        dp^\da {\mem+\,} \k\mem \Phi^\da{}_\dc\mem dq^\dc
    \,,\quad
    e^{\da1} 
    \,=\, 
        dq^\da
    \,.
\end{align}
In the meantime,
recall that
the ASD ``area'' two-forms of {\Pleb} 
\cite{Plebanski:1977zz,Capovilla:1991qb}
are defined as
\begin{align}
    \S^{\a\b}
    \,:=\,
        \te_\wrap{\da\db}\mem 
        e^{\da\a} \swedge e^{\db\b}
    \,.
\end{align}
The choice made in \eqref{tet-e} yields
\begin{align}
\begin{split}
    \label{Sigmas-qp}
    \S^{11} 
    \,&=\,
        \te_\wrap{\da\db}\mem dq^\da \swedge dq^\db
    \,,\quad
    \S^{01}
    \,=\,
        dq^\da \swedge dp_\da
    \,,\\
    \S^{00}
    \,&=\,
        \te_\wrap{\da\db}\mem dp^\da \swedge dp^\db
        \mem+\mem
        2\k\mem \Phi_\wrap{\da\db}\,
            dp^\da \swedge dq^\db
        \mem+\mem
        \frac{\k^2}{2}\,
            \Phi_\wrap{\dc\dd}\mem \Phi^{\dc\dd}
        \,
            \te_\wrap{\da\db}\mem dq^\da \swedge dq^\db
    \,.
\end{split}
\end{align}
It is immediate that
$d\S^{11} = d\S^{01} = 0$.
Moreover,
straightforward computation 
using the heavenly equation in \eqref{heavenly}
shows that
$d\S^{00} = 0$
holds as well.
Noting that
the ASD area two-forms are covariantly closed
by the torsion-free condition of the Levi-Civita connection,
one sees that the closure $d\Sigma^{\a\b} = 0$
implies vanishing of the ASD spin connection coefficients
associated with the coframe in \eqref{tet-e}.
This shows that the ASD spinor bundle $S^-\nem\M$ is flat.
On the other hand,
the SD spinor bundle $S^+\nem\M$ 
exhibits curvature
by the spin connection one-form
and its curvature two-form
\begin{align}
    \label{gamma-R}
    \gamma^\da{}_\db
    \,&=\,
        - \k\mem \Phi^\da{}_{\db\dc}\mem
        e^{\dc 1}
    \,,\quad
    R^\da{}_\db
    \,=\,
        \k\mem
        \Phi^\da{}_{\db\dc\dd}\mem
        e^{\dc 0} \swedge e^{\dd 1}
    \,,
\end{align}
which arises by using \eqref{heavenly}.
Clearly, the fourth-order derivative $\Phi_\wrap{\da\db\dc\dd}$
describes the SD Weyl tensor.

% \newpage

The second method works with vector fields.
The observation is that
the frame dual to the coframe in \eqref{tet-e}
is
\begin{align}
    \label{tet-E}
    E_\wrap{0\da}
    \,=\, 
        \frac{\partial}{\partial p^\da}
    \,,\quad
    E_\wrap{1\da}
    \,=\, 
        \frac{\partial}{\partial q^\da}
        \mem-\mem 
        \k\mem \Phi^\dc{}_\da\, \frac{\partial}{\partial p^\dc}
    \,,
\end{align}
which clearly describes a set of divergence-free vector fields:
a unimodular tetrad.
Explicitly, they preserve
the volume form
$\nu = -i\mem dq^\doz \swedge dq^\diz \swedge dp^\doz \swedge dp^\diz$.
Furthermore,
direct computation of Lie brackets shows that
\begin{align}
    \label{EEvanishes}
    \comm{E_\wrap{0\da}}{E_\wrap{0\db}}
    \,=\,
        0
    \,,\quad
    \comm{E_\wrap{0\da}}{E_\wrap{1\db}}
    \mem+\mem
    \comm{E_\wrap{1\da}}{E_\wrap{0\db}}
    \,=\,
        0
    \,,\quad
    \comm{E_\wrap{1\da}}{E_\wrap{1\db}}
    \,=\,
        0
    \,,
\end{align}
on the support of the heavenly equation in \eqref{heavenly}.
As a result, one finds that
the necessary conditions for
a theorem established by Mason and Newman \cite{MasonNewman:1989}
are locally fulfilled,
in which case
the frame in \eqref{tet-E}
defines a SD inverse metric 
$g^{-1} \in \Gamma(\Sym^2T\M)$
by
\begin{align}
    \label{imetg}
    g^{-1}
    \,&=\,
        \e^{\a\b}\mem \te^{\da\db}\mem
            E_{\a\da}
            {\:\odot\:}
            E_{\b\db}
    \,=\,
            2\mem \frac{\partial}{\partial p_\da}            
            {\:\odot\:}
            \frac{\partial}{\partial q^\da}
        - 2\hem \k\mem \Phi_\wrap{\da\db}\,
            \frac{\partial}{\partial p_\da}
            {\:\odot\:}
            \frac{\partial}{\partial p_\db}
    \,.
\end{align}
It is evident that $g^{-1}$ in \eqref{imetg}
is the pointwise inverse of $g$ in \eqref{metg}.

To sum up,
a SD spacetime $\M$ arises as a cotangent bundle $\Tstar\N$,
whose base $\N$ is itself a symplectic manifold
by 
$
    \S^{11} 
    =
        \te_\wrap{\da\db}\mem dq^\da \swedge dq^\db
$.
The soldering is encoded in
$
    \S^{01}
    =
        dq^\da \swedge dp_\da
$.

\medskip

A few remarks are in order.
Firstly,
the second heavenly equation in \eqref{heavenly}
reveals that a single scalar field $\Phi$
dictates the dynamics of four-dimensional gravity in the SD sector.
Notably, the right-hand side of \eqref{heavenly}
describes a doubling of a ``fiberwise Poisson bracket''
on $\Tstar\N$.
% (cf. \rcite{fedosov1994simple}).
This is the basis of 
the manifest double copy correspondence between
SD gravity and SD Yang-Mills (YM) theory
\cite{monteiro2011kinematic}.
See also
\rrcite{plebanski1995moyal,plebanski1996principal,plebanski1996lagrangian,przanowski1999nonlinear,park1990self,castro1993super}
for a related discussion.

Secondly,
it is easily seen from
\eqref{imetg}
and the unimodularity of the tetrad in \eqref{tet-E}
that
the Laplacian operator
due to the curved Riemannian metric in \eqref{metg} is
\begin{align}
    \label{BBox}
    \Box_\Phi
    \,:=\,
        \Box
        \mem-\mem 2\hem\k\mem \Phi_\wrap{\da\db}\mem
            \frac{\partial^2}{\partial p_\da \partial p_\db}
        % \mem-\mem 2\hem\k\mem \Phi^{\da\db}\hem
        %     \frac{\partial^2}{\partial p^\da \partial p^\db}
    \,.
\end{align}
Importantly,
it follows that
linear perturbations $\delta\Phi$ around 
a solution $\Phi$ of the heavenly equation in \eqref{heavenly}
are harmonic
with respect to this curved Laplacian
\cite{dunajski2000hyper}:
$\delta\Phi \in \ker(\square_\Phi\nem)$.

Thirdly,
it should be clear that
{\Pleb}'s second heavenly coordinates
in \eqref{xcoords}
can be locally constructed in a given SD spacetime
as follows.
First, 
trivialize the ASD spinor frame
within a local patch
such that the ASD spin connection coefficients vanish.
Second,
observe that 
the ASD area two-forms $\S^{\a\b}$ are closed,
and there is enough freedom
to bring $\S^{11}$ and $\S^{01}$
into Darboux forms as in \eqref{Sigmas-qp}.
Third,
demand closure on $\S^{00}$
with a choice of a coframe
that is equivalent to that in \eqref{tet-e}
modulo a local SD rotation
and find the heavenly equation in \eqref{heavenly}.
See, e.g., the exposition in \rcite{Adamo:2021bej}.

Fourthly,
on a related note,
the second heavenly coordinates
can be seen as a gauge-fixing for the metric perturbation.
The vanishing of
the ASD spin connection coefficients
endows one with a
constant ASD spinor dyad
$\o^\a \doteq \delta_0{}^\a$,
$\i^\a \doteq \delta_1{}^\a$.
With this understanding,
the metric perturbation in \eqref{h-in-qp} translates to
\begin{align}
    \label{hF}
    h_\wrap{\a\da\b\db}
    \,=\,
        \o_\a \o_\b\mem
        \Phi_\wrap{\da\db}
    \,.
\end{align}
This describes a lightcone gauge
using $\o_\a$ as a constant reference spinor.

\newpage

\subsubsection{Residual Diffeomorphisms and Redundancies}
\label{REVIEW>RESID}

To recapitulate,
the second heavenly coordinates
are defined by the Darboux formatting of the ASD area two-forms,
$
    \S^{11} 
    =
        \te_\wrap{\da\db}\mem dq^\da \swedge dq^\db
$ and $
    \S^{01}
    =
        dq^\da \swedge dp_\da
$.
With the understanding that
this
amounts to
a gauge-fixing for the metric perturbation field,
diffeomorphisms on the SD spacetime $\M$
will be restricted to
those preserving
the structures
$\S^{01}$ and $\S^{11}$,
or equivalently the ``$2+2$ splicing''
$\M \cong \Tstar\N$
by a symplectic base $\N$.

It is not difficult to see that
such restricted diffeomorphisms
describe a subgroup 
$\mathrm{SDiff}(\mathcal{N}) {\mem\ltimes\mem} \Gamma(\Tstar\mathcal{N}) \subset \Diff(\M)$,
spanned by
the uplift of symplectomorphisms on $\N$
and
local shifts of the zero section of $\Tstar\N$.
Explicitly,
\begin{subequations}
\label{Xuv}
\begin{align}
    \label{Xu}
    X_u
    \,&=\,
            - u^{\da}(q)\mem
            \frac{\partial}{\partial q^\da}
            + p_\db\mem 
            u^\db{}_\da(q)\mem
            \frac{\partial}{\partial p_\da}
    \,,\\
    \label{Xv}
    X_v
    \,&=\,
        v_\da(q)\mem
        \frac{\partial}{\partial p_\da}
    \,,
\end{align}
\end{subequations}
where $u,v \in \Cinfty(\N)$
are the parameters.
Here, we have denoted
$u_\da := \partial u/\partial q^\da$,
$u_\wrap{\da\db} := \partial^2 u/\partial q^\da \partial q^\db$,
$v_\da := \partial v/\partial q^\da$,
etc.
It is straightforward to characterize the corresponding Lie algebra explicitly
by computing Lie brackets between vector fields.
% \begin{align}
% \begin{split}
%     % [v^{0\da} \partial_{0\da} + v^{0\da}{}_\dc\hem z^{\dc1} \partial_{1\da} 
%     % , u^{1\db} \partial_{1\db}]
%     % &= - (v^{0\da}{}_\db u^{1\db} - u^{0\da}{}_\db v^{1\db}) \partial_{1\da}
%     % \,,\\
%     % [v^{0\da} \partial_{0\da} + v^{0\da}{}_\dc\hem z^{\dc1} \partial_{1\da} 
%     % , u^{0\db} \partial_{0\db} + u^{0\db}{}_\dd\hem z^{\dd1} \partial_{1\db} ]
%     % &= - 
%     % \,,\\
%     \comm{ X_{u} }{ X_{v} } \,=\, X_{
%         u_\da v^\da
%     }
%     \,,\quad
%     \comm{ X_{u} }{ Y_{v} } \,=\, Y_{
%         u_\da v^\da
%     }
%     \,,\quad
%     \comm{ Y_{u} }{ Y_{v} } \,=\, 0
%     \,.
% \end{split}
% \end{align}

Under
pullback by
the infinitesimal residual diffeomorphisms
described
in \eqref{Xuv},
the coframe in \eqref{tet-e}
transforms as
$e^{\da\a} \mapsto
e^{\da\a} + \pounds_{\k X} e^{\da\a}
=
(\delta^\da{}_\db \mminus \k\hem u^\da{}_\db)\mem e'^{\db\a}$,
where 
$(\delta^\da{}_\db \mminus \k\hem u^\da{}_\db)$
can be undone by a local SD rotation
while \smash{$e'^{\da\a} = dx^{\da\a} + \k\mem \Phi'^\da{}_\wrap{\db}\mem dx^{\db\b} \o_\b$}
indicates a change in
the {\Pleb} scalar
$\Phi' = \Phi + \delta\Phi$.
Explicitly, one finds
\begin{subequations}
\label{Phivar.uv}
\begin{align}
\label{Phivar.u}
    \delta_u \Phi
    \,&=\,
        \frac{1}{6}\,
            p_\da\mem p_\db\mem p_\dc\mem u^{\da\db\dc}(q)
        + \k\hem X_u\act{ \Phi }
    \,,\\
\label{Phivar.v}
    \delta_v \Phi
    \,&=\,
        \frac{1}{2}\,
            p_\da\mem p_\db\mem v^{\da\db}(q)
        + \k\hem X_v\act{ \Phi }
    \,.
\end{align}
\end{subequations}
Direct computation verifies that
the infinitesimal variations in \eqref{Phivar.uv}
are elements of $\ker(\Box_\Phi)$,
so
they indeed leave
the heavenly equation in \eqref{heavenly}
invariant.

The heavenly equation is also left invariant by
the shifts
\begin{align}
    \label{Phivar.ab}
    \delta_a \Phi
    \,=\,
        a(q)
    \,,\quad
    \delta_b \Phi
    \,=\,
        p_\da\mem b^\da(q)
    \,,
\end{align}
where $a,b \in \Cinfty(\N)$.
The variations in \eqref{Phivar.ab}
are elements of
not only $\ker(\Box_\Phi)$ 
but also $\ker(\Box)$,
so in this case
each side of the heavenly equation in \eqref{heavenly}
is separately left invariant.
Note that \eqref{Phivar.ab}
is reminiscent to the symmetry of Galileons \cite{hinterbichler2015hidden}.

Two solutions 
to the heavenly equation
that can be mapped to each other
via \eqrefs{Phivar.uv}{Phivar.ab}
should be identified as the same solution.
The transformation in \eqref{Phivar.u}
gives spacetime geometries
related by a coordinate transformation
and a SD local Lorentz transformation.
The transformation in \eqref{Phivar.v}
gives spacetime geometries
related by a coordinate transformation.
The transformations in \eqref{Phivar.ab}
% simply gives the same coframe
simply preserve the coframe
$e^{\da\a} = dx^{\da\a} + \k\mem \Phi^\da{}_\dc\mem dx^{\dc\c} \o_\c$.
% on the same spacetime.
In these senses,
the transformations in
\eqrefs{Phivar.uv}{Phivar.ab}
will be identified as
gauge redundancies of the heavenly equation in \eqref{heavenly}.
See \rcite{dunajski2000hyper}
for a further discussion.

% \newpage

\subsection{Curved Twistor Space}
\label{REVIEW>T}

\subsubsection[$\a$-Surfaces]{$\bm{\a}$-Surfaces}
\label{REVIEW>SRF}

The {\Pleb} description of SD spacetimes
employs a distinguished ASD direction $\o_\a$.
In an alternative approach,
one treats all ASD directions on an equal footing
by means of an auxiliary ASD spinor variable $\l_\a \inn \C^2$,
which will be eventually thought of as projective coordinates on a Riemann sphere $\CP^1$.

Again, there are two versions of this construction,
providing a dual view:
a differential-forms version,
due to 
Pleba\'nski and Gindikin \cite{plebanski1975some,gindikin1986construction},
and 
a vector-fields version,
due to 
Mason and Newman \cite{MasonNewman:1989,chakravarty1991canonical,chacon2019self}.

\medskip

The first version states the following.
Let $\M$ be a complex four-manifold 
equipped with a nondegenerate holomorphic four-form $\nu$.
Suppose
$\M$ admits a set of holomorphic two-forms $\S^{\a\b}$
such that
\begin{align}
    \label{S(l)}
    d\S(\lambda) \,=\, 0
    \:\:\:\:\text{and}\:\:\:\:
    \S(\lambda) \wedge \S(\lambda)
    \,=\, 0
    \,\quad
    \forall\mem \lambda_\a \in \C^2
    \,,
\end{align}
where $\S(\lambda) := \lambda_\a\lambda_\b \hem\S^{\a\b}$.
According to
Pleba\'nski and Gindikin \cite{plebanski1975some,gindikin1986construction},
\eqref{S(l)}
implies the existence of a coframe $e^{\da\a}$ 
such that
$\S^{\a\b} = \te_\wrap{\da\db}\mem e^{\da\a} \swedge e^{\db\b}$,
whose associated metric\footnote{
    Note that this metric can also be directly obtained from $\S^{\a\b}$
    via the Urbantke formula
    \cite{Urbantke:1984eb,Capovilla:1991qb}.
} 
$
    g
    =
        \e_\wrap{\a\b}\mem \te_\wrap{\da\db}\mem
            e^{\da\a} {\:\odot\:} e^{\db\b}
$
defines a SD Riemannian structure on $\M$.

The second version states the following.
Let $\M$ be a complex four-manifold,\footnote{
    Mason and Newman \cite{MasonNewman:1989} aimed to generalize
    an earlier result by
    Ashtekar, Jacobson, and Smolin
    \cite{Ashtekar:1987qx},
    so
    it was assumed that $\M$ is topologically
    a product of a three-manifold and a line.
    Here, we work locally.
}
equipped with a nondegenerate holomorphic four-form $\nu$.
Let $V_{\a\da}$ be four linearly independent holomorphic vector fields on $\M$.
Consider vector fields $L_\da := \lambda^\a\hem V_{\a\da}$ for $\lambda_\a\in\mathbb{C}^2$.
Suppose they satisfy
\begin{align}
    \label{L}
    [L_\da,L_\db] \,=\, 0
    \:\:\:\:\text{and}\:\:\:\:
    \pounds_{L_\da} \nu \,=\, 0
    \qquad
    \forall\mem \lambda_\a \in \C^2
    % \,,
    \,.
\end{align}
According to Mason and Newman \cite{MasonNewman:1989},
\eqref{L} implies that
the vector fields 
$E_{\a\da} = f^{-1}\hem V_{\a\da}$
form
an orthonormal frame
that defines
a SD Riemannian structure on $\M$,
where $f^2 = i\mem \nu(V_\wrap{0\doz},V_\wrap{0\diz},V_\wrap{1\doz},V_\wrap{1\diz})$.
Conversely,
an SD spacetime $(\M,g)$
locally admits an orthonormal frame $E_{\a\da}$
and a nonvanishing function $f$
such that
$L_\da = f\l^\a\hem E_{\a\da}$
satisfy \eqref{L}
for some volume form $\nu$.

\medskip

The above two approaches
are related via
$\te^{\da\db} (\i_{L_\da} \i_{L_\db} \hem\nu) = i\mem \S(\lambda)$
and 
$\i_{L_\da} \S(\lambda) = 0$,
while
the metric volume form is
$\ve = f^2\mem \nu$.
More explicitly,
\begin{align}
    \pounds_{L_\da} \hnem=\hem 0
    \qiq
    3i\mem \partial_\wrap{[\n} \S(\l)_\wrap{\r\s]}
    \,=\,
        \te^{\da\db}\hem
        \comm{L_\da}{L_\db}^\m\,
        \nu_{\m\n\r\s}
    \,.
\end{align}
For the sake of simplicity,
below we wish to presume cases in which
the scaling factor is trivial as $f = 1$,
so $L_\da = \l^\a\hem E_{\a\da}$.
This holds
in the second heavenly gauge,
in particular.

Crucially,
the vanishing Lie bracket
$[L_\da,L_\db] = 0$
in \eqref{L}
encodes the integrability of translations along SD null directions
in $\M$
for each fixed $\l_\a$,
upon a trivialization of the ASD spinor bundle.
Such translations form
SD null surfaces,
known as $\a$-surfaces.
The space of $\a$-surfaces
is the twistor space $\PT$,
which is a complex three-manifold.

\subsubsection{Double Fibration}
\label{REVIEW>DFIB}

To establish the precise relation between
the twistor space $\PT$ and the spacetime $\M$,
one employs the correspondence space.
This simply means to promote the auxiliary spinor variable $\l_\a$ to coordinates of an extended space.

The correspondence space
$\F \eqq {\P S^-\nem\M}$
is the projectivization 
of
the ASD spinor bundle.
This removes the zero section of $S^-\nem\M$
and imposes the identification
$\lambda_\a \sim r\mem \lambda_\a$
for $r \in \C^*$.
A function $f(x,\lambda)$ on $S^-\nem\M$
homogeneous in $\lambda_\a$
defines a function on $\F$.
It descends to the twistor space
$\PT$ iff $L_\da\act{
    f(x,\lambda)
} = 0$:
constancy along the $\a$-surfaces.
Here, $L_\da = \l^\a\hem E_{\a\da} \in \Gamma(T\F)$
is understood as a vector field on $\F$.

The Gindikin two-form
in \eqref{S(l)}
can be brought to a Darboux form as 
$\S(\lambda) = \te_{\smash{\da\db}\vphantom{\b}}$ $d_x F^\da\hnem(x,\lambda) \swedge d_x F^\db\hnem(x,\lambda)$,
on account of its closure.
Here, $d_x$ denotes the 
``partial'' exterior derivative 
% that discards $d\lambda_\a$.
holding $\lambda_\a$ constant.
Consequently, 
the earlier equation
$\i_{L_\dc} \S(\lambda) = 0$ implies
\begin{align}
    \label{LF}
    L_\dc\act{
        F^\da\hnem(x,\lambda)
    } \,=\, 0
    \,,
\end{align}
so $F^\da(x,\lambda)$ 
gives a function on $\F$
(with homogeneity one in $\lambda_\a$)
that descends to $\PT$.

The incidence relation
is
the equation
\begin{align}
    \label{incidence}
    \mu^\da
    \,=\,
        F^{\da}(x,\lambda)
    \,.
\end{align}
For each fixed $\lambda_\a$ and $\mu^\da$,
the locus of \eqref{incidence}
is an $\a$-surface in $\M$.
It follows that the twistor space $\PT$
is the space of $[Z_\rmA]$,
where $Z_\rmA = (\lambda_\a,\mu^\da)$.
Here, $[Z_\rmA]$ denotes the equivalence class of $Z_\rmA$
due to projective rescalings
$Z_\rmA \sim r\mem Z_\rmA$
for $r \in \C^*$.
In other words,
the pair $Z_\rmA = (\lambda_\a,\mu^\da)$
provides
homogeneous coordinates for the twistor space $\PT$.

\medskip
The incidence relation in \eqref{incidence}
is formalized as
the map $\pi_1 : \F \to \PT : (x^\m,[\lambda_\a]) \mapsto [(\lambda_\a,F^\da(x,\lambda))]$.
This establishes
the double fibration,
shown below.
\begin{align}
\label{fibration}
\adjustbox{valign=c}{\begin{tikzpicture}
    \node[b] (o) at (0,0) {};
    \node[b] (a) at (-1.5, -1.5) {};
    \node[b] (b) at ( 1.5, -1.5) {};
    %%%
    \node[b] (bundle) at ($(o)$) {\vphantom{|}};
    \node[b] (Bundle) at ($(bundle)$) {$\mathclap{
        \:\:
        \F
    }\phantom{|}$};
    \node[b] (base a) at ($(Bundle)+(a)$) {\vphantom{|}};
    \node[b] (Base a) at ($(base a)$) {$
    \mathclap{
        \PT
    }\phantom{|}$};
    \node[b] (base b) at ($(Bundle)+(b)$) {\vphantom{|}};
    \node[b] (Base b) at ($(base b)$) {$\mathclap{
        \M
    }\phantom{|}$};
    %%%
    \draw[->] (bundle)--(base a) node[midway,above,pos=0.62] 
    {\scriptsize $\pi_1\:\:$};
    \draw[->] (bundle)--(base b) node[midway,above,pos=0.62] {\scriptsize $\,\,\pi_2$};
\end{tikzpicture}}
\end{align}
The map
$\pi_2: \F \to \M : (x^\m,[\lambda_\a]) \mapsto x^\m$
is the bundle projection.
The double fibration facilitates a precise discussion on the relationship between spacetime and its twistor space.

First,
for each twistor $[Z] \in \PT$,
the corresponding $\a$-surface in spacetime is
\begin{align}
    \mathcal{S}_{[Z]}
    \,=\,
    \pi_2(\pi_1^{-1}([Z]))
    \,\subset\,
        \M
    \,.
\end{align}
Second,
for each spacetime point $x \inn \M$,
\begin{align}
    \label{twistorline}
    \LL_x
    \,:=\,
        \pi_1(\pi_2^{-1}(x))
    \,\subset\,
    \PT
\end{align}
defines a holomorphically embedded $\CP^1$ in the twistor space $\PT$,
known as the twistor line $\LL_x$.
The incidence relation in \eqref{incidence}
gives an explicit parameterization of the twistor line:
$[Z_\rmA] = [(\l_\a,F^\da(x,\l))]$
for $\l_\a \in \C^2$.

It follows that
\begin{align}
    [Z] \,\in\, \LL_x \,\subset\, \PT
    \qfq
        x \,\in\, \mathcal{S}_{[Z]} \,\subset\, \M
    \,,
\end{align}
which describes a duality between twistor space and spacetime.
An $\a$-surface $\mathcal{S}_{[Z]}$ is incident at a spacetime point $x$
iff
the twistor $[Z]$ lies on the twistor line $\LL_x$.
The vivid geometrical picture behind this statement
is that
the twistor line $\LL_x$
can be taken as
the lightcone---%
as the space $\CP^1 \cong \mathrm{S}^2$ of null directions, i.e., \textit{sky}---%
at $x$.
A twistor $[Z]$ appears on the sky $\LL_x$
for observer at $x$
iff $x$ lies on the pathway of the light ray $[Z]$.

In sum,
the twistor space $\PT$
of a self-dual spacetime $\M$ 
is a three-manifold
where each point $[Z_\rmA]$ corresponds to 
an $\a$-surface in $\M$.
The double fibration in \eqref{fibration}
explicates the precise relationship between
$\PT$ and $\M$
via the correspondence space $\F = \P S^-\nem\M$.
The twistor line $\LL_x$
is the twistor-space counterpart
of the spacetime point $x \in \M$,
whose precise definition 
arises via the correspondence space $\F$
as in \eqref{twistorline}.

\subsubsection{Deformed Incidence Relation}
\label{REVIEW>DINC}

It is instructive to describe
the twistor space of
flat spacetime.
Flat spacetime means
the complexified Minkowski space
$\mflat$,
which
is the linear space $\C^4$
equipped with a holomorphic flat metric $\eta$.
% Let $x^\m$ be its Cartesian coordinates.
% Let the lightcone coordinates be
% \begin{align}
%     x^{\da\a}
%     \,&\doteq\,
%         \frac{1}{\sqrt{2}}\mem
%         \TwoByTwoMatrixWide{
%             x^0 \mplus x^3
%         }{
%             x^1 \mminus ix^2
%         }{
%             x^1 \mplus ix^2
%         }{
%             x^0 \mminus x^3
%         }
%     \,.
% \end{align}
Let $x^\m = (t,x,y,z)$ be its Cartesian coordinates.
Let the lightcone coordinates be
\begin{align}
    x^{\da\a}
    \,&\doteq\,
        \frac{1}{\sqrt{2}}\mem
        \TwoByTwoMatrixMed{
            t + z
        }{
            x - iy
        }{
            x + iy
        }{
            t - z
        }
    % \,=\,
    %     \TwoByTwoMatrix{
    %         p^\doz
    %     }{
    %         q^\doz
    %     }{
    %         p^\diz
    %     }{
    %         q^\diz
    %     }
    \,.
\end{align}
The Lorentzian real section
is given by $[x^{\da\a}]^* = x^{\da\a}$,
on which the metric exhibits signature $(+,-,-,-)$
while
the metric volume form is oriented such that
$\ve_{0123} = +1$.
$\a$-planes in flat spacetime $\mflat$
are given by 
$x^{\da\a} \sim x^{\da\a} + y^\da\hem \lambda^\a$
for fixed $\lambda^\a$
and variable $y^\da$,
along which
\begin{align}
    \label{incidence.flat}
    \mu^\da \,=\, x^{\da\a}\hem \lambda_\a
\end{align}
are constant.
\eqref{incidence.flat} is the incidence relation for flat spacetime,
so the twistor space is
\begin{align}
    \label{pt}
    \pt
    \,=\,
        \Big\{\mem{
            [(\lambda_\a , \mu^\da)] \in \CP^3
        \,\Big|\,
            \lambda_\a \neq 0
        }\mem\Big\}
    \,.
\end{align}
The double fibration is given as follows,
where $\mathbb{F} = \P\hnem S^-\mflat$.
\begin{align}
\label{fibration.flat}
\adjustbox{valign=c}{\begin{tikzpicture}
    \node[b] (o) at (0,0) {};
    \node[b] (a) at (-1.5, -1.5) {};
    \node[b] (b) at ( 1.5, -1.5) {};
    %%%
    \node[b] (bundle) at ($(o)$) {\vphantom{|}};
    \node[b] (Bundle) at ($(bundle)$) {$\mathclap{
        \:\:
        \mathbb{F}
    }\phantom{|}$};
    \node[b] (base a) at ($(Bundle)+(a)$) {\vphantom{|}};
    \node[b] (Base a) at ($(base a)$) {$
    \mathclap{
        \pt
    }\phantom{|}$};
    \node[b] (base b) at ($(Bundle)+(b)$) {\vphantom{|}};
    \node[b] (Base b) at ($(base b)$) {$\mathclap{
        \mflat
    }\phantom{|}$};
    %%%
    \draw[->] (bundle)--(base a) node[midway,above,pos=0.62] 
    {\scriptsize $\pi_1\:\:$};
    \draw[->] (bundle)--(base b) node[midway,above,pos=0.62] {\scriptsize $\,\,\pi_2$};
\end{tikzpicture}}
\end{align}
Functions on the flat twistor space $\pt$
are transcribed to the correspondence space $\mathbb{F}$
by the pullback of
$\pi_1 : (x^{\da\a},[\lambda_\a]) \mapsto [(\lambda_\a,x^{\da\a}\lambda_\a)]$.
Conversely,
a function on $\mathbb{F}$
descends to $\pt$
iff it is annihilated by
$\lambda^\a\mem \partial_{\a\da}$.

Upon introduction of the spacetime curvature,
the incidence relation in \eqref{incidence.flat}
is deformed into the generic form in \eqref{incidence}.
In this sense, \eqref{incidence}
may be referred to as 
``deformed incidence relation''
for an explicit disambiguation,
following 
Mason and Skinner
\cite{GravityMHVTwistors}.
In the same context,
``curved twistor space'' will refer to $\PT$,
while ``flat twistor space'' refers to the undeformed $\pt$.

\newpage

The celebrated nonlinear graviton construction of Penrose \cite{Penrose:1976js} states that,
the curved twistor space $\PT$ of a SD spacetime $\M$ 
locally arises by
deforming the complex structure of 
flat twistor space $\pt$.
Such deformations utilize
the holomorphic Poisson structure in \eqref{I}.
In this way,
a one-to-one correspondence arises between
SD, Ricci-flat Riemannian metrics
and
curved twistor spaces
\cite{Penrose:1976js,Atiyah:1978wi}.

\subsection{Dunajski-Mason Recursion}
\label{REVIEW>DM}

A concrete understanding on the deformed incidence relation
is facilitated by 
a recursion operator
introduced by Dunajski and Mason \cite{dunajski2000hyper,dunajski1996heavenly,dunajski1997recursion}.
This analysis
examines the explicit content of
the deformed incidence relation
in Pleba\'nski's coordinates.

Intuitively speaking,
the idea is to understand
the deformed incidence relation $\m^\da = F^\da(x,\l)$,
which treats all ASD directions $\l_\a$ equally,
by expanding around
the fixed reference ASD direction $\o_\a$.
Said in another way,
the goal is to obtain
an explicit ``lightcone gauge'' expression
for the deformed incidence relation.

We recall the setup in \Sec{REVIEW>PLEB}.
Let $x^{\da0} \eqq p^\da$, $x^{\da1} \eqq q^\da$
be {\Pleb}'s second heavenly coordinates
on a SD spacetime $(\M,g)$,
and let $\Phi$ be the {\Pleb} scalar.
The orthonormal frame $E_{\a\da}$ is given in \eqref{tet-E}.
The curved Laplacian $\Box_\Phi$ is shown in \eqref{BBox}.
The Mason-Newman \cite{MasonNewman:1989}
self-duality equations in \eqref{EEvanishes}
is equivalent to
$\comm{L_\da}{L_\db} = 0$
for all $\l_\a \in \C^2$,
where $L_\da = \l^\a\hem E_{\a\da}$
form a Lax pair.

\subsubsection{Recursion Operator}
\label{REVIEW>DM>OP}

% We recall the setup in \Sec{REVIEW>PLEB}.
The starting point is to consider
a (formal) operator $\Rec : \ker(\Box_\Phi\nem) \to \ker(\Box_\Phi\nem)$ such that
\begin{align}
    \label{Rec}
    E_{0\da}\act{ \Rec\act{ \psi } }
    \,=\,
        E_{1\da}\act{ \psi }
    \,,
\end{align}
where $\psi$ denotes an element of $\ker(\Box_\Phi)$.
It is easy to check that
\begin{align}
    \psi \,\in\, \ker(\Box_\Phi)
    \qiq
        \Rec\act{\psi} \,\in\, \ker(\Box_\Phi)
    \,,
\end{align}
which uses
$\Box_\Phi\act{\psi} = 2\mem \te^{\da\db} E_\wrap{0\da}\act{ E_\wrap{1\db}\act{ \psi } }$
(cf. \eqref{imetg})
and 
$
    \comm{E_\wrap{0\da}}{E_\wrap{0\db}}
    = 0
$
from \eqref{EEvanishes}.
Similarly,
the existence of $\Rec\act{\psi}$
can be seen from
acting on
$E_1{}^\da$
to both sides
of \eqref{Rec}.
Of course, 
it should be understood that
the image $\Rec\act{\psi}$ is not unique
and is subject to the ambiguities of choosing a primitive for inverting $E_{0\da} = \partial/\partial p^\da$.

% 
% In particular, note that
% \begin{align}
%     \psi \in \ker(\Box_\Phi)
%     \qiq
%     \Box_\Phi\act{
%         \Rec\act{
%             \psi
%         }
%     }
%     \,=\,
%         2\mem \te^{\da\db}
%             E_{1\da}\act{
%                 E_{0\db}\act{
%                     \Rec\act{
%                         \psi
%                     }
%                 }
%             }
%     \,=\,
%         2\mem \te^{\da\db}
%             E_{1\da}\act{
%                 E_{1\db}\act{
%                     \psi
%                 }
%             }
%     \,=\,
%         0
% \end{align}

With the above definition of $\Rec$,
% With the definition of $\Rec$ in \eqref{Rec},
it follows that
\begin{align}
    \label{resol}
    L_\da\act{ \psi }
    \,=\,
        \lambda_0\,
        E_{0\da}\act{
            (\l_1/\l_0 \mminus \Rec)\act{
                \psi
            }
        }
    \,.
\end{align}
Crucially, \eqref{resol} implies that
a harmonic function 
$\psi \in \ker(\Box_\Phi)$
on $\M$
can be promoted to a function $\hat{\psi}$ on the correspondence space $\F = \P S^-\nem\M$
as a Laurent series in $\zeta$,
which descends to the curved twistor space $\PT$:
\begin{align}
    \label{Laurent}
    \hat{\psi}(x,\zeta)
    \,= 
        \sum_{n=-\infty}^{\infty} 
        % \frac{1}{\zeta^n}\,
        \zet{n}\,
            \Rec^n\act{
                \psi(x)
            }
    \qiq
    L_\da\act{
        \hat{\psi}(x,\zeta)
    } 
        \,=\, 0
    \,.
\end{align}
The operator $\Rec$ is referred to as 
the DM recursion operator.

% \subsubsection{Deformed Incidence Relation in {\Pleb} Coordinates}
\subsubsection{Recursive Construction of Deformed Incidence Relation}
\label{REVIEW>DM>CONS}

Next,
observe the following facts
about the {\Pleb} coordinates $q^\da$:
\begin{align}
    E_{0\dc}\act{
        q^\da
    }
    \,=\,
        0
    \,,\quad
    \Box_\Phi\act{
        q^\da
    }
    \,=\,
        0
    \,.
\end{align}
The first equation implies that
the deformed incidence relation for $\lambda^\a = \o^\a \doteq \delta_0{}^\a$
is given by
$F^\da(x,\o) = q^\da$.
In the meantime,
the second equation 
implies that $q^\da$
is a harmonic function on $\M$
so that the recursion operator $\Rec$
can be acted on.
This means that
$q^\da$ can be lifted
to a function on $\PT$
through the Laurent series in \eqref{Laurent}.

Multiplying
this Laurent series
% associated with $q^\da$
by $\lambda_1$
gives
a function
that has homogeneity one in $\lambda$.
By setting $\Rec^n\act{ q^\da } = 0$
for $n<0$,
this function is found as
$
    q^\da\hem\lambda_1
    + p^\da\hem \lambda_0
    + \k\mem \Phi^\da\mem {\lambda_0}^2\nem/\lambda_1
    + \cdots
$,
which indeed 
reproduces the flat incidence relation in \eqref{incidence.flat}
in the limit $\k \to 0$.
In this manner,
the deformed incidence relation
$F^\da(x,\lambda)$
is
concretely constructed in Pleba\'nski's 
second heavenly
coordinate system
by ``expanding around $\lambda_\a \propto \o_\a$.''

To elaborate,
\eqref{Rec}
translates to
% \begin{align}
%     \frac{\partial}{\partial p_\da}\BB{
%         \Rec\act{\psi}
%     }
%     \,=\,
%     \bb{
%         \te^{\db\da}\mem \frac{\partial}{\partial q^\db}
%         -\k\mem \Phi^{\da\db}\mem \e_\wrap{\db\dc}\mem
%             \frac{\partial}{\partial p_\dc}
%     }\mem \psi
%     % \qiq
%     % \ve_{\da\db}
%     % \frac{\partial}{\partial p_\db}
%     % \frac{\partial}{\partial p_\da}\BB{
%     %     \Rec\act{\psi}
%     % }
%     % \,=\,
%     %     \frac{1}{2}\, \Box_\Phi \psi
%     % \,\,\implies\,\,
%     % \frac{\partial}{\partial p^\da}
%     % \frac{\partial}{\partial p_\da}\BB{
%     %     \Rec\act{\psi}
%     % }
%     % \,=\,
%     %     -\frac{1}{2}\, \Box_\Phi \psi
%     % \,.
%     % \,,
%     \,,
% \end{align}
\begin{align}
    \label{rec-unpacked}
    \frac{\partial}{\partial p_\dc}\BB{
        \Rec\act{\psi}
    }
    \,=\,
    \bb{
        \te{}^{\db\dc}\mem \frac{\partial}{\partial q^\db}
        + \k\mem \Phi^\dc{}_\wrap{\db}\mem
            \frac{\partial}{\partial p_\db}
    }\mem \psi
    \,,
\end{align}
where both sides are annihilated by $\partial/\partial p^\dc$.
It follows that
\begin{align}
\begin{split}
    \frac{\partial}{\partial p_\dc}\BB{
        p^\da
    }
    \,&=\,
        \delta^\da{}_\dc
    \,=\,
    \bb{
        \te{}^{\db\dc}\mem \frac{\partial}{\partial q^\db}
        + \k\mem \Phi^\dc{}_\wrap{\db}\mem
            \frac{\partial}{\partial p_\db}
    }\mem 
        q^\da
    \,,\\
    \frac{\partial}{\partial p_\dc}\BB{
        \k\hem \Phi^\da
    }
    \,&=\,
        \k\hem \Phi^{\da\dc}
    \,=\,
    \bb{
        \te{}^{\db\dc}\mem \frac{\partial}{\partial q^\db}
        + \k\mem \Phi^\dc{}_\wrap{\db}\mem
            \frac{\partial}{\partial p_\db}
    }\mem 
        p^\da
    \,,\\
    \frac{\partial}{\partial p_\dc}\bb{
        \k\mem
            \te^{\db\da}\mem
            \frac{\partial \Phi}{\partial q^\db}
    }
    \,&=\,
        % \k\mem
        %     \te^{\db\da}\mem
        %     \frac{\partial \Phi^\dc}{\partial q^\db}
        % -\k\mem
        %     \te^{\da\dc}\mem
        %     \frac{\partial \Phi^\db}{\partial q^\db}
        -\k\mem
            \te^{\dc\db}\mem
            \frac{\partial \Phi^\da}{\partial q^\db}
        - \frac{\k}{2}\, \te^{\da\dc}\mem
            \Box\hem \Phi
    \,=\,
    \bb{
        \te{}^{\db\dc}\mem \frac{\partial}{\partial q^\db}
        + \k\mem \Phi^\dc{}_\wrap{\db}\mem
            \frac{\partial}{\partial p_\db}
    }\mem 
        \k\hem \Phi^\da
    \,,
\end{split}
\end{align}
where the last line uses
the Schouten identity and
the heavenly equation in \eqref{heavenly}.
In this way,
one obtains
\cite{dunajski2000hyper}
\begin{align}
    \label{deformed-incidence}
    F^{\da}\hnem(x,\lambda)
    \,=\,
        x^{\da\a}\hem \lambda_\a
        \mem+\mem \k\mem \Phi^\da
            \, \frac{{\lambda_0}^2}{\lambda_1}
            % \,\frac{\lp{\o\lambda}^2}{\lp{\i\lambda}}
        \mem+\mem \k\mem
            \te^{\db\da}\mem
            \frac{\partial \Phi}{\partial q^\db}
            \, \frac{{\lambda_0}^3}{{\lambda_1}^2}
            % \,\frac{\lp{\o\lambda}^3}{\lp{\i\lambda}^2}
        \,+\, 
        % \O(\k^2)
        \cdots
    \,,
\end{align}
Of course,
this expression is valid in the $\lambda_1 \neq 0$ patch of $\CP^1$.
% while $\o^\a \doteq \delta_0{}^\a$ and $\i^\a \doteq \delta_1{}^\a$.
% We have denoted $\lp{\a\b} := \a^\a \b_\a$.

The physical interpretation of \eqref{deformed-incidence} should be clear:
in the lightcone gauge
by the reference $\o^\a$,
the null geodesics
along the directions $\tpi^\da \o^\a$
for any $\tpi^\da$
remain intact 
as straight lines
under gravitational perturbations,
while deviations
exhibited by null geodesics with generic $\lambda^\a \neq \o^\a$
are perturbatively computed
in the orders of $\l_0/\l_1 = \lp{\o\l}/\lp{\i\l}$.
% and the gravitational coupling $\k$.

To obtain concrete expressions for
% the $\O(\k^2)$ terms omitted 
the omitted higher-order terms
in \eqref{deformed-incidence}
for generic configurations of $\Phi$,
one can
utilize a homotopy operator
in the $p$-direction.
% due to an inherent nonlocality of $\Rec$:
% the differential operator $E_{1\da}$
% in \eqref{Rec}
% has to be inverted.
In this paper,
we simply specialize in specific examples
when proceeding into higher orders.

% \medskip
The formula in \eqref{deformed-incidence}
provides an explicit expression for the deformed incidence relation
when the spacetime $\M$ is described with {\Pleb}'s second heavenly coordinates.
In a sense, {\Pleb}'s coordinates
are the closest cousins of
the null Cartesian coordinates
in curved SD spacetimes.

\subsubsection{Canonical Transformations on Fibers}
\label{REVIEW>DM>CAN}

The curved twistor space $\PT$
fibers over $\CP^1$
by $\pi_0 : \PT \to \CP^1 : [(\lambda_\a,\mu^\da)] \mapsto [\lambda_\a]$,
in which case
each fiber is equipped with 
the symplectic form 
$\te_\wrap{\da\db}\mem d\mu^\da \swedge d\mu^\db$
as the counterpart of the Gindikin two-form
$\S(\lambda) = \lambda_\a\lambda_\b\hem \S^{\a\b}$
on $\F$.
Accordingly,
each fiber of $\PT$
with respect to $\pi_0$
is equipped with
the holomorphic Poisson structure
\begin{align}
    \label{I}
    I
    \,=\,
        \te^{\da\db}\mem
        \frac{\partial}{\partial \m^\da}
        \mwedge
        \frac{\partial}{\partial \m^\db}
    \,.
\end{align}

Given the incidence relation $\m^\da = F^\da(x,\l)$ in \eqref{incidence},
the pullback of the symplectic form
$\te_\wrap{\da\db}\mem d\mu^\da \swedge d\mu^\db$
by $\pi_1$
gives
$\te_\wrap{\da\db}\mem dF^\da(x,\l) \swedge dF^\db(x,\l)$.
By identifying its spacetime part
as the Gindikin two-form,
$\te_{\smash{\da\db}\vphantom{\b}}\mem d_x F^\da\hnem(x,\lambda) \swedge d_x F^\db\hnem(x,\lambda) = \S(\lambda)$,
one extracts the {\Pleb} ASD area two-forms $\S^{\a\b}$
and reconstructs the SD spacetime metric $g$,
either 
directly by the Urbantke formula
\cite{Urbantke:1984eb,Capovilla:1991qb}
or
through deducing an orthonormal coframe
$e^{\da\a}$
via
$d_x F^\da\hnem(x,\lambda) = H^\da{}_\wrap{\db}(x,\lambda)\mem e^{\db\b} \l_\b$,
where 
$H^\da{}_\wrap{\db}(x,\l)$
is an $\SL(2,\C)$-valued function
of homogeneity zero in $\l$.
In this way,
one reconstructs spacetime geometry
from the twistor lines.

More specifically,
from a given deformed incidence relation $\m^\da = F^\da(x,\l)$,
the SD spacetime
is reconstructed
in the second heavenly description
as follows
(see, e.g., \rcite{Adamo:2021bej}).
First of all,
the expansion of
$F^\da(x,\lambda)$ around
$\l_0 \eqq 0$
will take the form
$F^\da(x,\lambda)
=
    q^\da(x)\hem\lambda_1
    + p^\da(x)\hem \lambda_0
    + \phi^\da(x)\, {\lambda_0}^2\nem/\lambda_1
    + \cdots
$.
Certainly,
$q^\da(x)$ and $p^\da(x)$
extracted in this way
are none other than the second heavenly coordinates.
The {\Pleb} scalar and its heavenly equation
are then deduced by
demanding regularity of
$\te_\wrap{\da\db}\mem d_xF^\da(x,\lambda) \swedge d_xF^\db(x,\lambda) = \Sigma(\lambda)$
at $\l_1 \eqq 0$
as well as its closure.
In a similar fashion,
the orthonormal coframe $e^{\da\a}$
in \eqref{tet-e}
arises as
$d_x F^\da(x,\l) = \smash{\bigbig{
    \delta^\da{}_\db 
    - \k\hem \Phi^\da{}_\db\mem (\l_0/\l_1)
    + \cdots
}}\mem e^{\db\b} \l_\b$.

% More specifically,
% from a given deformed incidence relation $\m^\da = F^\da(x,\l)$,
% the SD spacetime
% is reconstructed
% in the second heavenly description
% as follows
% (see, e.g., \rcite{Adamo:2021bej}).
% First of all,
% the expansion of
% $F^\da(x,\lambda)$ around
% $\l_0 \eqq 0$
% will take the form
% $F^\da(x,\lambda)
% =
%     q^\da(x)\hem\lambda_1
%     + p^\da(x)\hem \lambda_0
%     + \phi^\da(x)\, {\lambda_0}^2\nem/\lambda_1
%     + \cdots
% $,
% % where $\Phi^\da(x)$ is a generic spinor field on $\M$
% % for the moment.
% where
% $q^\da(x)$, $p^\da(x)$, and $\phi^\da(x)$
% are spinor fields on $\M$.
% Certainly,
% $q^\da(x)$ and $p^\da(x)$
% extracted in this way
% are none other than the second heavenly coordinates.
% The {\Pleb} scalar and its heavenly equation
% are then deduced by
% demanding regularity of
% $\te_\wrap{\da\db}\mem d_xF^\da(x,\lambda) \swedge d_xF^\db(x,\lambda) = \Sigma(\lambda)$
% at $\l_1 \eqq 0$
% as well as its closure.
% In a similar fashion,
% the orthonormal coframe $e^{\da\a}$
% in \eqref{tet-e}
% arises as
% $d_x F^\da(x,\l) = $ $\smash{\bigbig{
%     \delta^\da{}_\db 
%     - \k\hem \Phi^\da{}_\db\mem (\l_0/\l_1)
%     + \cdots
% }}\mem e^{\db\b} \l_\b$
% by working
% in the expansion around $\l_0 = 0$.

Lastly, 
the explicit formula in \eqref{deformed-incidence}
enables us to
directly investigate how
$F^\da(x,\l)$ transforms under
the redundancies identified
in \Sec{REVIEW>RESID}.
For instance,
the infinitesimal spacetime diffeomorphism described in \eqref{Xu}
entails an active variation of the {\Pleb} scalar 
as shown in \eqref{Phivar.u}.
Accordingly,
$F^\da(x,\l)$
also exhibits an extra variation
that is not captured by
$X_u\act{ F^\da(x,\l) }$.
Direct computation
in the Laurent expansion
verifies that
the total variation $\delta_u F^\da(x,\l)$
defines an exact one-form on
the correspondence space
by
% \begin{align}
%     \te_\wrap{\da\db}\mem
%         \BB{
%             X_u\act{ F^\da(x,\l) } + \delta_u F^\da(x,\l)
%         }
%     \mem
%         dF^\db(x,\l)
%     \,=\,
%         d\BB{
%             u(q)
%             - p_\da\mem u^\da(q)\mem \zeta
%             + \O(\zeta^2)
%         }
%     \,,
% \end{align}
$
    \te_\wrap{\da\db}\mem
        \delta_u F^\da(x,\l)
    \mem
        dF^\db(x,\l)
$.
In general,
the transformations
identified in 
\eqrefss{Xuv}{Phivar.uv}{Phivar.ab}
are reproduced by 
stipulating that the $\m$-fibers of $\pi_0 : \PT \to \CP^1$
are transformed by Poisson diffeomorphisms
with respect to the Poisson structure in \eqref{I}:
\begin{align}
    \te_\wrap{\da\db}\,
        \delta\hnem\m^\da\mem
        d\m^\db
    \,=\,
        dh(\m)
    \,.
\end{align}
Explicitly,
one can verify that
the Hamiltonian exhibits the expansion
$
    h(F(x,\lambda)\hhnem) =
    \k\hem
    {\lambda_1}^2 $ $ \smash{\bigbig{
        h^{(0)}(x)
        - h^{(1)}(x)\mem \zeta
        + h^{(2)}(x)\mem \zeta^2
        - h^{(3)}(x)\mem \zeta^3
        + \cdots
    }}
$
when pulled back by $\pi_1: \F \too \PT$,
where
\begin{align}
    \label{TwistorHams}
    h^{(0)}
    \,&=\,
        u(q)
    \,,\quad
    h^{(1)}
    \,=\,
        p_\da\mem u^\da(q) + v(q)
    \,,\\
    \nonumber
    h^{(2)}
    \,&=\,
        \tfrac{1}{2}\, p_\da p_\db\hem u^{\da\db}(q)
        + \k\mem u_\da(q)\mem \Phi^\da
        + p_\da\mem v^\da(q)
        - b(q)
    \,,\\
    \nonumber
    h^{(3)}
    \,&=\,
    \begin{aligned}[t]
        \tfrac{1}{6}\, p_\da p_\db p_\dc\mem u^{\da\db\dc}(q)
        + \k\hem X_u\act{ \Phi }
        + \tfrac{1}{2}\, p_\da p_\db\mem v^{\da\db}(q)
        + \k\hem X_v\act{ \Phi }
        - a(q)
        - p_\da\mem b^\da(q)
    \end{aligned}
    \,.
\end{align}
Note that these are quickly deduced by
series expanding $u(F(x,\l))$, $v(F(x,\l))$, etc.
\section{A Type N Primer}
\label{TYPEN}

\subsection{{\Pleb} Scalar}

For a preliminary demonstration of the recursive construction of a deformed incidence relation,
we would like to first consider
a type N spacetime
\cite{Sparling:1981nk,sparling-TN1-14,tod1981asymptotically}
which Dunajski and Mason \cite{dunajski2000hyper} themselves used as an example
(see their Sec.\,3.7).
Our goal is to reproduce the same result as
\rcite{dunajski2000hyper}
and present a resummed formula.

This type N example
arises by
taking the {\Pleb} second heavenly scalar as
\begin{align}
    \Phi
    \,=\,
        \frac{1}{\rp{pq}}
    \,,
\end{align}
where $\rp{pq} = p_\da\hem q^\da$.
It follows that
\begin{align}
    \label{typeN:derivs}
    \Phi^\da
    \mem=\mem
        - \frac{q^\da}{\rp{pq}^2}
    \,,\quad
    \Phi^{\da\db}
    \mem=\mem
        2\,
            \frac{q^\da\hnem q^\db}{\rp{pq}^3}
    \,,\quad
    \Phi^{\da\db\dc}
    \mem=\mem
        -6\,
            \frac{q^\da\hnem q^\db\hnem q^\dc}{\rp{pq}^4}
    \,,\quad
    \Phi^{\da\db\dc\dd}
    \mem=\mem
        24\,
            \frac{q^\da\hnem q^\db\hnem q^\dc\hnem q^\dd}{\rp{pq}^5}
    \,.
\end{align}
The metric certainly takes a KS form;
$\Phi^{\da\db}$ in \eqref{typeN:derivs} gives
the graviton field
\begin{align}
    h
    \,=\,
        \frac{2}{\rp{pq}^3}\,
            \rp{q\mem dq}^2
    \,.
\end{align}
The Weyl tensor $\Phi^{\da\db\dc\dd}$
clearly describes a repeated null direction $q^\da$ of full degeneracy.

\subsection{Incidence Relation}

% Recall that the DM recursion operator is a map
% $\Rec : \ker(\Box_\Phi) \to \ker(\Box_\Phi)$.
% By using the second derivative in \eqref{typeN:derivs},
% the curved Laplacian 
% boils down to
% \begin{align}
%     \label{typeN:BBox}
%     \Box_\Phi
%     \,=\,
%         2\mem
%             \frac{\partial}{\partial q^\da}
%             \frac{\partial}{\partial p_\da}
%         -
%             \frac{4\hem\k}{\rp{pq}^3}\mem
%             \bb{\nem{
%                 q_\da\mem \frac{\partial}{\partial p_\da}
%             }\nem}^{\nem\nem2}
%     \,.
% \end{align}
% The defining formula
% for the DM recursion operator $\Rec$,
% spelled out in \eqref{rec-unpacked},
% boils down to 
% \begin{align}
%     \label{typeN:rec}
%     \frac{\partial}{\partial p_\dc}\BB{
%         \Rec\act{\psi}
%     }
%     \,=\,
%     \bb{
%         % \te{}^{\dd\dc}\mem \frac{\partial}{\partial q^\dd}
%         \frac{\partial}{\partial q_\dc}
%         + \frac{2\k}{\rp{pq}^3}\mem 
%             q^\dc q_{\dd}\mem
%             \frac{\partial}{\partial p_\dd}
%     }\mem \psi
%     \,.
% \end{align}
% Preliminary computation shows that
% the repeated action of $\Rec$
% can be given as
% \begin{align}
%     q^\da
% \next
%     p^\da
% \next
%     - \frac{\k}{\rp{pq}^2}\,
%         q^\da
% \next
%     \frac{\k}{\rp{pq}^2}\,
%         p^\da
% \next
%     - \frac{\k}{\rp{pq}^2}\,
%     \frac{p^\da}{q^\da}\,
%         p^\da
%     - \frac{1}{2}\mem \frac{\k^2}{\rp{pq}^2}\,
%         q^\da
% \next
%     \cdots
%     \,,
% \end{align}
% for each value of $\da$.

Recall that the DM recursion operator is a map
$\Rec : \ker(\Box_\Phi) \to \ker(\Box_\Phi)$.
By using the second derivative in \eqref{typeN:derivs},
the curved Laplacian 
boils down to
\begin{align}
    \label{typeN:BBox}
    \Box_\Phi
    \,=\,
        \Box 
        \mem-\mem \frac{4\hem\k}{\rp{pq}^3}\mem
            \CD^2
    \,.
\end{align}
The defining formula
for the recursion operator $\Rec$,
spelled out in \eqref{rec-unpacked},
boils down to 
\begin{align}
    \label{typeN:rec}
    \frac{\partial}{\partial p_\dc}\BB{
        \Rec\act{\psi}
    }
    \,=\,
    \bb{
        % \te{}^{\dd\dc}\mem \frac{\partial}{\partial q^\dd}
        \frac{\partial}{\partial q_\dc}
        + \frac{2\k}{\rp{pq}^3}\mem 
            q^\dc\hem \CD
    }\mem \psi
    \,.
\end{align}
Here, we have identified a useful combination,
\begin{align}
    \label{Dee}
    \CD
    \,:=\,
        q_\da\mem \frac{\partial}{\partial p_\da}
    \,\,\,\qiq\,\,\,
    \CD\hem q^\da
    \,=\,
        0
    \,,\quad
    \CD\hem \rp{pq}
    \,=\,
        0
    \,.
\end{align}

Suppose an auxiliary constant spinor $\eta_\da$.
Preliminary computation shows that
the repeated action of $\Rec$
on $\refq = \eta_\da\hem q^\da$
can be given as
\begin{align}
\begin{split}
\label{typeN:seq-prelim}
{}&{}
    \refq
\next
    -\pref
\next
    -\n\mem \refq
\next
    -\n\mem \pref
\\
{}&{}
\next
    -\tfrac{1}{2}\mem \n^2\mem \refq
    -\n \xi\mem \pref
\next
    -\tfrac{3}{2}\mem \n^2\mem \pref
    -\n \xi^2\mem \pref
\next
    \cdots
\,,
\end{split}
\end{align}
where 
one learns that
it is convenient to define
\begin{align}
    \n
    \,:=\,
        \frac{\k}{\pq^2}
    \,,\quad
    \xi
    \,:=\,
        \frac{\pref}{\refq}
    \,.
\end{align}

\newpage

Based on this preliminary exploration,
we consider the ansatz
\begin{subequations}
\label{typeN:ansatz}
\begin{align}
\label{typeN:ansatz.even}
    \psi_{2n}
    \mem\,&=\,\mem
    %     \sum_{k=0}^{n-1}\,
    %     c_k\,
    %         \nu^{n-k}\hem
    %         \xi^{2k-1}
    %         \mem \pref
    % \,=\,
        \nu^n
            \mem \refq
        \,+\, 
        \sum_{k=1}^{n-1}\,
        c_{n,k}\,
            \nu^{n-k}\hem
            \xi^{2k-1}
            \mem \pref
    \,,\\
\label{typeN:ansatz.odd}
    \psi_{2n+1}
    \mem\,&=\,\mem
    %     \sum_{k=0}^{n-1}\,
    %     c'_k\,
    %         \nu^{n-k}\hem
    %         \xi^{2k}
    %         \mem \pref
    % \,=\,
        \nu^n
            \mem \pref
        \,+\,
        \sum_{k=1}^{n-1}\,
        c'_{n,k}\,
            \nu^{n-k}\hem
            \xi^{2k}
            \mem \pref
    \,.
\end{align}
\end{subequations}
With $c_{n,0} = 1 = c'_{n,0}$
and
$c_{n,n} = 0 = c'_{n,n}$,
direct computation shows that
% \begin{subequations}
% \label{typeN:kers}
% \begin{align}
% \kern-0.5em
%     \Box_\Phi\act{
%         \psi_{2n}
%     }
%     \,&=\,
%     \frac{\refq}{\pq}\,
%         \sum_{k=0}^{n-1}\,
%         c_{n,k}\mem
%         \BB{
%             8(n\mminus k)(n\mminus 1\mminus k)\,
%                 \nu^{n-k}\hem \xi^{2k}
%             - 8k(2k\mminus1)\,
%                 \nu^{n-k+1}\hem \xi^{2k-2}
%         }
%     \nonumber
%     \,,\\
%     \,&=\,
%     \frac{\refq}{\pq}\,
%         \sum_{k=0}^{n-1}\,
%         8\mem
%         \BB{
%             (n\mminus k)(n\mminus 1\mminus k)\mem c_{n,k}
%             - (2k\mplus1)(k\mplus1)\mem c_{n,k+1}
%         }\mem
%         \nu^{n-k}\hem \xi^{2k}
%     \label{typeN:ker.even}
%     \,,\kern-0.4em\\
% \kern-0.5em
%     \Box_\Phi\act{
%         \psi_{2n+1}
%     }
%     \,&=\,
%     \frac{\pref}{\pq}\,
%         \sum_{k=0}^{n-1}\,
%         c'_{n,k}\mem
%         \BB{
%             8(n\mminus k)(n\mminus 1\mminus k)\,
%                 \nu^{n-k}\hem \xi^{2k}
%             - 8k(2k\mplus1)\,
%                 \nu^{n-k+1}\hem \xi^{2k-2}
%         }
%     \nonumber
%     \,,\\
%     \,&=\,
%     \frac{\pref}{\pq}\,
%         \sum_{k=0}^{n-1}\,
%         8\mem
%         \BB{
%             (n\mminus k)(n\mminus 1\mminus k)\mem c'_{n,k}
%             - (2k\mplus3)(k\mplus1)\mem c'_{n,k+1}
%         }\mem
%         \nu^{n-k}\hem \xi^{2k}
%     \,.\kern-0.4em
%     \label{typeN:ker.odd}
% \end{align}
% \end{subequations}
$\psi_{2n}$ and $\psi_{2n+1}$
are annihilated by $\Box_\Phi$ in \eqref{typeN:BBox} if 
\begin{align}
    \label{typeN:crec}
    c_{n,k+1}
    \,=\,
        \frac{
            (n\mminus k)(n\mminus 1\mminus k)
        }{
            (2k\mplus1)(k\mplus1)
        }\,
            c_{n,k}
    \,,\quad
    c'_{n,k+1}
    \,=\,
        \frac{
            (n\mminus k)(n\mminus 1\mminus k)
        }{
            (2k\mplus3)(k\mplus1)
        }\,
            c'_{n,k}
    \,.
\end{align}
To derive these,
one would use the Schouten identity
in the form
\begin{align}
    \label{eta-p-q}
    \frac{\eta^\da}{\refq}
    \,=\,
        \frac{p^\da}{\pq}
        +
        \ratio{}\,
            \frac{q^\da}{\pq}
    \,.
\end{align}
With our prescribed boundary conditions,
\eqref{typeN:crec} is uniquely solved by
\begin{align}
    c_{n,k}
    \,=\,
        % 2^k\hem
        %     \frac{
        %         \binom{n}{k}
        %         \binom{n-1}{k}
        %     }{
        %         \binom{2k}{k}
        %     }
            \frac{
                1
            }{
                % (-2)^k\mem
                % \binom{-\frac{1}{2}}{k}
                (2k\mminus1)!!
            }\mem
                \binom{n}{k}\hnem
                \binom{n-1}{k}
    \,=\,
        (2k\mplus1)\mem c'_{n,k}
    \,.
\end{align}

In the meantime,
plugging in \eqref{typeN:ansatz} to \eqref{typeN:rec}
shows that
% \begin{subequations}
% \begin{align}
%     {}&{}
%     \frac{\partial}{\partial p_\dc}\mem
%         \psi_{2n+1}
%     \,=\,
%         \frac{\refq}{\pq}\mem
%         \sum_{k=0}^{n-1}\,
%             \nu^{n-k}\mem
%             \BB{
%                 - (2n\mminus 4k\mminus 1)\mem
%                     c'_{n,k}\mem
%                     \xi^{2k+1}\mem
%                     q^\dc
%                 + (2k\mplus 1)\mem
%                     c'_{n,k}\mem
%                     \xi^{2k}\mem
%                     p^\dc
%             }
%     \,,\\
% \begin{split}
%     {}&{}
%     \frac{\partial}{\partial q_\dc}
%     \mem \psi_{2n}
%     \,=\,
%         \frac{\refq}{\pq}\mem
%         \sum_{k=0}^{n-1}\,
%             \nu^{n-k}\mem
%             \BB{
%                 (2k\mminus 1)\mem
%                     c_{n,k}\mem
%                     \xi^{2k+1}\mem
%                     q^\dc
%                 + (2n\mminus 1)\mem
%                     c_{n,k}\mem
%                     \xi^{2k}\mem
%                     p^\dc
%             }
%     \,,\\
%     {}&{}
%     \frac{2\k}{\rp{pq}^3}\mem 
%             q^\dc\hem \CD
%     \mem \psi_{2n}
%     \,=\,
%     %     \frac{\refq}{\pq}\mem
%     %     \sum_{k=0}^{n-1}\,
%     %         \nu^{n-k+1}\mem
%     %         \BB{
%     %             -4k\mem c_{n,k}\mem \xi^{2k-1}\mem
%     %             q^\dc
%     %         }
%         \frac{\refq}{\pq}\mem
%         \sum_{k=0}^{n-1}\,
%             \nu^{n-k}\mem
%             \BB{
%                 -4(k\mplus 1)\mem c_{n,k+1}\mem \xi^{2k+1}\mem
%                 q^\dc
%             }
%     \,.
% \end{split}
% \end{align}
% \end{subequations}
% Hence
$\Rec\act{\psi_{2n}} = a_n\mem \psi_{2n+1}$ if
\begin{align}
\begin{split}
    -(2n\mminus 4k\mminus 1)\mem a_n\mem c'_{n,k}
    \,&=\,
        (2k\mminus 1)\mem
            c_{n,k}
        -4(k\mplus 1)\mem c_{n,k+1}
    \,,\\
    (2k\mplus 1)\mem a_n\mem c'_{n,k}
    \,&=\,
        (2n\mminus 1)\mem c_{n,k}
    \,.
\end{split}
\end{align}
It is straightforward to check that
$a_n = 2n\mminus 1$ serves as a solution.
Similarly, it holds that
$\Rec\act{\psi_{2n-1}} = n^{-1}\mem \psi_{2n}$.
Consequently,
since \eqref{typeN:seq-prelim} has described
$\Rec^2 \refq = -\psi_2$,
one finds
\begin{align}
    % \Rec^2 \refq
    % \,=\,
    %     -\psi_2
    % \qiq
    \Rec^{2n} \refq
    \,=\,
        % - \frac{1}{2^{n-1} n}\mem
        % \binom{2n\mminus2}{n\mminus1}\mem
        - \frac{(2n\mminus3)!!}{n!}\mem
            \psi_{2n}
    \,,\quad
    \Rec^{2n+1} \refq
    \,=\,
        % - \frac{1}{2^n}\mem
        % \binom{2n}{n}\mem
        - \frac{(2n\mminus1)!!}{n!}\mem
            \psi_{2n+1}
    \,.
    % \,,
\end{align}
% Therefore,
% recalling that
% \eqref{typeN:seq-prelim} has described
% $\Rec^2 \refq = -\psi_2$,
% one finds that
% \begin{align}
%     \Rec^n \refq
%     \,=\,
%         \refq\mem
%         \sum_{k=0}^{\lfloor n/2 \rfloor}
%             \binom{\frac{1}{2}}{k}
%             \binom{1\mminus 2k}{n\mminus 2k}
%             \mem \nu^k\hem \xi^{n-2k}
%             (-1)^{n-k}
%     \,.
% \end{align}

In this way,
the completion of $\refq$ in $\F$
that descends to $\PT$ is found as
\begin{align}
\begin{split}
    \sum_{n=0}^\infty\,
        \Rec^n \refq\,
            \zet{n}
    \,=\,
        \refq\mem
        \sqrt{
            \bb{
                1 - \xi\mem \frac{\l_0}{\l_1}
            }^{\nem\nem\hnem2}
            - 2\n\hem \zet{2}
        }
    \,.
\end{split}
\end{align}
This implies that the deformed incidence relation
can be realized as
\begin{align}
\begin{split}
    \label{typeN:incidence-etaform}
    F^\da(x,\l)
    \,=\,
\begin{aligned}[t]
{}&{}
        \eta_1^\da\mem
        \frac{\rp{\eta_0 q}\l_1}{\rp{\eta_0\eta_1}}
        \mem\sqrt{
            \bb{
                1 - 
                    \frac{\rp{p\eta_0}}{\rp{\eta_0 q}}
                    \frac{\l_0}{\l_1}
            }^{\nem\nem\hnem2}
            - \frac{2\k}{\pq^2}\hhem \zet{2}
        }
\\
{}&{}
        -
        \eta_0^\da\mem
        \frac{\rp{\eta_1\hnem q}\l_1}{\rp{\eta_0\eta_1}}
        \mem\sqrt{
            \bb{
                1 - \frac{\rp{p\eta_1}}{\rp{\eta_1\hnem q}} \frac{\l_0}{\l_1}
            }^{\nem\nem\hnem2}
            - \frac{2\k}{\pq^2}\hhem \zet{2}
        }
    \,,
\end{aligned}
\end{split}
\end{align}
with the aid of two constant reference spinors
$\eta_0^\da$, $\eta_1^\da$.
In particular, one can simply have
\begin{align}
    \label{typeN:incidence}
    F^\da(x,\lambda)
    \,&=\,
        q^\da\hem \lambda_1
        \,
        \sqrt{
            \bb{\nem{
                1 + \frac{p^\da}{q^\da} \frac{\l_0}{\l_1}
            }\nem}^{\nem\nem2}
            - \frac{2\k}{\lsq pq \rsq^2}\hhem
            \bb{\nem{
                \frac{\l_0}{\l_1}
            }\nem}^{\nem\nem2}
        \mem}
    \quad\text{for each $\da = \doz,\diz$}
    \,.
\end{align}
% Note that this describes
% \begin{align}
%     \Rec^n \refq
%     \,=\,
%         \refq\mem
%         \sum_{k=0}^{\lfloor n/2 \rfloor}
%             \binom{\frac{1}{2}}{k}
%             \binom{1\mminus 2k}{n\mminus 2k}
%             \mem (-\nu)^k\hem (-\xi\zeta)^{n-2k}
%     \,.
% \end{align}

\newpage

% Note that this result could have been also derived by
% directly employing the ansatz
% $\refq\mem f(\xi,\nu;\l_0/\l_1)$
% for the completion of $\refq$,
% appealing to homogeneity.
% Note also that,
% provided a branch cut prescription,
% \eqrefsor{typeN:incidence-etaform}{typeN:incidence}
% nicely reduce to the flat incidence relation 
% in the limit $\k\to0$:
% $F^\da(x,\l) \to q^\da\hem \l_1 + p^\da\hem \l_0$.

\section{Eguchi-Hanson}
\label{EH}

The EH solution
\cite{Eguchi:1978xp,Eguchi:1978gw,Eguchi:1979yx}
describes the metric
\begin{align}
    \label{EH:line0}
    g
    \,=\,
        \frac{dr^2}{
            1 - a^4\hnem/r^4
            \vphantom{\big|}
        }  
        + \frac{r^2}{4}\mem
        \bb{
            (\Theta^1)^2 + (\Theta^2)^2
            + \bigbig{
                1 - a^4\hnem/r^4
            }\mem 
               (\Theta^3)^2
        }
    \,.
\end{align}
Here,
$a$ is a constant
while
$\Theta^1$, $\Theta^2$, $\Theta^3$ describe the Maurer-Cartan forms,
\begin{align}
\begin{split}
    \Theta^1
    \,&=\,
        \sin\psi\mem
            d\theta
        - \cos\psi\mem
            \sin\theta\mem d\phi
    \,,\\
    \Theta^2
    \,&=\,
        \cos\psi\mem
            d\theta
        +
        \sin\psi\mem
            \sin\theta\mem d\phi
    \,,\\
    \Theta^3
    \,&=\,
        d\psi + \cos\theta\mem d\phi
    \,.
\end{split}
\end{align}
The EH solution is an instance of
Gibbons-Hawking \cite{hawking1977gravitational,Gibbons:1978tef} gravitational instanton,
asymptotically locally Euclidean
and
two-centered with equal masses
\cite{Prasad:1979kg}.

\subsection{Kerr-Schild Metric}

It is known \cite{Tod:1982mmp,BurnettStuart:1979sparlingtod,Berman:2018hwd}
that the line element in \eqref{EH:line0}
can locally\footnote{
    The $\mathrm{SU}(2)$ parametrization in \eqref{x=rU}
    naturally posits that the Eulerian angle $\psi$ is $4\pi$-periodic.
    % The angle variables $\phi \in [0,2\pi[$ and $\theta \in [0,\pi]$ describe a two-sphere
    % in the meantime.
    However,
    the EH line element in \eqref{EH:line0}
    avoids a singularity at $r=a$
    iff $\psi$ is $2\pi$-periodic;
    see, e.g., \rrcite{Berman:2018hwd,AAS}.
} be brought to a KS form:
\begin{align}
    \label{EH:line1}
    g
    \,=\,
        \eta
    \mem+\mem
        \frac{8\k}{\rp{pq}^3}\,
            \rp{p\mem dq}^2
    \,,
\end{align}
where $x^{\da1} = q^\da$, $x^{\da0} = p^\da$.
To show this, we consider the identities
\begin{align}
\begin{aligned}[t]
    x^2
    \,=\,
        \e_\wrap{\a\b} \te_\wrap{\da\db}\mem
            x^{\da\a} x^{\db\b}
    \,=\,
        2\hem \pq
    \,,\quad
    \e_\wrap{\a\b} \te_\wrap{\da\db}\mem
        x^{\db\b}
    \,=\,
        \frac{x^2}{2}\mem
        (x^{-1})_{\a\da}
    \,,
\end{aligned}
\end{align}
where $(x^{-1})_{\a\da}$ denotes the inverse of $x^{\da\a}$
as a $2\mtimes2$ matrix.
Then \eqref{EH:line1} boils down to
\begin{align}
    \label{EH:linexix}
    g
    \,=\,
        dx^{\da\a} \odot
            d\hhem\bb{
                \frac{x^2}{2}\mem (x^{-1})_{\a\da}
            }
    \mem+\mem
        \frac{64\k}{x^2}\,
            \BB{
                \i^\a\hem 
                    (x^{-1} dx)_\a{}^\b
                \o_\b
            }^{\nem\hnem2}
    \,,
\end{align}
where
$(x^{-1} dx)_\a{}^\b = (x^{-1})_{\a\da}\mem dx^{\da\b}$.
Provided a Euclidean reality structure 
$\delta^{\da\a}$,
we take
\begin{align}
    \label{x=rU}
    x^{\da\a}(r,\phi,\theta,\psi)
    \,=\,
        \frac{r}{\sqrt{2}}\,
            \delta^{\da\b}\mem
                U_\b{}^\a(\phi,\theta,\psi)
    \,,
\end{align}
where $U_\b{}^\a(\phi,\theta,\psi)$
computes a product of exponentiated sigma matrices such that
\begin{align}
    \label{ixdx}
    (x^{-1} dx)_\a{}^\b
    \,=\,
        \frac{dr}{r}\,
            \delta_\a{}^\b
        + \frac{1}{2i}\,
            \Theta^a\mem (\s_a)_\a{}^\b
    \,.
\end{align}
With this parametrization, \eqref{EH:linexix} yields
\begin{align}
    \label{EH:linerb}
    g
    \,=\,
        % 2\b^2\mem 
        % \BB{
        %     dr\mem U_\a{}^\b
        %     + r\mem dU_\a{}^\b
        % } 
        % \modot
        % \BB{
        %     dr\mem (U^{-1})_\b{}^\a
        %     + r\mem d(U^{-1})_\b{}^\a
        % }
        dr^2
        % - 2\b^2 r^2\mem (U^{-1} dU)_\a{}^\b \modot (U^{-1} dU)_\b{}^\a
        + \frac{r^2}{4}\mem 
            % \delta_{ab}\mem \Theta^a \modot \Theta^b
            \BB{
                (\Theta^1)^2
                +
                (\Theta^2)^2
                +
                (\Theta^3)^2
            }
    \mem+\mem
        \frac{64\k}{r^2}\,
            \bb{
                \frac{dr}{r}
                - \frac{1}{2i}\,
                    \Theta^3
            }^{\nem\hnem2}
    \,.
\end{align}
Upon performing a shift
\begin{align}
    \psi
    \,\,\,\mapsto\,\,\,\,
        \psi
        \mem+\mem
        \frac{1}{2i}\mem
            \log\bb{
                1 - \frac{64\k}{r^4}
            }
    \,,
\end{align}
\eqref{EH:linerb}
gets diagonalized as \eqref{EH:line0}
with $64\k = a^4$.

\subsection{{\Pleb} Scalar}

The KS metric in \eqref{EH:line1} describes the graviton field
\begin{align}
    \label{EH:meth}
    h
    \,=\,
        \frac{4}{\rp{pq}^3}\,
            \rp{p\mem dq}^2
    \,.
\end{align}
It can be seen that \eqref{EH:meth}
precisely arises when
$x^{\da1} = q^\da$, $x^{\da0} = p^\da$
are Pleba\'nski's second heavenly coordinates
while the {\Pleb} scalar is given by
\begin{align}
    \label{EH:Pleb}
    \Phi
    \,=\,
        \frac{2}{\pq}
        \frac{\pref^2}{\refq^2}
    \,,
\end{align}
where $\eta_\da$ is a constant reference spinor.
Explicitly,
\begin{align}
\begin{split}
    \label{EH:derivs}
    {}&{}
    \Phi^\da
    \,=\,
        2\mem
        \frac{q^\da}{\pq^2}
        \frac{\pref^2}{\refq^2}
        +
        4\mem
        \frac{p^\da}{\pq^2}
        \frac{\pref}{\refq}
    \,,\quad
    \Phi^{\da\db}
    \,=\,
        4\mem \frac{p^\da p^\db}{\pq^3}
    \,,\\
    {}&{}
    \Phi^{\da\db\dc}
    \,=\,
        -12\mem \frac{p^{(\da} p^\db q^{\dc)}}{\pq^4}
    \,,\quad
    \Phi^{\da\db\dc\dd}
    \,=\,
        48\mem \frac{p^{(\da} p^\db q^\dc q^{\dd)}}{\pq^5}
    \,.
\end{split}
\end{align}
It is easy to check that
$\Phi^{\da\db}$ in \eqref{EH:derivs}
defines a null Maxwell field.
On a related note,
the heavenly equation is
``effectively linearized''
as $\Box\hem\Phi = 0 = \k\mem \Phi_\wrap{\da\db} \Phi^{\da\db}$,
i.e.,
in the sense of Tod \cite{Tod:1982mmp}.
The Weyl tensor $\Phi^{\da\db\dc\dd}$ 
in \eqref{EH:derivs}
shows that the solution is type D.

\subsection{Incidence Relation}

Now let us implement the DM recursion.
The curved Laplacian is
\begin{align}
    \label{EH:BBox}
    \Box_\Phi
    \,=\,
        \Box 
        \mem-\mem \frac{8\k}{\rp{pq}^3}\mem
            \Pdeg (\Pdeg - \id)
    \,,
\end{align}
while the recursion operator $\Rec$ is defined via
\begin{align}
    \label{EH:rec}
    \frac{\partial}{\partial p_\dc}\BB{
        \Rec\act{\psi}
    }
    \,=\,
    \bb{
        \frac{\partial}{\partial q_\dc}
        + \frac{4\k}{\rp{pq}^3}\mem 
            p^\dc\hem \Pdeg
    }\mem \psi
    \,.
\end{align}
Here, we have defined
\begin{align}
    \Pdeg
    \,:=\,
        p_\da\mem \frac{\partial}{\partial p_\da}
    \,,
\end{align}
which measures the degree of the $p$-variable.

With hindsight,
we study the repeated action of $\Rec$ on
the quadratic quantity $q^\da q^\db$:
\begin{align}
\Rec\,\,\,:\,\,\,
    q^\da q^\db
\next
    2q^{(\da} p^{\db)}
\next
    \bb{
        1 - \frac{4\k}{\pq^2}
    }\mem
    p^\da p^\db
\next
    0
\,.
\end{align}
Remarkably, the sequence terminates at a finite order.
This implies quadratic holomorphic coordinates
on $\PT$
\cite{Bittleston:2023bzp,hitchin1979polygons},
parametrized as
\begin{align}
\begin{split}
    \label{EH:quadX}
    X^{\da\db}(x,\l)
    \,&=\,
        \BB{
            q^\da\hem \l_1 + p^\da\hem \l_0
        }
        \BB{
            q^\db\hem \l_1 + p^\db\hem \l_0
        }
        - \frac{4\k}{\pq^2}\mem
            p^\da p^\db\mem {\l_0}^2
    \,,\\
    % X^{\da\db}(x,\l)
    \,&=\,
        x^{\da\a}\hem x^{\db\b}\mem
        \bb{
            \l_\a\hhem \l_\b
            - \frac{16\k}{(x^2)^2}\mem
                \i_\a\hhem \i_\b
                \mem \lp{\o\l}^2
        }
    \,.
\end{split}
\end{align}
Note the relation
$
    X_\wrap{\da\db}(x,\l)\mem
    X^{\da\db}(x,\l)
    = -8\k\mem {\l_0}^2
$.
% \begin{align}
%     X_\wrap{\da\db}(x,\l)\mem
%     X^{\da\db}(x,\l)
%     \,=\,
%         -4\k\mem {\l_0}^2
%     \,.
% \end{align}
\newpage

Note that $X^{\da\db}(x,\l)$
already provides a natural object 
for describing the holomorphic twistor lines.
To explicitly construct
a deformed incidence relation,
however,
we may
suppose a constant reference spinor $\eta_\da$
and take
\begin{align}
    \label{EH:incidenceX}
    F^\da(x,\l)
    \,=\,
        \frac{
            X^{\da\db}(x,\l)\mem \eta_\db
        }{
            \sqrt{
                X^{\dc\dd}(x,\l)\mem \eta_\dc\hhem \eta_\dd
            \mem}
        }
    \,.
\end{align}
Since $L_\dc$ annihilates $X^{\da\db}(x,\l)$,
$F^\da(x,\l)$ in \eqref{EH:incidence} is also annihilated by $L_\dc$.
To clarify,
this is one possible simple realization of $F^\da(x,\l)$
using $X^{\da\db}(x,\l)$.
More explicitly,
\begin{align}
    \label{EH:incidence}
    F^\da(x,\l)
    \,=\,
    \frac{\displaystyle
        % \bb{
            % q^\da + p^\da\mem \frac{\l_0}{\l_1}
        % }\nem
        \BB{
            q^\da\hem \l_1 + p^\da\hem \l_0
            % x^{\da\a}\hem \l_\a
        }\nem
        \bb{
            1 - \frac{\pref}{\refq} \frac{\l_0}{\l_1}
        }
        + \frac{4\k\hem p^\da}{\pq^2}\mem
            \frac{\pref}{\refq}
            % \zet{2}
            \frac{{\l_0}^2}{\l_1}
    }{
        \sqrt{\displaystyle
            \bb{
                1 - \frac{\pref}{\refq} \frac{\l_0}{\l_1}
            }^{\nem\nem\hnem2}
            - \frac{4\k}{\pq^2}\mem
            \bbn{\frac{\pref}{\refq} \frac{\l_0}{\l_1}}^{\nem\nem\hnem2}
        \mem}
    }
    \,.
\end{align}
It should be clear that \eqref{EH:incidence} nicely reduces to the flat incidence relation,
$x^{\da\a}\hem \l_\a = q^\da\hem \l_1 + p^\da\hem \l_0$,
in the limit $\k \too 0$.
Also, taking $\eta^\da = \delta^\da{}_0$
gives rise to \eqref{summary:EH}.
The Taylor expansion of
% $F^\da(x,\l)$
\eqref{EH:incidence}
around $\l_\a \propto \o_\a$
conforms to the formula in \eqref{deformed-incidence},
so one should be able to obtain the same result by
running the DM recursion directly on $q^\da$.

% Note that this is one possible simple realization of $F^\da(x,\l)$
% using $X^{\da\db}(x,\l)$.
% In general, the choice of $F^\da(x,\l)$
% is subject to the canonical transformations
% described in \Sec{REVIEW>DM>CAN}.

In \rcite{Bittleston:2023bzp},
the authors factorize
$X^{\da\db}(x,\l)$
as
\begin{align}
    M_\pm^\da(x,\l)
    \,=\,   
        x^{\da\a}\hem \l_\a
        \pm \frac{4\k}{\pq^2}\mem
            p^\da
            \mem \l_0
    \qiq    
    X^{\da\db}(x,\l)
    \,=\,
        M_+^{(\da}(x,\l)\mem M_-^{\db)}(x,\l)
    \,.
\end{align}
Note that 
these $M^\da_\pm(x,\l)$ are not individually annihilated by $L_\dc$:
$L_\dc\act{ M^\da_\pm } \propto M^\da_\pm$.
\eqref{EH:incidenceX}
describes their particular combination,
\begin{align}
    F^\da
    \,=\,
        \frac{1}{2}\lrp{\hnem
            M_+^\da\mem
            \sqrt{\frac{\rp{\eta M_-}}{\rp{\eta M_+}}}
        +
            M_-^\da\mem
            \sqrt{\frac{\rp{\eta M_+}}{\rp{\eta M_-}}}
        \,}
    \,,
\end{align}
such that $L_\dc\act{F^\da} = 0$.

\section{Self-Dual Taub-NUT}
\label{SDTN}

The Gibbons-Hawking \cite{hawking1977gravitational,Gibbons:1978tef}
ansatz reads
\begin{align}
    \label{gh}
    g
    \,=\,
	\frac{1}{V}\mem
    	\BB{
    		dt + a
    	}^{\nem2}
	- V\mem
        \BB{
            dx^2
            + dy^2
            + dz^2
        }
    \,,
\end{align}
which describes a SD spacetime if
$V$ and $a$
are zero- and one-forms
on the three-dimensional space of $(x,y,z)$
such that
$-dV = i\mem {*}_3\mem da$
% \begin{align}
%     -dV \,=\, i\mem {*}_3\mem da
% \end{align}
with respect to the three-dimensional Hodge star ${*}_3$.

The SDTN solution describes
a single-centered, asymptotically locally flat instance
of the Gibbons-Hawking metric:
\begin{align}
    \label{SDTN:ghdata}
    V \,=\, 1 + \frac{\k}{R}
    \,,\quad
    a \,=\,
        \frac{\k}{R}\mem\bb{
            \frac{x+iy}{R+z}\mem dx 
            + \frac{y-ix}{R+z}\mem dy
            + dz
        }
    \,,
\end{align}
where $R^2 = x^2 + y^2 + z^2$.

\subsection{Kerr-Schild Metric}

The KS metric of the SDTN solution reads \cite{note-sdtn}
\begin{align}
    \label{SDTN:ks}
    g
    \,=\,
        dt^2 \mminus dx^2 \mminus dy^2 \mminus dz^2
        \mem+\mem 
        \frac{\k}{R}\mem
            \bb{
                d(t \mminus z)
                - \frac{x +\nem iy}{R +\nem z}\mem
                d(x \mminus iy)
            }^{\nem\nem\hnem2}
    \,,
\end{align}
where $R^2 = x^2 + y^2 + z^2$.
An elegant formula is revealed
in the spinor notation
\cite{note-sdtn}:
\begin{align}
    \label{SDTN:g1}
    g
    \,=\,
        \eta
        \mem+\mem
        \frac{2\k}{R}\,
            \BB{
                \o_\a\hhem \ta_\da\mem dx^{\da\a}
            }^{\nem\nem2}
    % \qfq
    % h
    % \,=\,
    %     \frac{1}{R}\,
    %         \BB{
    %             \o_\a\hhem \ta_\da\mem dx^{\da\a}
    %         }^{\nem\nem2}
    \,.
\end{align}
The ASD spinor $\o_\a \doteq \e_{\a0}$
is a constant reference,
controlling the direction of the Misner string
(``gauge information'').
The SD spinor $\ta_\da$ is
one of the principal spinors
(``physical information''),
encoding the ``square root'' of the three-vector $\vex = (x,y,z)$:
\begin{align}
    \label{SDTN:alpha}
    \ta^\da
    \,\,=\,\,
    \lrp{
        \begin{array}{c}
            1
        \\[2pt]
        \displaystyle
            \frac{x +\nem iy}{R +\nem z}
        \end{array}
    }^{\kern-3.5pt\da}
    \,.
\end{align}
\rcite{note-sdtn} explicitly shows that
flowing along the null congruence
of 
$\ta^\da \o^\a$
brings the line element in \eqref{SDTN:g1}
to the well-known Gibbons-Hawking metric 
due to
\eqrefs{gh}{SDTN:ghdata}.
That is,
\rcite{note-sdtn} found
the explicit diffeomorphism between 
the KS and Gibbons-Hawking coordinates
for the SDTN solution.

\subsection{{\Pleb} Scalar}

\rcite{note-sdtn} also shows that
the above KS coordinates 
serve as
the second heavenly coordinates
at the same time.
\rcite{note-sdtn} explicitly constructs the {\Pleb} scalar as
\begin{align}
    \label{SDTN:Pleb}
    \Phi
    \,=\,
        \frac{2R + z}{3}\mem
        \bb{
            \frac{x +\nem iy}{R +\nem z}
        }^{\nem\nem\hnem2}
    \,.
\end{align}
Straightforward computation shows that
\begin{align}
    \label{SDTN:derivs}
    \Phi^{\da\db}
    \,=\,
        \frac{1}{R}\,
            \ta^\da \ta^\db
    \,,\quad
    \Phi^{\da\db\dc\dd}
    \,=\,
        \frac{3}{2R^5}\,
            K^{(\da\db} K^{\dc\dd)}
    \,=\,
        \frac{1}{R^3}\,
            \frac{
                6\hem
                \ta^{(\da}
                \ta^\db
                \tb^\dc
                \tb^{\dd)}
            }{
                \rp{\ta\tb}{}^2
            }
    \,,
\end{align}
% where the other principal spinor
% arises as
% \begin{align}
%     \label{SDTN:beta}
%     \tb^\da
%     \,\,=\,\,
%     \lrp{\nem\hnem
%         \begin{array}{c}
%         \displaystyle
%             - \frac{x - iy}{R + z}
%         \\[7.5pt]
%             1
%         \end{array}
%     \hhnem
%     }^{\kern-3.5pt\da}
%     \,.
% \end{align}
where the other principal spinor
arises through
\begin{align}
    K^\da{}_\db
    \,=\,
        \sqrt{2}\mem x^{\da\a} \delta_\wrap{\a\db}
        - \frac{1}{\sqrt{2}}\mem
            \delta^\da{}_\db\mem
            (x^{\dc\c} \delta_\wrap{\c\dc})
    \,\doteq\,
        (\vex \mdot \vec{\sigma})^\da{}_\db
    \,,\quad
    K^{\da\db}
    \,=\,
        R\: \frac{
            2\hem \ta^{(\da} \tb^{\db)}
        }{\rp{\tb\ta}}
        % \bb{
        %     \begin{array}{cc}
        %     \displaystyle
        %         x-iy
        %     &
        %         -z
        %     \\
        %     \displaystyle
        %         -z
        %     &
        %         -(x+iy)
        %     \end{array}
        % }^{\kern-3pt\da\db}
    \,.
\end{align}
Here, $\delta_\wrap{\a\db} \doteq \diag(1,1)$
encodes the Killing vector defining the stationary direction of the solution. 
Again, $\Phi^{\da\db}$ in \eqref{SDTN:derivs} defines a null Maxwell field
while
the Weyl tensor
$\Phi^{\da\db\dc\dd}$
shows that the solution
is of algebraic type D.
% The graviton field is
% \begin{align}
%     h
%     \,=\,
%         \frac{1}{R}\,
%             \rp{\ta\hem dq}^2
%     \,.
% \end{align}

% As before, we denote
% $q^\da = x^{\da1}$ and $p^\da = x^{\da0}$:
% \begin{align}
%     q^\doz
%     \,=\,
%         \frac{
%             x - iy
%         }{\sqrt{2}}
%     \,,\quad
%     q^\diz
%     \,=\,
%         \frac{
%             t - z
%         }{\sqrt{2}}
%     \,,\quad
%     p_\doz
%     \,=\,
%         - \frac{
%             x + iy
%         }{\sqrt{2}}
%     \,,\quad
%     p_\diz
%     \,=\,
%         \frac{
%             t + z
%         }{\sqrt{2}}
%     \,.
% \end{align}
% It is also convenient to denote
% \begin{align}
%     \label{SDTN:zeta}
%     \zeta
%     \,:=\,
%         \frac{x+iy}{R+z}
%     \,=\,
%         \frac{p_\doz}{R+z}
%     \,.
% \end{align}

It is also convenient to denote
\begin{align}
    \label{SDTN:zeta}
    \zeta
    \,:=\,
        \frac{x +\nem iy}{R +\nem z}
    \,,
\end{align}
which satisfies
\begin{align}
    \label{SDTN:zeta-derivs}
    \frac{\partial}{\partial q^\dc}\mem
        \zeta
    \,=\,
        \frac{\zeta}{\sqrt{2}R}\mem
            \ta_\dc
    \,,\quad
    \frac{\partial}{\partial p_\dc}\mem
        \zeta
    \,=\,
        -\frac{1}{\sqrt{2}R}\mem
            \ta^\dc
    \,,\quad
    \frac{\partial^2}{\partial p_\da \partial p_\db}\mem
        \zeta
    \,=\,
        -\frac{1}{2R^3}\mem
            K^{\da\db}
    \,.
\end{align}

\newpage

\subsection{Incidence Relation}

Now let us implement the DM recursion.
The curved Laplacian is
\begin{align}
    \label{SDTN:BBox}
    \Box_\Phi
    \,=\,
        \Box 
        \mem-\mem
        \frac{2\k}{R}\,
            \ta_\da \ta_\db\mem
            \frac{\partial^2}{\partial p_\da \partial p_\db}
    \,.
\end{align}
The recursion operator $\Rec$ is defined via
\begin{align}
    \label{SDTN:rec}
    \frac{\partial}{\partial p_\dc}\BB{
        \Rec\act{\psi}
    }
    \,=\,
    \bb{
        \frac{\partial}{\partial q_\dc}
        + \frac{\k}{R}\,
            \ta^\dc\hem \ta_\db\hem \frac{\partial}{\partial p_\db}
    }\mem \psi
    \,.
\end{align}

To proceed,
we make three observations.
Firstly,
the $p$-derivatives of the {\Pleb} scalar in \eqref{SDTN:Pleb}
describe
\begin{align}
\label{SDTN:obs1}
    \Phi^\doz
    \,=\,
        -\sqrt{2}\mem \zeta
    \,,\quad
    \Phi^\diz
    \,=\,
        -\frac{\zeta^2}{\sqrt{2}}
    \,.
\end{align}
Secondly,
$\zeta^n$ is harmonic with respect to 
both $\Box$ and $\Box_\Phi$
for any value of $n$:
\begin{align}
\begin{split}
\label{SDTN:obs2}
{}&{}
    \Box\mem \zeta^n
    \,=\,
        -\sqrt{2}n\,
            \frac{\partial}{\partial q^\dc}
            \bb{
                \frac{\zeta^{n-1}}{R}\mem
                    \ta^\dc
            }
    \,=\,
        \frac{n\zeta^n}{R^3}\mem
        \BB{
            (nR\mplus z) - (nR\mplus z)
        }
    \,=\,
        0
    \,,\\
{}&{}
        \frac{\partial^2}{\partial p_\da \partial p_\db}\mem
        \zeta^n
    \,=\,
        \frac{n}{2R^3}\hem
        \bb{
            \begin{array}{cc}
                \bigbig{(n\mminus2)R+z}\mem \zeta^{n-2}
            &
                \bigbig{(n\mminus1)R+z}\mem \zeta^{n-1}
            \\
                \bigbig{(n\mminus1)R+z}\mem \zeta^{n-1}
            &
                \bigbig{nR+z}\mem \zeta^n
            \end{array}
        }^{\kern-3pt\da\db}
    \,.
\end{split}
\end{align}
Thirdly,
the image of $\zeta^n\hnem/n$ under the recursion operator
can be given as
\begin{align}
\label{SDTN:obs3}
    \Rec
    \,\,\,:\,\,\,
        \frac{\zeta^n}{n}
\next
        \frac{\zeta^{n+1}}{n+1}
    \,.
\end{align}
To see this,
we use \eqref{SDTN:zeta-derivs} to observe
\begin{align}
\begin{split}
    \label{SDTN:zetaEs}
    \frac{\partial}{\partial p_\dc}\bb{
        \frac{\zeta^{n+1}}{n+1}
    }
    \,&=\,
        -\frac{\zeta^n}{\sqrt{2}R}\mem
            \ta^\dc
    \,,\\
    \bb{
        \frac{\partial}{\partial q_\dc}
        + \frac{\k}{R}\,
            \ta^\dc\hem \ta_\db\hem \frac{\partial}{\partial p_\db}
    }\nem\bb{
        \frac{\zeta^n}{n}
    }
    \,&=\,
        -\frac{\zeta^n}{\sqrt{2}R}\mem
            \ta^\dc
        \mem+\mem
        \frac{\k\hem \zeta^{n-1}}{R}\,
            \ta^\dc\hem \ta_\db\hem 
            \bb{
                -\frac{1}{\sqrt{2}R}\mem
                    \ta^\db
            }
    \,.
\end{split}
\end{align}

The three observations in \eqrefss{SDTN:obs1}{SDTN:obs2}{SDTN:obs3}
together establishes that
the repeated action of $\Rec$
on $q^\doz$ and $q^\diz$
can be given as
\begin{subequations}
\label{SDTN:sequences}
\begin{align}
\label{SDTN:sequence.a}
    q^\doz
&\next
    p^\doz
\next
    -\sqrt{2}\k\mem \zeta
\next
    -\sqrt{2}\k\mem \frac{\zeta^2}{2}
\next
    -\sqrt{2}\k\mem \frac{\zeta^3}{3}
\next
    \cdots
\,,\\
\label{SDTN:sequence.b}
    q^\diz
&\next
    p^\diz
\next
    -\sqrt{2}\k\mem \frac{\zeta^2}{2}
\next
    -\sqrt{2}\k\mem \frac{\zeta^3}{3}
\next
    -\sqrt{2}\k\mem \frac{\zeta^4}{4}
\next
    \cdots
\,.
\end{align}
\end{subequations}
Noting that
\begin{align}
    \lrp{
        \begin{array}{c}
            \s
        \\
            1
        \end{array}
    }\mem
    \log\nem\bigbig{
        1 - \s\hem\zeta
    }
    +
    \lrp{
        \begin{array}{c}
            0
        \\
            1
        \end{array}
    }\mem
    \s\hem\zeta
    \,=\,
        - \lrp{
            \begin{array}{c}
                \s^2\zeta
                + \s^3\zeta^2/2
                + \s^4\zeta^3/3
                + \cdots
            \\[1.5pt]
                \s^2\zeta^2/2
                + \s^3\zeta^3/3
                + \s^4\zeta^4/4
                + \cdots
            \end{array}
        }
    \,,
\end{align}
we find that
the deformed incidence relation
due to \eqref{SDTN:sequences} is
\begin{align}
    \label{SDTN:incidence}
    F^\da(x,\l)
    \,&=\,
        x^{\da\a} \l_\a
        + \sqrt{2}\hem\k\mem
            \delta^{\da\a}\mem
            \bbsq{
                \l_\a
                \hem
                \log\nem\bb{\nem{
                    1- 
                        \frac{x +\nem iy}{R +\nem z}
                        \mem 
                        \frac{\l_0}{\l_1}
                }\nem}
                + \o_\a\hem
                    \l_0\mem
                    % \frac{\l_0}{\l_1}\mem 
                    \frac{x +\nem iy}{R +\nem z}
            }
    \,,
\end{align}
where $\delta^{\da\a} \doteq \diag(1,1)$
is the inverse of $\delta_{\a\da}$.
% Of course, it reduces to $x^{\da\a}\l_\a$
% in the limit $\k\to0$.

\newpage

Remarkably, \eqref{SDTN:incidence}
is only linear in the parameter $\k$.
To our knowledge,
the exact linearization in the parameter $\k$
is not a property manifested in
the deformed incidence relations obtained in previous literature,
e.g., \rcite{Adamo:2025fqt}.

Note that
\eqref{SDTN:incidence} also establishes
the deformed incidence relation for the SD Kerr-Taub-NUT solution,
since 
the SD Kerr-Taub-NUT solution
is none other than
the SDTN solution
subject to a $+ i\vea$ shift of the center
\cite{nja,note-sdtn,Crawley:2021auj,Ghezelbash:2007kw}.

\subsection{Periodicity and Misner String}
\label{MISNER}

On account of the $2\pi i$ periodicity of logarithm,
the deformed incidence relation in 
\eqref{SDTN:incidence}
indicates that the $\m$-coordinates
of the curved twistor space
for the SDTN solution
will
exhibit the periodicity
% \begin{align}
%     \label{mu-period}
%     \mu^\da
%     \,\,\,\sim\,\,\,\,
%         2\sqrt{2}\hem \pi i\k\,
%             \delta^{\da\a} \l_\a
%     \,.
% \end{align}
\begin{align}
    \label{mu-period}
    \sqrt{2}\hem \mu^\da
    \,\,\,\sim\,\,\,\,
        4\pi i\k\,
            \delta^{\da\a} \l_\a
    \,.
\end{align}
This periodicity seems to reflect
the Misner string geometry of the SDTN solution.
The Misner string
is the semi-infinite line defect
which the Taub-NUT metric involves
\cite{Misner:1963flatter,Bonnor:1969ala,sackfield1971physical}.
As is well-known,
the Misner string
is the gravitational analog of the Dirac string
\cite{Misner:1963flatter,Bonnor:1969ala,sackfield1971physical,Mazur:1986gb,GP_2009_Ch341}
when the NUT charge is seen as the gravitomagnetic charge 
\cite{demianski1966combined,plebanski1975some}.
The Misner string
entails
a multivaluedness of 
the time coordinate
% a spacetime coordinate
\cite{Misner:1963flatter,Bonnor:1969ala,sackfield1971physical,Mazur:1986gb,GP_2009_Ch341,%
dowker1974nut,dowker1967gravitational,Ramaswamy:1981JMP,samuel1986gravitational%
}.
% see, e.g., \rrcite{LL:1975vol2,dowker1974nut,dowker1967gravitational,Ramaswamy:1981JMP,zee1985gravitomagnetic,samuel1986gravitational,zimmerman1989geodesics,maartens1998gravito,bunster2006monopoles,magnon1987mass,Alfonsi:2020lub,note-sdtn}
% for a comprehensive reference.
With this understanding,
it seems natural to argue that
\eqref{mu-period}
is the twistor-space imprint of the Misner string.

To substantiate this claim explicitly,
we first reproduce
the analysis carried out in \rcite{note-sdtn}.
Namely,
we zoom to the vicinity of the point $\vex = (0,0,-\e^{-1})$
for small enough $\e>0$.
Plugging in $-\e^{-1}+z$ to $z$ in \eqref{SDTN:ks}
returns
the metric
\begin{align}
    \label{Misner:geps}
    % g
    % \,=\,
        \bb{
            dt - 2\k\mem \frac{
                d(x\mminus iy)
            }{
                x\mminus iy
            }
        }^{\nem\nem\hnem2}
        - d\bb{\nem
            (x\mplus iy)
            + \frac{
                4\k\hem (\e^{-1}\hnem\mminus z) - 4\k^2
            }{
                x\mminus iy
            }
        }\mem d(x\mminus iy)
        + dz^2
        \mem+\mem
            \O(\e^1)
    \,.
\end{align}
By a single-valued holomorphic coordinate transformation
that shifts $x \mminus iy$
while fixing $x \mplus iy$,
\eqref{Misner:geps} can be brought to
\begin{align}
    \label{Misner:g}
    g
    \,=\,
        \BB{
            dt - 2\k\mem d\log (x\mminus iy)
        }^{\nem\nem2}
        - dx^2 - dy^2 -dz^2
    \,,
\end{align}
where we have finally taken
the $\epsilon \to 0$ limit.
\eqref{Misner:g}
is the metric of an infinite Misner string
inserted along the $t$-$z$ plane
\cite{Griffiths_Podolsky_2009_Ch3}.
Clearly, it is the image of the flat metric
$dt^2 - dx^2 - dy^2 - dz^2$
under a multivalued diffeomorphism
$t \mapsto t - 2\k \log (x\mminus iy)$,
which entails $4\pi i\k$ periodicity in time:
\begin{align}
    \label{xu-period}
    \sqrt{2}\hem x^{\da\a}
    \,\,\,\sim\,\,\,
        4\pi i\k\mem \delta^{\da\a}
    \,.
\end{align}
This 
$4\pi i\k$ periodicity
corresponds to gravitomagnetic flux
$i\k/2$ in the $G_\text{N} \eqq 1$ unit,
flowing in the $-z$ direction in each equal-$t$ snapshot.\footnote{
    Physically, 
    this measurement of the NUT flux describes a
    Sagnac interferometery
    \cite{Sagnac:1913ether,Sagnac:1913preuve}
    as the gravitational Aharonov-Bohm effect
    \cite{Ashtekar:1975Sagnac,sakurai1980comments,Anandan:1981Sagnac}.
    We also clarify that 
    we are more or less adopting the
    Bonnor \cite{Bonnor:1969ala,sackfield1971physical}
    interpretation of the Misner string
    in the current analysis.
}
This is consistent with the fact that
the SDTN solution described in \eqref{SDTN:g1}
carries mass $\k/2$ and NUT charge $-i\k/2$.

\newpage

Next, we repeat
the same limiting procedure
on the incidence relation in \eqref{SDTN:incidence}.
Plugging in $-\e^{-1}+z$ to $z$ in \eqref{SDTN:incidence}
and then performing the same
single-valued holomorphic coordinate transformation
as in \eqref{Misner:geps}
gives
\begin{align}
\begin{split}
    \label{Misner:incidence.eps}
    F^\da(x,\l)
    \,&=\,
\begin{aligned}[t]
    {}&{}
        \BB{
            x^{\da\a}
            - \sqrt{2}\hem\k\hem
            \delta^{\da\a}
            \log(x\mminus iy)
        }\mem \l_\a
    \\[-0.08\baselineskip]
    {}&{}
        - \frac{
            (\s_3)^{\da\a} \l_\a
        }{\sqrt{2}\hem\e}
        + \sqrt{2}\hem\k\mem
        \delta^{\da\a} \l_\a
        \log\nem\bb{\hnem\nem{
            - \frac{\l_0}{\l_1}\mem
                \frac{2}{\e}
        }\nem\hnem}
        \mem+\mem
        \O(\e^1)
    \,.
\end{aligned}
\end{split}
\end{align}
The divergent parts of \eqref{Misner:incidence.eps}
can be removed by
a linear transformation on the twistor space
representing 
% a Poincar\'e transformation.
% the constant translation $z \mapsto -\e^{-1} + z$.
a constant translation in spacetime.
We then see that
the periodicity 
entailed in
the multivalued diffeomorphism
$t \mapsto t - 2\k \log (x\mminus iy)$,
$\sqrt{2}\hem x^{\da\a} \sim \sqrt{2}\hem x^{\da\a} + 4\pi i\k\mem \delta^{\da\a}$,
% directly
boils down to
the periodicity of $\m$-coordinates
stated in \eqref{mu-period},
$\sqrt{2}\hem \m^\da \sim \sqrt{2}\hem \m^\da + 4\pi i \k\mem \delta^{\da\a} \l_\a$.
In this way,
the spacetime multivaluedness of the Misner string
translates to
a multivaluedness in twistor space.\footnote{
    It will be interesting
    if this comment can be connected to
    defect operators
    in twistor space or spacetime.
}

\section{Self-Dual {\PD}}
\label{SDPD}

The Plebański–Demiański solution \cite{PD}
describes a seven-parameter family of
Einstein-Maxwell geometries,
characterized by
mass $M$, magnetic mass $\Mm$,
electric charge $Q$, magnetic charge $\Qm$,
spin length $a$,
acceleration $1/b$,
and
cosmological constant $\L$.
The SDPD solution in \rcite{AAS}
refers to the {\PD} solution
for the case of
$M = i\Mm \neq 0$,
$Q = \Qm = 0$,
$a = 0$,
$b \neq 0$,
and
$\L = 0$.
Namely,
the SDPD solution
is a vacuum, SD spacetime
characterized by two parameters:
mass $M$ and acceleration $1/b$.

It can be shown that
the SDPD solution
describes
an asymptotically locally Euclidean
instance of a Gibbons-Hawking gravitational instanton,
two-centered with unequal masses
\cite{Casteill:2001zk,AAS}.
In particular,
\rcite{AAS} found
the explicit diffeomorphism between 
the KS and Gibbons-Hawking coordinates
for the SDPD solution.

\subsection{Kerr-Schild Metric}

In our notations and conventions,
\rcite{AAS}'s KS metric for the SDPD solution
describes
\begin{align}
    \label{SDPD:ks}
    g
    \,&=\,
    \eta \mem+\mem
        \frac{\k}{S}\mem \bb{
            \frac{z+t}{U \nem+\hnem S}\, d(z-t)
            + \frac{x+\nem iy}{S+\nem V}\, d(x-iy)
        \hnem}^{\nem\nem2}
    \,,
\end{align}
where $U$, $V$, and $S$ are
functions of
$t,x,y,z$ such that
\begin{subequations}
\begin{align}
    &
    U - V \,=\, b
    \,,\quad
    U + V \,=\,
        % \frac{
        %     -t^2 + x^2 + y^2 + z^2
        % }{b}
        \bigbig{-t^2 + x^2 + y^2 + z^2}\hhnem / b
    \,,\\
    &
    % S \,=\, 
    %     \sqrt{
    %         U^2 - (z^2 - t^2)
    %     } 
    % \,=\, 
    %     \sqrt{
    %         V^2 + (x^2 + y^2)
    %     }    S \,=\, 
    S^2
    \,=\,
        U^2 - (z^2 - t^2)
    \,=\, 
        V^2 + (x^2 + y^2)
    \,.
\end{align}
\end{subequations}
It is useful to note that
\begin{subequations}
\begin{align}
    dU \,&=\, dV
    \,=\,
        \frac{1}{b}\,\BB{
            - t\mem dt
            + x\mem dx
            + y\mem dy
            + z\mem dz
        }
    \,,\\
    dS \,&=\,
        \frac{1}{b}\,\bb{
            \frac{V}{S}\mem (
                t\mem dt + z\mem dz
            )
            + \frac{U}{S}\mem (
                x\mem dx + y\mem dy
            )
        \nem}
    \,.
\end{align}
\end{subequations}
The parameter $\k$
will encode half the mass $M$,
while $1/b$ is the aforementioned acceleration.

\newpage

% As will be discussed more in \Sec{SINGLECOPY},
The SDPD metric
could be understood as representing
a causally disconnected pair of
SDTN black holes,
accelerating  along the worldlines
$z^2 - t^2 = b^2$.
To see this,
consider replacing $z$ with $b+z$ in \eqref{SDPD:ks}
and then taking the limit $b \to \infty$.
Via this series of operations,
one shows that the SDPD metric in
the vicinity of the point $(t,x,y,z) = (0,0,0,b)$
approaches to the SDTN metric in \eqref{SDTN:ks}.
See also \App{SINGLECOPY}.
% as $S \to R$,
% $U \to b + z$,
% and
% $V \to z$.

In the spinor notation,
\eqref{SDTN:ks} boils down to
\begin{align}
    \label{SDPD:g1}
    g
    \,=\,
        \eta
        \mem+\mem
        \frac{2\k}{R}\,
            \BB{
                \o_\a\hhem \ta_\da\mem dx^{\da\a}
            }^{\nem\nem2}
    % \qfq
    % h
    % \,=\,
    %     \frac{1}{R}\,
    %         \BB{
    %             \o_\a\hhem \ta_\da\mem dx^{\da\a}
    %         }^{\nem\nem2}
    \,.
\end{align}
Again,
the ASD spinor $\o_\a \doteq \e_{\a0}$
is a constant reference
while 
the SD spinor $\ta_\da$
turns out to be
one of the principal spinors:
\begin{align}
    \label{SDPD:alpha}
    \ta^\da
    \,\,=\,\,
    \lrp{
        \begin{array}{c}
        \displaystyle
            \frac{z+ t}{U \nem+\hnem S}
        \\[12pt]
        \displaystyle
            \frac{x+\nem iy}{S+\nem V}
        \end{array}
    }^{\kern-3.5pt\da}
    \,.
\end{align}

\subsection{{\Pleb} Scalar}

Consider the {\Pleb} scalar
\begin{align}
    \label{SDPD:Pleb}
    \Phi
    \,=\,
        \frac{2S \hem\mplus\hhnem V}{3}\mem
        \bb{
            \frac{b}{U\mminus\hem S}
            \frac{z+t}{U \nem+\hnem S}
            \frac{x+\nem iy}{S+\nem V}
        }^{\nem\nem\hnem2}
    \,=\,
        \frac{2S \hem\mplus\hhnem V}{3}\mem
        \bb{
            \frac{b\mem\ta^\doz \ta^\diz
            }{U\mminus\hem S}
        }^{\nem\nem\hnem2}
    \,.
\end{align}
Straightforward computation shows that
\begin{align}
    \label{SDPD:derivs}
    \Phi^{\da\db}
    \,=\,
        \frac{1}{S}\,
            \ta^\da \ta^\db
    \,,\quad
    \Phi^{\da\db\dc\dd}
    \,=\,
        \frac{3}{2S^5}\,
            K^{(\da\db} K^{\dc\dd)}
    \,=\,
        \frac{1}{S^3}\,
            \frac{
                6\hem
                \ta^{(\da}
                \ta^\db
                \tb^\dc
                \tb^{\dd)}
            }{
                \rp{\ta\tb}{}^2
            }
    \,,
\end{align}
% where the other principal spinor
% arises as
% \begin{align}
%     \label{SDTN:beta}
%     \tb^\da
%     \,\,=\,\,
%     \lrp{\nem\hnem
%         \begin{array}{c}
%         \displaystyle
%             - \frac{x - iy}{R + z}
%         \\[7.5pt]
%             1
%         \end{array}
%     \hhnem
%     }^{\kern-3.5pt\da}
%     \,.
% \end{align}
where the other principal spinor
arises through
\begin{align}
    \label{SDPD:K}
    K^{\da\db}
    \,=\,
        x^{\da\a}\hem x^{\db\b}\mem
            \bb{
                \begin{array}{cc}
                \displaystyle
                    0
                &
                    b^{-1}
                \\
                \displaystyle
                    b^{-1}
                &
                    0
                \end{array}
            }{\vphantom{\bb{}}}_{\kern-3.5pt\a\b}
            +
            \frac{1}{2}\mem
            \bb{
                \begin{array}{cc}
                \displaystyle
                    0
                &
                    b
                \\
                \displaystyle
                    b
                &
                    0
                \end{array}
            }^{\kern-3pt\da\db}
    \,=\,
        S\: \frac{
            2\hem \ta^{(\da} \tb^{\db)}
        }{\rp{\tb\ta}}
    \,.
\end{align}
Explicitly,
\begin{align}
    K^{\da\db}
    \,=\,
    \frac{1}{b}\mem
    \lrp{
        \begin{array}{cc}
        \displaystyle
            (x-iy)(z+t)
        &
            % S^2 - UV
            \tfrac{1}{2}\mem
            \bigbig{
                b^2 + t^2 - z^2 + x^2 + y^2
            }
        \\[4pt]
        \displaystyle
            % S^2 - UV
            \tfrac{1}{2}\mem
            \bigbig{
                b^2 + t^2 - z^2 + x^2 + y^2
            }
        &
            (x+iy)(t-z)
        \end{array}
    }^{\kern-3.5pt\da\db}
    \,.
\end{align}
Again, $\Phi^{\da\db}$ in \eqref{SDPD:derivs} defines a null Maxwell field
while
the Weyl tensor
$\Phi^{\da\db\dc\dd}$
shows that the solution
is of algebraic type D.
% The graviton field is
% \begin{align}
%     h
%     \,=\,
%         \frac{1}{R}\,
%             \rp{\ta\hem dq}^2
%     \,.
% \end{align}

% As before, we denote
% $q^\da = x^{\da1}$ and $p^\da = x^{\da0}$:
% \begin{align}
%     q^\doz
%     \,=\,
%         \frac{
%             x - iy
%         }{\sqrt{2}}
%     \,,\quad
%     q^\diz
%     \,=\,
%         \frac{
%             t - z
%         }{\sqrt{2}}
%     \,,\quad
%     p_\doz
%     \,=\,
%         - \frac{
%             x + iy
%         }{\sqrt{2}}
%     \,,\quad
%     p_\diz
%     \,=\,
%         \frac{
%             t + z
%         }{\sqrt{2}}
%     \,.
% \end{align}
% It is also convenient to denote
% \begin{align}
%     \label{SDTN:zeta}
%     \zeta
%     \,:=\,
%         \frac{x+iy}{R+z}
%     \,=\,
%         \frac{p_\doz}{R+z}
%     \,.
% \end{align}

\subsection{Incidence Relation}

Now let us implement the DM recursion.
The curved Laplacian is
\begin{align}
    \label{SDPD:BBox}
    \Box_\Phi
    \,=\,
        \Box 
        \mem-\mem
        \frac{2\k}{S}\,
            \ta_\da \ta_\db\mem
            \frac{\partial^2}{\partial p_\da \partial p_\db}
    \,.
\end{align}
The recursion operator $\Rec$ is defined via
\begin{align}
    \label{SDPD:rec}
    \frac{\partial}{\partial p_\dc}\BB{
        \Rec\act{\psi}
    }
    \,=\,
    \bb{
        \frac{\partial}{\partial q_\dc}
        + \frac{\k}{S}\,
            \ta^\dc\hem \ta_\db\hem \frac{\partial}{\partial p_\db}
    }\mem \psi
    \,.
\end{align}
\newpage

Brute-force calculation shows that
the repeated action of $\Rec$ on $q^\diz$
can be given as
\begin{align}
\begin{split}
{}&{}
    q^\diz
\next
    p^\diz
    \mem=\mem
        - q^\diz\hem\zeta^1\mem \omega
\next
    q^\diz\hem\zeta^2\mem \frac{\tk}{4}
\\
{}&{}
\next
    q^\diz\hem\zeta^3\mem 
    \tk\mem
    \bb{
        \frac{2}{3} + \omega
    }
\next
    q^\diz\hem\zeta^4\mem 
    \tk\mem
    \bb{
        \frac{1}{2} - \frac{\tk}{2}
        + \frac{4}{3}\mem \omega
        + \omega^2
    }
\next
    \cdots
\,,
\end{split}
\end{align}
from which we learn that
it is convenient to define
dimensionless variables
\begin{align}
    \label{SDPD:abbrv1}
    \tk
    \,:=\,
        \frac{\k}{b}
    \,,\quad
    \zeta
    \,:=\,
        \frac{b\mem\ta^\doz \ta^\diz
        }{U\mminus\hem S}
    \,,\quad
    \omega
    \,:=\,
        \frac{S \hem\mplus V}{b}
    \,.
\end{align}
This preliminary exploration motivates the ansatz
\begin{align}
    \label{SDPD:ansG1}
    F^\diz(x,\l)
    \,=\,
        q^\diz \l_1\mem
        G(\t,\omega)
    \transition{where}
    \t
    \,:=\,
        \zeta\, \frac{\l_0}{\l_1}
    \,.
\end{align}
A tedious calculation then shows that
this $F^\diz(x,\l)$
descends to the twistor space
if
\begin{align}
\begin{split}
    \label{SDPD:G1pde}
    G(\t,\omega)
    \,&=\,
        \bbsq{
            \t
            - \bb{
                \omega
                + \frac{2\tk\mem \t}{
                    \t\omega + \t - 1
                }
            }^{\nem\nem-1}
        }\mem
        \frac{\partial G(\t,\omega)}{\partial\t}
    \,,\\
    \,&=\,
        \bbsq{
            \bb{
                \omega
                + \frac{2\tk\mem \t}{
                    \t\omega + \t - 1
                }
            }
            - \t^{-1}
        \nem}\mem
        \frac{\partial G(\t,\omega)}{\partial\omega}
    \,.
\end{split}
\end{align}
\eqref{SDPD:G1pde} implies
\begin{align}
\begin{split}
    \label{SDPD:dlogG1}
    d\log G
    % \,&=\,
    %     \frac{d\t}{\displaystyle
    %         \t
    %         - \bb{
    %             \omega
    %             + \frac{2\tk\mem \t}{
    %                 \t\omega + \t - 1
    %             }
    %         }^{\nem\nem-1}
    %     }
    %     +
    %     \frac{d\omega}{\displaystyle
    %         \bb{
    %             \omega
    %             + \frac{2\tk\mem \t}{
    %                 \t\omega + \t - 1
    %             }
    %         }
    %         - \t^{-1}
    %     }
    % \,,\\
    \,&=\,
        \frac{d\t}{\t}
        + \frac{\omega'\hem d\omega'}{
            2\tk - \omega' + \omega'^2
        }
    % \,,
    \transition{where}
    \omega' \,:=\,
        \frac{\t\omega + \t - 1}{\t}
    \,.
\end{split}
\end{align}
Let us denote
\begin{align}
    \Delta
    \,:=\,
        \sqrt{1-8\tk\vphantom{0^0}}
    \,,\quad
    \rho_\pm
    \,:=\,
        \frac{1\pm\Delta}{2}
    \qiq
    2\tk - \omega' + \omega'^2
    \,=\,
        \bigbig{\r_+ \hnem- \omega'}
        \bigbig{\r_- \hnem- \omega'}
    \,.
\end{align}
Then \eqref{SDPD:dlogG1} is integrated as
\begin{align}
    G 
    \,=\,
        \t\mem
        \bigbig{\r_+ \hnem- \omega'}^{\r_+/\Delta}
        \bigbig{\r_- \hnem- \omega'}^{-\r_-/\Delta}
    \,,
\end{align}
where the integration constant is fixed by
examining the limit $\k \to 0$.
Noting that $\r_+/\Delta - -\r_-/\Delta = 1$,
we arrive at
\begin{align}
    \label{SDPD:F1}
    F^\diz
    \,&=\,
        Y_+^{\frac{\Delta+1}{2\Delta}}
        \hem
        Y_-^{\frac{\Delta-1}{2\Delta}}
    % \,,
    \transition{where}
    Y_\pm
    \,:=\,
        \t\hem q^\diz \l_1
            \bigbig{\r_+ \hnem- \omega'}
    \,.
\end{align}
% where
% \begin{align}
%     Y_\pm
%     \,:=\,
%         \t\hem q^\diz \l_1
%             \bigbig{\r_+ \hnem- \omega'}
%     \,=\,
%     \bb{
%         1 - \t\mem \bb{
%             \omega + \frac{1 \mp \Delta}{2}
%         }
%     }\mem
%         q^\diz \l_1
%     \,.
%     % \,=\,
%     %     - q^\diz \zeta \l_0\mem
%     %         \bigbig{
%     %             \omega' \hnem\mminus \r_\pm
%     %         }
%     % \,=\,
%     %     - q^\diz \l_1\mem
%     %         \bigbig{
%     %             \t\omega + \t - 1- \t \r_\pm
%     %         }
% \end{align}

% \newpage

By working similarly,
it can be shown that
the solution for $F^\doz$ can be taken as
\begin{align}
\begin{split}
    \label{SDPD:F0F1.first}
    {}&{}
    F^\doz F^\diz
    \,=\,
        \frac{\ta^\doz}{\ta^\diz}\mem Y_+ Y_-
        + \frac{b}{\sqrt{2}\hem \ta^\diz}\mem 
        % x^{\diz\a} \l_\a\mem
        \bigbig{
            q^\diz \l_1 + p^\diz \l_0
        }\hem
        \l_1
    \,,\\
    {}&{}
    \iq
    L_\dc\act{
        F^\doz F^\diz
    }
    \,=\,
        0
    \,,\quad
    F^\doz
    \,=\,
        q^\doz \l_1 + p^\doz \l_0
        + \O\bigbig{(\l_0/\l_1)^2}\mem \l_1
    \,.
\end{split}
\end{align}
To this end, one considers another dimensionless combination
$(S \mminus V)/b$.
The details will be omitted for reasons of space,
but
we note that
it is easy to verify
$L_\dc\act{ F^\doz F^\diz } = 0$
by direct computation.

\newpage

It follows that $F^\doz F^\diz$ in \eqref{SDPD:F0F1.first}
admits a further simplification:
\begin{align}
    \label{SDPD:F0F1}
    F^\doz F^\diz
    \,=\,
        \bigbig{x^{\doz\a}\l_\a}\nem
        \bigbig{x^{\diz\a}\l_\a}
        + \k\hem b\mem \ta^\doz \ta^\diz {\l_0}^2
    \,.
\end{align}
This expression uses both components of the principal spinor in \eqref{SDPD:alpha},
\smash{$\ta^\doz$ and $\ta^\diz$}.

In sum,
we conclude that
the deformed incidence relation for the SDPD solution
can be given as
\begin{align}
\begin{split}
    \label{SDPD:incidence}
    F^\doz
    \,&=\,
        \frac{\ta^\doz}{\ta^\diz}\mem 
            Y_+^{\frac{\Delta-1}{2\Delta}}
            \hem
            Y_-^{\frac{\Delta+1}{2\Delta}}
        + \frac{b}{\sqrt{2}\hem \ta^\diz}\mem 
            % x^{\diz\a} \l_\a\mem 
            \bigbig{
                q^\diz \l_1 + p^\diz \l_0
            }\hem
            \l_1
        \,
            Y_+^{-\frac{\Delta+1}{2\Delta}}
            \hem
            Y_-^{-\frac{\Delta-1}{2\Delta}}
    \,,\\
    F^\diz
    \,&=\,
        Y_+^{\frac{\Delta+1}{2\Delta}}
        \hem
        Y_-^{\frac{\Delta-1}{2\Delta}}
    \,,
\end{split}
\end{align}
which gives rise to \eqref{summary:SDPD} 
in the introduction.
More succinctly, one can state the product relation in \eqref{SDPD:F0F1}
while presenting $F^\diz$.
The notations we have defined are
\begin{align}
\begin{split}
    \label{SDPD:notations}
    Y_\pm
    \,&=\,
    %     q^\diz \l_1
    %     + \bb{
    %         1 + \frac{b\hem(1 \mp \Delta)}{2\hem(S \mplus\hnem V)}
    %     }\mem p^\diz \l_0
    % \,=\,
        q^\diz \l_1 + p^\diz \l_0
        % x^{\diz\a} \l_\a
        + \frac{b\hem(1 \mp \Delta)}{2\sqrt{2}}\mem \ta^\diz \l_0
    \,,\\[0.2\baselineskip]
    \ta^\doz
    \,&=\,
        \frac{
            \sqrt{2}\hem 
                p^\doz
                % x^{\doz0}
        }{U \nem+\hnem S}
    \,,\quad
    \ta^\diz
    \mem=\,
        \frac{
            \sqrt{2}\hem 
                p^\diz
                % x^{\doz0}
        }{S+\nem V}
    \,,\\[0.2\baselineskip]
    \Delta
    \,&=\,
        \sqrt{1-8\k/b\vphantom{0^0}}
    \,,
\end{split}
\end{align}
while we remind ourselves that
$\ta^\da$ describes one of the principal spinors
of the solution.

\subsection{Non-Accelerating Limit}

As we have remarked earlier,
zooming to the point $(t,x,y,z) = (0,0,0,b)$,
which means to
replace $z$ with $b+z$
and then take the limit $b \to \infty$,
produces the SDTN solution from the SDPD solution
at the metric level.
% In this subsection,
Here,
we want to establish this relation
at the level of the incidence relation as well.

% For any $\Xi_1$ and $\Xi_2$,
% the following holds as an identity in
% the expansion in small $b^{-1}$:
% \begin{align}
%     \Xi_1^{\frac{\Delta+1}{2\Delta}}
%     \hem
%     \Xi_2^{\frac{\Delta-1}{2\Delta}}
%     \,&=\,
%         \Xi_1
%         \mem+\mem
%             2\k\mem \frac{\Xi_1}{b}\,
%             \log\hnem\bb{\nem{
%                 \frac{\Xi_1}{\Xi_2}
%             }\nem}
%         \mem+\mem
%         \cdots
%     \,.
% \end{align}
% In this way, 
% the small acceleration limit
% generates
% a logarithm.

Direct computation shows that
$F^\da$ in \eqref{SDPD:incidence},
when zoomed to the point $(t,x,y,z) = (0,0,0,b)$,
becomes
\begin{align}
\begin{split}
    \label{SDPD:NAlimit}
    F^\da
    \,=\,
    {}&{}
        \frac{
            b\hhem (\s_3)^{\da\a} \l_\a
        }{\sqrt{2}}
        + x^{\da\a} \l_\a
    \\[-0.12\baselineskip]
    {}&{}
        + \sqrt{2}\hem\k\mem
            \delta^{\da\a}\mem
            \bbsq{
                \l_\a
                \hem
                \log\nem\bb{\nem{
                    1- \frac{x +\nem iy}{R +\nem z}\mem
                     \frac{\l_0}{\l_1}
                }\nem}
                + \o_\a\mem
                    % \frac{\l_0}{\l_1}\mem 
                    \l_0\mem
                    \frac{x +\nem iy}{R +\nem z}
            }
    \mem+\mem
        \O(b^{-1})
    \,.
\end{split}
\end{align}
The divergent term,
$b\hhem (\s_3)^{\da\a} \l_\a /\nem \sqrt{2}$,
can be compensated by
a linear transformation on the twistor space
that precisely represents the
constant spacetime translation by
$(0,0,0,b)$.
With this understanding,
one extracts a finite result in the non-accelerating limit,
$b \to \infty$,
which precisely reproduces the SDTN result in
\eqref{SDTN:incidence}.

% Lastly, it can be also noted that
% $F^\diz$ in \eqref{SDPD:incidence}
% can be written as
% \begin{align}
%     F^\diz
%     \,=\,
%         \sqrt{Y_+Y_-}\mem
%         \exp\bbsq{
%             \frac{1}{2\Delta}
%             \log \frac{Y_-}{Y_+}
%         }
%     \,.
% \end{align}

To elucidate the key mechanism that
generates the logarithm in \eqref{SDPD:NAlimit},
note that
$F^\diz$ in \eqref{SDPD:incidence}
can be written as
\begin{align}
    \label{SDPD:exp-log}
    F^\diz
    \,=\,
        \sqrt{Y_+Y_-}\mem
        \exp\nem\bbsq{
            \frac{1}{2\Delta}
            \log\nem\bb{\nem\frac{Y_-}{Y_+}\nem}
        }
    \,=\,
        \sqrt{Y_+Y_-}\mem
        \bbsq{
            1
            + \log\nem\bb{\nem\frac{Y_-}{Y_+}\nem}
            + \cdots
        }
    \,.
\end{align}
It could be interesting if the SDPD incidence relation,
when written like \eqref{SDPD:exp-log},
could be approached in terms of
\rcite{Araneda:2023tiv}'s
conformal transformation
relating SDTN to SDPD.

\newpage

\section{Applications}
\label{STATES}

In this last section,
we would like to leave some brief remarks on the 
physical applications of the exact deformed incidence relations
obtained in \Secss{EH}{SDTN}{SDPD}.

By definition,
the deformed incidence relations
will be used in studying the geometry of null geodesics
in SD spacetimes.
By definition,
$F^\da(x,\l) = \m^\da$
gives each $\a$-surface
in the SD spacetime.
Two spacetime points, $x$ and $y$,
lie on the same null ray
iff $F^\da(x,\l) = F^\da(y,\l)$
for some $\l_\a$.
Solving this equation by
eliminating $\l_\a$
will give a function of $x$ and $y$
that provides an indicator of null separation
(like the Synge \cite{ruse1931taylor,synge1931characteristic,Poisson:2011nh} world function).
% that could generalize
% the Synge \cite{ruse1931taylor,synge1931characteristic,Poisson:2011nh} world function
% for null-separated points.
The deformed incidence relation also give the eikonal phase for massless waves
in the curved background
as will be shown shortly.

On a related note,
the explicit deformed incidence relations
might help obtaining
exact Green's functions on SD black hole backgrounds
(cf. the closed-form Green's function of Page \cite{Page:1979ga}).
More broadly,
applications to
scattering theory,
black hole perturbation theory,
or celestial holography
could be envisioned
\cite{Adamo:2023fbj,Adamo:2024xpc,Adamo:2025fqt,Guevara:2023wlr,Bittleston:2023bzp,Crawley:2023brz,Kmec:2025ftx}.

Invariantly,
a definite use of the exact deformed incidence relation
is the construction of 
massless on-shell fields
on SD backgrounds
through Penrose transform.
They can be scattering, plane-wave states
as in 
\rrcite{Adamo:2023fbj,Adamo:2024xpc,Adamo:2025fqt},
or in principle be
Coulombic states as well.
Let us sketch these directions briefly.

\subsection{Plane Wave States}
\label{STATES>PW}

Consider the description of SD spacetimes
in the precise setup of \Sec{REVIEW},
where the deformed incidence relation reads
$\m^\da = F^\da(x,\l)$.
% To begin with,
% let us describe how plane wave states are
% realized on SD black hole backgrounds.
% for scattering problems.
For constant spinors $\k_\a$ and $\tk_\da$,
consider
\begin{align}
    \label{eikS}
    S(x)\,=\,
        \tk_\da\hem F^\da(x,\k)
    \,.
\end{align}
It is easily seen that this solves
the Hamilton-Jacobi equation
and thus 
gives the classical eikonal of a massless scalar wave:
% propagating on the curved background spacetime:
$\k^\a E_{\a\da}\act{S} = 0$
implies
$g^{-1}\hnem(dS,dS) = 0$.
Note also that this $S(x)$
is ignorant of the rescaling
$(\k,\tk) \mapsto (\t\k,\t^{-1}\tk)$
because $F^\da(x,\l)$
carries projective weight $+1$ on $\l$.
Hence its invariant parameter
is rather
the null momentum,
$k_{\a\da} = \k_\a \tk_\da$.

With this understanding,
one sees that
\begin{subequations}
\label{psi}
\begin{align}
    \label{psi0}
    \psi(x)
    \,=\,
        \exp\hnem\bigbig{
            i\hem \tk_\da F^\da(x,\k)
        \hnem}
\end{align}
gives a massless scalar wave
propagating on the curved background,
generalizing
the flat spacetime formula
$\mathe^{ikx} =
\exp\hnem\bigbig{i\hem \tk_\da\hem x^{\da\a} \k_\a \hnem}$.
A spin-$s$ version reads
\begin{align}
    \label{psi-s}
    \psi_{\a_1\cdots\a_{2s}}\hnem(x)
    \,=\,
        \k_{\a_1} {\nem\cdots\mem} \k_{\a_{2s}}
        \hem
        \exp\hnem\bigbig{
            i\hem \tk_\da F^\da(x,\k)
        \hnem}
    \,.
\end{align}
\end{subequations}
% where $s = 1,2,\cdots$.
Direct computation 
in the setup of \Sec{REVIEW}
shows that
\eqrefs{psi0}{psi-s}
satisfy the zero-rest-mass equations
in the SD background.
In fact,
the twistor representatives
for these plane wave states are given by
\cite{Adamo:2021bej}
\begin{align}
    \label{rep-s}
    h_{(-2s-2)}\hhnem(\l,\m)
    \,=\mem
        \int_{\C^*} d\t\,\hem
            \t^{2s+1}\,\,
            \bar{\delta}^2\hnem(\k \mminus \t\l)\,
            \mathe^{i\t[\tk\m]}
    \,,
\end{align}
where $s = 0,1,2,\cdots$.
That is, the Penrose transform \cite{Penrose:1969ae}
of \eqref{rep-s}
yields \eqref{psi}.
% 
% More details about this Penrose transform
% can be found in \App{PTF>C}.

\newpage

For a concrete demonstration,
let us specialize in the case of the SDTN solution.
By plugging in
the exact formula in \eqref{SDTN:incidence}
to \eqref{psi-s}
with $s = 0,1,2,\cdots$,
we obtain
\begin{align}
\label{SDTN:psi-s}
    \psi_{\a_1\cdots\a_{2s}}
    \,=\,
    \bbsq{
        \k_{\a_1} {\nem\cdots\mem} \k_{\a_{2s}}
        \mem
        \mathe^{ikx}\,
        \exp\nem\bb{\nem{
            2\sqrt{2}\hem iM\,
                k_\wrap{0\diz}\mem
                \frac{x +\nem iy}{R +\nem z}
        }\nem}
    \hnem}
        \,
        \bb{\nem{
            1- 
                \frac{x +\nem iy}{R +\nem z}
                \mem 
                \frac{\k_0}{\k_1}
        }\nem}^{\nem\nem
            4i\hnem M\omega
        }
    \,,
\end{align}
where we have made the substitution $\k = 2M$
so that $M$ describes the mass.
By construction,
\eqref{SDTN:psi-s} gives
the spin-$s$ fields
representing
massless on-shell
% plane wave perturbations
plane waves
propagating on the SDTN background
in KS coordinates,
which are ASD for $s>0$.
When compared to the flat-spacetime formula,
we see that two deformation factors are in play.
First,
the wave gains an extra eikonal phase
in the $k_\wrap{0\diz}$ momentum direction.
Second,
the amplitude of the wave
is modulated by an extra factor
that carries the power
$4i\hnem M\omega$,
where
$\omega = k_0$
is the energy of the wave
with respect to the isometry direction $\delta^{\da\a}$.
Note that this factor
traces back
to the Misner string periodicity
discussed in \eqref{MISNER}
and is also reported in \rcite{Adamo:2023fbj};
the Dirac quantization condition
for the NUT charge $-iM$
describes $8\pi M\omega \in 2\pi i\hem \Z$.

In the same way,
it is left as a straightforward exercise
to explicitly obtain
the plane wave states
on the EH or SDPD backgrounds
in KS coordinates,
by plugging our exact formulae in 
\eqrefs{EH:incidence}{SDPD:incidence}.
It should be also clear that
our demonstration in \eqref{SDTN:psi-s}
also covers the case of
SD Kerr-Taub-NUT background
via a simple translation
(the Newman-Janis \cite{Newman:1965tw-janis} shift),
since
the SD Kerr-Taub-NUT solution
in the holomorphic category
is none other than
the SDTN solution
yet
centered off the origin
\cite{Ghezelbash:2007kw,Crawley:2021auj,nja}.

\subsection{Coulombic or Black Hole States}
\label{STATES>BH}

A more interesting application
could be constructing
perturbations on SD black hole backgrounds
% that represent
% ASD black holes
% of infinitesimal mass.
that spawns ``mini'' ASD black holes.

It has been known that
the twistor representatives for
the linearized fields of
ASD black holes
around the Minkowski background
are constructed from 
quadrics in \textit{flat} twistor space.
See, e.g., \rrcite{Chacon:2021wbr,white2021twistorial,Guevara:2021yud,Araneda:2022xii,Araneda:2023tiv,AAS}
for a recent literature.
This is realized
either 
at the level of
linearized curvature
or
at the level of
KS metric perturbation.

In principle,
one can consider computing
the Penrose transforms
of the known twistor representatives of ASD black holes
with the deformed twistor lines
of SD black holes.
This means to employ
holomorphic\footnote{
    This approach
    differs from that of
    the recent work \cite{AASS};
    see \eqref{holQ} and \App{AASS}
    for clarifications.
} quadrics in \textit{curved} twistor spaces.
% Based on the flat limit,
The resulting zero-rest-mass fields
will be physically interpreted as
the curvature or metric perturbations
on the SD black hole backgrounds
due to the insertion of 
an ASD black hole
of infinitesimal mass.
% Let us sketch this direction below.

% Let $Q(Z)$ be a holomorphic quadratic polynomial in twistor space
% that is not a perfect square.
% The zero locus of such a polynomial
% will be called a twistor quadric.
% It can be seen that
% a twistor quadric defines
% a spin-$s$ ASD zero-rest-mass field
% by the Penrose transform
% \begin{align}
%     \label{IP7}
%     \phi_{\a_1\cdots\a_{2s}}\nem(x)
%     \,=\,
%         \oint_{\Gamma \subset \LL_x}\nem\nem
%             \frac{\ldl}{2\pi i}\,
%             \l_{\a_1} {\cdots} \l_{\a_{2s}}\,
%             \rho_x\bbsq{\hnem
%                 \bb{\nem\hnem
%                     \frac{-2}{Q}
%                 \hnem\nem}^{\hnem\nem\nem 1+s}
%             \hhhem}
%     \,,
% \end{align}
% where $\rho_x$ denotes the restriction onto the twistor line $\LL_x$.
% Namely,
% $\nabla^2 \phi = 0$
% for $s=0$
% while
% $\nabla_{\nem\a\da}\hem \phi^\a{}_{\a_2\cdots\a_{2s}} = 0$
% and
% $\nabla^2 \phi_{\a_1\cdots\a_{2s}}$
% for $s>0$.
% See \App{PTF} for more details.

Let $Q(Z)$ be a holomorphic quadratic polynomial in twistor space
that is not a perfect square.
The zero locus of such a polynomial
will be called a twistor quadric.

% For instance,
% the ASD EH instanton,
% the ASDTN solution,
% and the ASDPD solution
% can be obtained
% by using the following twistor quadrics,
% respectively
% \cite{AAS}:
% \begin{subequations}
% \label{QZclasses}
% \begin{align}
% \label{QZclass.a}
%     \vphantom{\Big|}
%     \overline{\text{EH}}&:\quad
%     Q(Z)
%     \,=\,
%         \rp{\tdo\m}
%         \rp{\ti\m}
%     \,,\\
% \label{QZclass.b}
%     \vphantom{\Big|}
%     \text{ASDTN}&:\quad
%     Q(Z)
%     \,=\,
%         - \sqrt{2}\mem 
%             \langle \l | \delta | \m \rsq
%             % \l^\a \delta_{\a\da}\hem \m^\da
%     \,,\\
% \label{QZclass.c}
%     \vphantom{\Big|}
%     \text{ASDPD}&:\quad
%     Q(Z)
%     \,=\,
%         - 2b^{-1}\hem
%             \rp{\tdo\m}
%             \rp{\ti\m}
%         -
%         b\mem
%             \lp{\o\l}
%             \lp{\i\l}
%     \,.
% \end{align}
% \end{subequations}
% Here, $\tdo_\da$, $\ti_\da$, $\o^\a$, $\i^\a$,
% and $\delta_{\a\da}$
% are reference spinors
% such that
% $\rp{\tdo\hem\ti} = 1 = \lp{\i\o}$.

Suppose a curved SD spacetime $\M$
and its curved twistor space $\PT$.
A holomorphic twistor quadric in $\PT$
defines a spin-$s$ zero-rest-mass field on $\M$ as
\begin{align}
    \label{Ip}
    \phi_{\a_1\cdots\a_{2s}}\nem(x)
    \,=\,
        \oint_{\Gamma \subset \LL_x}\nem\nem
            \frac{\ldl}{2\pi i}\,
            \l_{\a_1} {\cdots} \l_{\a_{2s}}\,
            \rho_x\bbsq{\hnem
                \frac{
                    % (-2)^{1+s}
                    1
                }{Q^{1+s}}
            \hhhem}
    \,,
\end{align}
up to some customary overall normalization constant.
Here,
$\rho_x$ denotes the restriction onto 
% \newpage\noindent
the holomorphic twistor line $\LL_x \subset \PT$.
This plugs in
the deformed incidence relation,
so
\begin{align}
    \label{holQ}
    \rho_x Q
    \,=\,
        Q\bigbig{
            \l,F(x,\l)\hhnem
        }
    \qiq
    L_\da\act{
        \rho_x Q
    }
    \,=\,
        0
    \,.
\end{align}
% \newpage

In the flat limit where
$F^\da(x,\l)$ approaches to $x^{\da\a} \l_\a$,
the formula in \eqref{Ip} for $s\eqq2$
gives the Weyl curvature
of any ASD black hole solution
in KS coordinates.\footnote{
    This fact can be established by
    a proposition presented in \eqref{nd},
    combined with
    \rcite{AAS}'s result reproduced in \Sec{AAS>DER};
    see also
    \rrcite{Chacon:2021wbr,white2021twistorial,Guevara:2021yud,Araneda:2023tiv}
    for related discussions.
}
Based on this observation,
\eqref{Ip} for $s \eqq 2$
will be interpreted
as the ASD Weyl curvature perturbation
on the SD background $\M$
due to the insertion of
an ASD black hole
of infinitesimal mass.
For $s = 1$ and $0$,
one obtains
the single and zeroth copy counterparts of this statement.

In principle,
one may insert any ASD black hole mode
on any SD background.
Yet,
the most amenable and physically relevant case
is when
an infinitesimal ASDTN black hole
is summoned
on the SDTN background.
Suppose a SDTN black hole
centered around the origin.
With respect to its KS coordinates,
we give an infinitesimal ASDTN mode,
centered at an arbitrary point $x_0$
for full generality.
This means to use the following twistor quadric:
\begin{align}
    \label{ASDTNQ}
    Q(Z)
    \,=\,
        - \sqrt{2}\hem 
            \l^\a\hem \delta_{\a\da}\hem
            \BB{
                \m^\da - x_0^{\da\b} \l_\b
            }
    \,.
\end{align}
See \App{AAS>DER}.
With $\k = 2M$,
plugging in \eqref{SDTN:incidence} to \eqref{ASDTNQ} gives
\begin{align}
\begin{split}
    \label{ASDTNQ!}
    \rho_x Q
    % \,&=\,
    %     - \sqrt{2}\hem 
    %         \l^\a\hem \delta_{\a\da}\hem
    %         \BB{
    %             F^\da(x,\l) - x_0^{\da\b} \l_\b
    %         }
    % \,,\\
    \,&=\,
        - \sqrt{2}\hem
            \l^\a\hhem \delta_{\a\da}\hem
            % \BB{
            %     x^{\da\b} - x_0^{\da\b}
            % }\mem \l_\b
            \bigbig{x - x_0}{}^{\hnem\da\b}
        + 4M\zeta\hem
            \lp{\o\l}^2\mem
            % \bb{\nem\hnem{
            %     \frac{x +\nem iy}{R +\nem z}
            % }\hnem\nem}
    % \,,\\
    % \,&=\,
    \,=\,
        \K^{\a\b}\hnem(x)\mem \l_\a \l_\b
    \,,
\end{split}
\end{align}
% where we have set $\k = 2M$.
on the holomorphic twistor line.
We have denoted 
$\zeta := (x \mplus iy)/(R \mplus\hnem z)$ as in \eqref{SDTN:zeta}.

Notably,
the $\log$ term disappears
by virtue of $\l^\a\hhem \l_\a = 0$.
As a result, 
\eqref{ASDTNQ!}
is simply quadratic in the $\l$-variable.
Hence it can be written as
$\rho_x Q = \K^{\a\b}\hnem(x)\mem \l_\a \l_\b$,
where
% \begin{align}
%     \label{cK}
%     \K^{\a\b}\hnem(x)
%     \,=\,
%         \sqrt{2}\hem
%             \delta_{\c\dc}\hem\hhem
%             \BB{
%                 x^{\dc(\a} + 2a^{\dc(\a}
%             }\mem
%             \e^{\b)\c}
%         + 4M\zeta\mem \o^\a \o^\b
%     % \,.
% \end{align}
$\K^{\a\b}\hnem(x)$ can be shown to define
a valence-$2$ Killing spinor in the SDTN background.
Explicitly,
\begin{align}
    \label{cKvec}
    \vec{\K}
    \,=\,
        \vex - \vex_0
        + 2M\zeta\mem \bigbig{1,-i,0}
    \qiq
    \K_\a{}^\b\hnem(x)
    \,=\,
        (\vec{\K}\mdot\vec{\sigma})_\a{}^\b
    \,.
\end{align}
The magnitude \smash{$\AK$} of the three-vector $\vK$ in \eqref{cKvec} 
also admits a simple expression.
% 
% Note that
% the magnitude of the three-vector $\vK$ in \eqref{cKvec} is
% \begin{align}
%     \label{AK}
%     \AK
%     \,=\,
%         \big|\hem{
%             \vex + 2\vea
%         }\hem\big|
%         + 4M\mem\bigbig{
%             |\vex| - z
%         }
%     \,.
% \end{align}
% % 
% % Note that
% % \begin{align}
% %     \cR^2
% %     \mem:=\,
% %         x
% %     \qiq
% %         \K_\a{}^\c\mem \K_\c{}^\b
% %         \,=\,
% %             \cR^2\mem \delta_\a{}^\b
% %         \,,\quad
% %         \K^{\a\b} \K_\wrap{\a\b}
% %         \,=\,
% %             -2\hem \cR^2
% %     \,.
% % \end{align}
%
% Since the setup of \Sec{REVIEW}
% sets the ASD spin connection coefficients to zero,
% this boils down to
% \begin{align}
%     \nabla_\wrap{(\c|\dc} \K_\wrap{|\a\b)}
%     \,=\,
%         0
%     \,.
% \end{align}
% 
% 
% 
A straightforward computation then shows that
\eqref{Ip} evaluates to
\begin{align}
    \label{IP7.eval}
    \phi_{\a_1\cdots\a_{2s}}
    \,=\,
        \frac{(2s\mminus\hnem1)!!}{s!}\,
        \frac{
            \K_\wrap{(\a_1\a_2}
            {\cdots\mem}
            \K_\wrap{\a_{2s-1}\a_{2s})}
            \vphantom{|_0}
        }{
            \AK^{1+2s}
            \vphantom{\big|^{0^o}}
        }
    \,.
\end{align}
By construction, \eqref{IP7.eval} gives
zero-rest-mass fields on the SDTN background.
For $s=1$ and $2$,
respectively,
we expect that
\eqref{IP7.eval}
will give
the field strength of an ASD dyon
and
the linearized Weyl curvature of an ASDTN black hole,
inserted
on the SDTN background.

It is now well-understood
to an explicit extent
that
the Kerr-Taub-NUT family of metrics
represent a pair of SDTN and ASDTN black holes,
in the holomorphic category \cite{nja}.
% 
% Specifically,
% the Kerr-Taub-NUT black hole
% with mass $M$ and NUT charge $N$
% arises by the nonlinear superposition of
% a SDTN black hole of mass $M$
% \cite{nja}.
% 
In the above,
we have described
a tangent vector
at SDTN
that generates
a ``geodesic'' segment 
joining SDTN and Kerr
in the space of Einstein manifolds,
so to speak
(cf. \rrcite{Guevara:2023wlr,Adamo:2023fbj}).

% Similarly, 
One may also be able to
give metric perturbations of
ASD black holes
by Penrose transform.
It will be interesting if this can lead to
a twistorial construction of the Kerr-Taub-NUT family
% of metrics
that literally and mechanically adds an ASDTN mode 
% on the SDTN background.
on SDTN.
% on top of SDTN.
In this case, one expects
taking $x_0^{\da\a} = -2a^{\da\a}$
up to overall translation or Wick rotation
so that $|a|$ controls the ring radius,
based on the explicit metric-level revelation of \rcite{nja}.

% Concretely,
% the Kerr-Taub-NUT metric with mass $M$ and NUT charge $N$
% represents
% the nonlinear superposition of
% a SDTN solution of mass $(M \mplus\hem iN)/2$
% and
% an ASDTN solution of mass $(M \mminus\hem iN)/2$
% \cite{nja}.

A reverse pathway can be concretely given as follows.
The nonlinear superposition of
a SDTN solution of mass $M/2$
and
an \text{ASDTN} solution of mass $M\e/2$
gives
the Kerr-Taub-NUT metric with mass $M(1+\e)/2$ and NUT charge $iM(\e-1)/2$
\cite{nja}.
In particular,
$\e \eqq 0$ gives pure SDTN
while $\e \eqq 1$ gives Kerr.
The derivative of this Kerr-Taub-NUT metric
with respect to $\e$
should yield a linearized metric perturbation
on the SDTN background
when evaluated
around $\e \eqq 0$.
Up to an appropriate linearized gauge transformation,
% around the SDTN background,
one will be able to generate this perturbation data
from a Penrose transform,
from which one can identify the twistor representative
in the SDTN twistor space.

Another physically interesting case
is the addition of ASDPD mode
on the SDPD background,
in which case
one could envision obtaining
the C-metric \cite{Weyl:1917gp,Kinnersley:1970zw}
or 
a four-parameter subset of
the {\PD} class
from twistor space.

\section{Summary and Conclusion}

The incidence relation
is a key object in twistor theory.
In this paper,
we explicitly obtained
the exact incidence relations
for all SD black hole solutions
in KS coordinates:
EH, SDTN, and SDPD.
Our results are closed-form.
They are summarized in 
\eqref{summary},
which gathers
\eqrefss{EH:incidence}{SDTN:incidence}{SDPD:incidence}.
Our methodology is the DM recursion
in the framework of
{\Pleb}'s second heavenly equation,
for which we explicitly constructed the {\Pleb} scalar.
Although this approach to the incidence relation
has been known for a long time,
our results on the SDTN and SDPD solutions are new.
The essential novelty is that
we obtained explicit closed-form formulae
in KS coordinates.

Some notable features of our results are as follows.
For the EH solution,
the parame\-trization of
the quadratic holomorphic coordinates
is linear in the gravitational coupling
and could be considered more fundamental.
For the SDTN solution,
the exact incidence relation
is linear in the gravitational coupling
and exhibits a periodicity
that encodes the Misner string geometry.
For the SDPD solution,
the exact incidence relation
describes fractional powers
and reduces to the SDTN answer
via a proper limiting procedure
in which the acceleration is sent to zero.
In all cases,
the exact incidence relation
reduces to the flat formula
when the gravitational coupling vanishes.
For the EH and SDPD solutions,
the product $F^\doz F^\diz$
can give a simpler expression.
See also \App{APP>IF}
for a related discussion.

The incidence relation
is of a foundational importance
in SD spacetime geometries,
as it gives explicit parameterizations to
the SD null surfaces
and
the deformed twistor lines.
Physically,
our incidence relation
for SD black holes
may find applications to
exact Green's functions,
scattering theory,
black hole perturbation theory,
or celestial holography.
Concretely,
we provided a brief demonstration
of the construction of zero-rest-mass fields
on SD black hole backgrounds
through Penrose transform.
In particular,
the scalar or ASD plane-wave states on the SDTN background
was obtained in \eqref{SDTN:psi-s},
whose amplitude is modulated by a power factor
encoding the Dirac quantization condition for NUT charge.

% physical implication? where can this be used?
% compute amplitudes?
% Compton amplitudes?

The Penrose transforms using
% the deformed holomorphic twistor lines
holomorphic twistor quadrics
also suggest
adding infinitesimal
curvature or metric
modes
on SD black hole backgrounds
that generates ``tiny'' ASD black holes.
It will be interesting to investigate
if this physical idea
can lead to
constructions of real black hole solutions
from twistor space
such as
the Kerr-Taub-NUT class
or a subclass of the {\PD} family.
The insight that
real black holes may be viewed as compositions of SD and ASD parts
is an old one
\cite{hawking1977gravitational,%
plebanski1998linear,robinson-TN44-03,robinson1987some,%
Gross:1983hb}.
A twistor theorist's aspiration has been
expanding around SD backgrounds
to add the ASD modes.
It will be nice if
the exact incidence relations obtained in this paper
can be helpful
to some extent
in progressing toward this direction,
in connection with
exciting recent developments \cite{nja,AASS}.

% \vfill
\bigskip
{
    % \small
    \noindent\textbf{Acknowledgements.}
    The author would like to thank
    Tim Adamo, Bernardo Araneda, Sean Seet, Atul Sharma, and Lionel Mason
    for interesting conversations
    during the programs
    ``Twistors in Geometry \& Physics'' 
    (Isaac Newton Institute for Mathematical Sciences,
    2024)
    or
    ``Simons Collaboration on Celestial Holography Annual Meeting''
    (Simons Foundation,
    2026).
    J.-H.K. was supported by the Department of Energy (Grant No.~DE-SC0011632) and by the Walter Burke Institute for Theoretical Physics.
}
\medskip

\newpage
\appendix

\section{Appendices}

\subsection{Penrose Transforms}
\label{PTF}

As is well-known,
the Penrose transform \cite{Penrose:1969ae}
encodes 
zero-rest-mass
fields in homogeneous meromorphic functions 
$f(Z)$
on twistor space
in terms of a contour integral,
or, more precisely,
in representatives of twistor cohomology classes
\cite{penrose:maccallum,Atiyah:2017erd,penrose:1986spinors2}.
This appendix records some details about Penrose transforms
in curved and flat backgrounds,
for reference.

\subsubsection{On Curved Background}
\label{PTF>C}

Firstly,
suppose a SD spacetime $(\M,g)$
equipped with
the second heavenly coordinate system $x^{\da\a}$,
% and {\Pleb} scalar $\Phi$,
for which the curved twistor space $\PT$ is constructed
exactly
as in \Sec{REVIEW}.
% The double fibration is $\PT \leftarrow \F \rightarrow \M$
% as in \eqref{fibration}.

Consider the Penrose transforms
\begin{subequations}
\label{phi}
\begin{align}
    \label{phi0}
    \phi(x)
    \,\,&=\,\,
        \oint_{\Gamma \subset \LL_x}\nem
            \frac{\ldl}{2\pi i}\,
            % f
            % \hem\big|_{\LL_x}
            % f(\l,F(x,\l)\hnem)
            \rho_x f_{(-2)}
    \,,\\
    \label{phiASD}
    \phi_{\a_1\cdots\a_{2s}}\nem\hhnem(x)
    \,\,&=\,\,
        \oint_{\Gamma \subset \LL_x}\nem
            \frac{\ldl}{2\pi i}\,
            \l_{\a_1} {\nem\cdots\mem} \l_{\a_{2s}}
            \,
            % f
            % \hem\big|_{\LL_x}
            % f(\l,F(x,\l)\hnem)
            \rho_x f_{(-2s-2)}
    % \,,\\
    % \label{phiSD}
    % \phi_{\da_1\cdots\da_{2h}}\nem\hhnem(x)
    % \,\,&=\,\,
    %     \oint_{\Gamma \subset \LL_x}\nem
    %         \frac{\ldl}{2\pi i}\,
    %         \frac{\partial}{\partial\m^{\da_1}}
    %         {\cdots}
    %         \frac{\partial}{\partial\m^{\da_{2h}}}
    %         \,
    %         f
    %         \mem\bigg|_{\LL_x}
    \,.
\end{align}
\end{subequations}
Here, $\rho_x f = f(\l,F(x,\l)\hhnem)$ 
denotes the restriction of $f(Z)$
on the twistor line $\LL_x$,
while the subscripts $(-2)$ and $(-2s\mminus2)$ denote the homogeneity.

The following computation explicitly confirms that
the spin-$s$ ASD field in \eqref{phiASD} is massless
in the specific setup of \Sec{REVIEW}:
\begin{align}
    \nabla_{\nem\c\dc}\hem \phi^\c{}_{\a_2\cdots\a_{2s}}
    \,&=\,
        E_{\c\dc}\act{
            \phi^\c{}_{\a_2\cdots\a_{2s}}
        }
    \,=\,
        \oint
            \frac{\ldl}{2\pi i}\,
            \l_{\a_2} {\nem\cdots\mem} \l_{\a_{2s}}
            \,
            L_\dc\act{
                \rho_x f_{(-2s-2)}\hhnem
            }
    \,=\,
        0
    \,.
\end{align}
The first equality holds by 
% the vanishing of the ASD spin connection coefficients.
the gauge-fixing of the ASD spin connection coefficients
to zero.
The last equality holds by 
% constancy along the $\a$-surface.
$L_\dc\act{F^\da(x,\l)} = 0$.
To clarify,
$\nabla_{\c\dc} \eqq \nabla_{\nem\hnem E_{\c\dc}}$ 
denotes the covariant derivative 
% along the tetrad
due to the Levi-Civita connection,
while
$E_{\c\dc}\act{ \phi^\c{}_{\a_2\cdots\a_{2h}} }
= \cont{d\phi^\c{}_{\a_2\cdots\a_{2h}}}{E_{\c\dc}}
$.

Similarly,
\eqref{phi0} defines a massless scalar field
on the SD, Ricci-flat background:
\begin{align}
    \Box_\Phi\hem \phi
    \,=\,
        2\hem \te^{\da\db} \e^{\a\b} E_{\a\da}\act{
            E_\wrap{\b\db}\act{
                \phi
            }
        }
    \,=\,
        4\hem \te^{\da\db} 
        \oint
            \frac{\ldl}{2\pi i}\,
            \frac{\eta^\a}{\lp{\l\eta}}
            E_{\a\da}\act{
            L_\db\act{
                \rho_x f_{(-2)}\hhnem
            }
            }
    \,=\,
        0
    \,.
\end{align}
This computation employs a constant reference spinor $\eta^\a$
and uses \eqref{EEvanishes},
i.e.,
the self-duality of the anholonomy
$\comm{E_\wrap{\a\da}}{E_\wrap{\b\db}} \propto \e_\wrap{\a\b}$.

\subsubsection{On Flat Background}
\label{PTF>F}

Secondly, suppose flat spacetime $\mflat$
and its associated flat twistor space $\pt$
via $\m^\da = x^{\da\a} \l_\a$.
Suppose a nondegenerate holomorphic quadratic polynomial in twistor space,
\begin{align}
    \label{twistorQ}
    Q(Z)
    \,=\,
        \frac{1}{2}\mem
            Q^{\rmA\rmB}\hem Z_\rmA Z_\rmB
    \qiq
    \rho_x Q
    \,=\,
        K^{\a\b}\hhnem(x)\mem \l_\a \l_\b
    \,,
\end{align}
where $Q^{\rmA\rmB}$ is a constant symmetric matrix,
and $Z_\rmA \eqq (\l_\a,\m^\da)$.
By nondegenerate, it means that
$Q(Z)$ is not a perfect square.
% $Q^{\rmA\rmB} = A^{(\rmA} B^{\rmB)}$
% for some $A^\rmA {\:\neq\:} B^\rmB$.
It is easy to see that
$Q(Z)$ defines
a rank-$2$ symmetric spinor field $K^{\a\b}\hhnem(x)$
on spacetime
via restricting to twistor lines:
$\rho_x Q = Q(\l,x\l)$.
In fact, 
$L_\dc\act{ \rho_x Q } = 0$
shows that
this $K^{\a\b}\hhnem(x)$ is a Killing spinor,
as is remarked by Penrose \cite{penrose1967twistoralgebra}.
Based on the nondegeneracy of $Q^{\rmA\rmB}$,
this Killing spinor can be represented as
\begin{align}
    K^{\a\b}
    \,=\,
        \rho\,
        \frac{
            2\hem \a^{(\a} \b^{\b)}
        }{
            \lp{\a\b}
        }
    \qiq
    \rho_x Q
    \,=\,
        \rho\,
        \frac{
            2\hem \lp{\a\l} \lp{\b\l}
        }{
            \lp{\a\b}
        }
    \,,
\end{align}
where $\rho(x)$, $\a^\a(x)$, $\b^\a(x)$ are scalar and spinor fields.
It follows that 
$\a^\a(x)$, $\b^\a(x)$
define shear-free null geodesic congruences
on $\mflat$,
on account of the Kerr theorem \cite{penrose1967twistoralgebra}.

With hindsight,
we are interested in
a certain class of Penrose transforms:
% associated with $Q(Z)$:
% \begin{align}
%     \label{Is}
%     \phi^{(n,m)}_{\a_1\cdots\a_{2s}}\nem(x)
%     \,=\,
%         -(-2)^n\hem 2^{m/2}
%         \oint 
%             \frac{\ldl}{2\pi i}\,
%             \l_{\a_1} {\cdots} \l_{\a_{2s}}\,
%             \rho_x\bbsq{
%                 \frac{1}{Q^n}
%                 \frac{1}{
%                     {(
%                         \tdo^\da Q_{,\da}
%                     )}^{\nem m}
%                     \vphantom{|^0}
%                 }
%             }
%     \,.
% \end{align}
% \begin{align}
%     \label{Is}
%     \phi^{(n,m)}_{\a_1\cdots\a_{2s}}\nem(x)
%     \,=\,
%         (-2)^{1+n}\hem 2^{m/2}
%         \oint 
%             \frac{\ldl}{2\pi i}\,
%             \l_{\a_1} {\cdots} \l_{\a_{2s}}\,
%             \rho_x\bbsq{
%                 \frac{1}{Q^{1+n}}
%                 \frac{1}{
%                     {(
%                         Q_{\nem,\da}\hem \tdo^\da 
%                     )}^{\nem m}
%                     \vphantom{|^0}
%                 }
%             }
%     \,.
% \end{align}
\begin{align}
    \label{Is}
    \phi^{(n,m)}_{\a_1\cdots\a_{2s}}\nem(x)
    \,=\,
        \oint_{\Gamma \subset \LL_x}\nem\nem
            \frac{\ldl}{2\pi i}\,
            \l_{\a_1} {\cdots} \l_{\a_{2s}}\,
            \rho_x\bbsq{\hnem
                \bb{\nem\hnem
                    \frac{-2}{Q}
                \hnem\nem}^{\hnem\nem\nem 1+n}
                \hnem
                \bb{\nem\hnem
                    \frac{-\sqrt{2}}{
                        Q_{\nem,\da}\hhem \tdo^\da
                    }
                \hnem\nem}^{\hnem\nem\nem m}
            \hhhem}
    \,.
\end{align}
Here, $\tdo^\da$ is a constant reference spinor,
$Q_{\nem,\da}$ denotes $\partial Q/\partial \m^\da$,
and $n,m,2s \in \NN$.

\begin{subequations}
In particular, we compute two specific subclasses.
The first subclass is
\begin{align}
    \label{Is:D}
    \phi^{(n,0)}_{\a_1\cdots\a_{2n}}
    \,=\,
        \binom{2n}{n}\,
        \frac{1}{\rho^{1+n}}\,
        \frac{
            \a_{(\a_1} {\nem\cdots\mem} \a_{\a_n}
            {\nem\cdots\mem}
            \b_{\a_{n+1}} {\nem\cdots\mem} \b_{\a_{2n})}
        }{
            \lp{\a\b}^n
            \vphantom{\big|}
        }
    \,,
\end{align}
where $n \inn \NN$.
The contour $\Gamma$ is such that,
in the parametrization
$\lambda_\a = \alpha_\a \mplus \beta_\a\hem \s$,
$\s$ wraps around $\s=0$ counterclockwise.
The second subclass is
\begin{align}
    \label{Is:N}
    \phi^{(0,m)}_{\a_1\cdots\a_{m}}
    \,=\,
        {}+\nem \frac{1}{\rho}\,
            \bb{\nem\nem\frac{\sqrt{2}}{
                \lsq\tdo|\xi|\a\rangle
                \vphantom{\big|}
            }\nem\nem}^{\nem\nem\hnem m}
        \,
            \a_{(\a_1} {\nem\cdots\mem} \a_{\a_m)}
    \transit{or}
        {}-\nem \frac{1}{\rho}\,
            \bb{\nem\nem\frac{\sqrt{2}}{
                \lsq\tdo|\xi|\b\rangle
                \vphantom{\big|}
            }\nem\nem}^{\nem\nem\hnem m}
        \,
            \b_{(\a_1} {\nem\cdots\mem} \b_{\a_m)}
    \,,
\end{align}
\end{subequations}
where $m \inn \NN$.
The evaluation is presumed to be on simple poles at
$\s = 0$ or $\s = \infty$,
while $\xi^{\da\a}\hnem(x)$ denotes the Killing vector
associated with the Killing spinor:
\begin{align}
    \partial_{\c\dc} K_{\a\b}
    \,=\,
        - \epsilon_{\c(\a} \xi_{\b)\dc}
    \qiq
    \rho_x\act{
        Q_{\nem,\da}\hem \tdo^\da 
    }
    \,=\,
        - \tdo_\da\hem \xi^{\da\a} \l_\a
    \,.
\end{align}

For nonzero spin,
\eqref{Is:D}
describes ``algebraic type D'' ASD massless fields
of spin $n$
while
\eqref{Is:N}
describes ``algebraic type N'' (i.e., null) ASD massless fields
of spin $m/2$.

% The zero-rest-mass fields
% in \eqrefs{Is:D}{Is:N}
% exhibit certain relations.
% % 
% First of all, the
% The above
These
``type D'' and ``type N''
zero-rest-mass fields 
exhibit the relation
\begin{align}
    \label{NDrelation}
    \tdo^{\da_1} \partial_{\a_1\da_1}
    {\nem\cdots\mem}
    \tdo^{\da_m} \partial_{\a_m\da_m}
    \phi^{(0,m)}_{\a_{m+1}\cdots\a_{2m}}
    \hem\,=\,\hem
        \frac{m!}{\bigbig{\hhnem\nem{-\hnem\sqrt{2}}}^{\nem\hnem m}}\,\hem
        \phi^{(m,0)}_{\a_1\cdots\a_{2m}}
    \,,
\end{align}
provided that one chooses to stipulate a condition
\begin{align}
    \label{auxQ}
    Q_{\nem,\da\db}\hem \tdo^\da \tdo^\db
    \,=\,
        0
    \,.
\end{align}
To see this, note that
\begin{align}
    \tdo^\da \partial_{\a\da}\mem
    \rho_x\act{
        f
    }
    \,=\,
        \rho_x\act{
            \l_\a\mem f_{\nem,\da}\hem \tdo^\da
        }
    \,,\quad
    \bb{\nem{
        \tdo^\da\hem \frac{\partial}{\partial\m^\da}
    }\nem\nem}^{\nem\nem\hnem m}
    \bbsq{
        \frac{1}{Q} \frac{1}{
            \bigbig{
                Q_{,\da}\tdo^\da
            }^{\nem m}
        }
    }
    \,=\,
        \frac{(-1)^m\hem m!}{Q^{1+m}\vphantom{\big|}}
    \,,
\end{align}
where the second equation holds
only when the customary condition in \eqref{auxQ}
is assumed.
It is worth noting that, for $m=1,2$,
\eqref{NDrelation} 
% gives
describes
% \begin{align}
%     \tdo^\da \partial_{\a\da} 
%         \phi^{(0,1)}_{\b}
%     \,=\,
%         -\frac{1}{\sqrt{2}}\, \phi^{(1,0)}_{\a\b}
%     \,,\quad
%     \tdo^\da \partial_{\a\da}\hem
%     \tdo^\dc \partial_{\c\dc}
%         \phi^{(0,2)}_{\b\d}
%     \,=\,
%         \phi^{(2,0)}_{\a\b\c\d}
%     \,.
% \end{align}
% \begin{align}
% \begin{split}
%     \te^{\da\db} 
%     \partial_{\a\da} 
%     \BB{
%         \sqrt{2}\hem
%         \phi^{(0,1)}_{\b} 
%         \tdo_\db
%     }
%     \,=\,
%         -\phi^{(1,0)}_{\a\b}
%     &\,,\quad
%     \te^{\da\db} \te^{\dc\dd} 
%     \partial_{\a\da} \partial_{\c\dc}
%     \BB{
%         \phi^{(0,2)}_{\b\d}
%         \tdo_\db \tdo_\dd
%     }
%     \,=\,
%         \phi^{(2,0)}_{\a\b\c\d}
%     \,,\\
%     \e^{\a\b} 
%     \partial_{\a\da} 
%     \BB{
%         \sqrt{2}\hem
%         \phi^{(0,1)}_{\b} 
%         \tdo_\db
%     }
%     \,=\,
%         0
%     &\,,\quad
%     \te^{\da\db} \e^{\c\d} 
%     \partial_{\a\da} \partial_{\c\dc}
%     \BB{
%         \phi^{(0,2)}_{\b\d}
%         \tdo_\db \tdo_\dd
%     }
%     \,=\,
%         0
%     \,,\\
%     &\,,\quad
%     \e^{\a\b} \e^{\c\d} 
%     \partial_{\a\da} \partial_{\c\dc}
%     \BB{
%         \phi^{(0,2)}_{\b\d}
%         \tdo_\db \tdo_\dd
%     }
%     \,=\,
%         0
%     \,.
% \end{split}
% \end{align}
\begin{subequations}
\label{nd}
\begin{align}
\label{nd.1}
    {}&{}
    \partial_\wrap{\a\da} 
        a_\wrap{\b\db}
    -
    \partial_\wrap{\b\db} 
        a_\wrap{\a\da}
    \,=\,
        \phi^{(1,0)}_{\a\b}\mem 
            \te_\wrap{\da\db}
    \,,\\
\label{nd.2}
    {}&{}
    \partial_\wrap{\a\da} \partial_\wrap{\c\dc}
        h_\wrap{\b\db\d\dd}
    - \partial_\wrap{\b\db} \partial_\wrap{\c\dc}
        h_\wrap{\a\da\d\dd}
    + \partial_\wrap{\b\db} \partial_\wrap{\d\dd}
        h_\wrap{\a\da\c\dc}
    - \partial_\wrap{\a\da} \partial_\wrap{\d\dd}
        h_\wrap{\b\db\c\dc}
    \,=\,
        \phi^{(2,0)}_{\a\b\c\d}\mem 
            \te_\wrap{\da\db}\hem
            \te_\wrap{\dc\dd}
    \,,
\end{align}
\end{subequations}
where
\smash{$a_{\a\da} =
    \sqrt{2}\hem
        \phi^{(0,1)}_{\a} 
        \tdo_\da
$}
and
\smash{$h_\wrap{\a\da\b\db} =
    \phi^{(0,2)}_{\a\b}
    \tdo_\da \tdo_\db
$}
serve as
an abelian gauge potential
and a linearized metric perturbation
on $\mflat$.
% exhibiting ASD, ``type D''
% field strength and Weyl curvature.

% First of all,
% it has been noted 
% \cite{Chacon:2021wbr,white2021twistorial,Guevara:2021yud}
% that
% the ``type D'' fields
% exhibit
% a double copy relation
% in a Kawai-Lewellen-Tye \cite{KLT} fashion:
% \begin{align}
%     \phi^{(2,0)}_\wrap{\a\b\c\d}
%     \,\,\propto\,\,
%         \phi^{(1,0)}_\wrap{(\a\b}
%         \mem
%         \frac{1}{
%             \phi^{(0,0)}
%         }\,
%         \phi^{(1,0)}_\wrap{\c\d)}
%     \,.
% \end{align}
% In fact,
% their twistor representatives
% also satisfy the relation
% \begin{align}
%     \label{Mrel}
%     \M^{(s)}
%     \,=\,
%         \frac{1}{Q^{1+s}}
%     \qiq
%     \M^{(2)}
%     \,=\,
%         \M^{(1)}\hem
%         \frac{1}{\M^{(0)}}\mem
%         \M^{(1)}
%     \,.
% \end{align}
% Notably,
% Guevara \cite{Guevara:2021yud}
% showed that these twistor representatives
% can be derived as
% the half-Fourier transforms of 
% three-point scattering amplitudes
% with respect to the massless leg,
% thus justifying the identification of
% the relation in \eqref{Mrel}
% as double copy.\footnote{
%     Guevara \cite{Guevara:2021yud}'s
%     result is yet limited to
%     the ASDTN quadric,
%     while
%     we see that the relation in \eqref{Mrel}
%     can be stated for any ASD black holes, in fact.
% }

\newpage

% \begin{align}
%     \sqrt{2}\hem
%         \phi^{(0,1)}_{\b} 
%         \tdo_\db
%         \phi^{(0,2)}_{\b\d}
%         \tdo_\db \tdo_\dd
% \end{align}

% \begin{align}
%     \te^{\da\db} 
%     \partial_{\a\da} 
%         a_\wrap{\b\db}
%     \,=\,
%         -\phi^{(1,0)}_{\a\b}
%     &\,,\quad
%     \e^{\a\b} 
%     \partial_{\a\da} 
%         a_\wrap{\b\db}
%     \,=\,
%         0
%     \,,\\
%     \te^{\da\db} \te^{\dc\dd} 
%     \partial_{\a\da} \partial_{\c\dc}
%         h_\wrap{\b\db\d\dd}
%     \,=\,
%         \phi^{(2,0)}_{\a\b\c\d}
%     &\,,\quad
%     \te^{\da\db} \e^{\c\d} 
%     \partial_{\a\da} \partial_{\c\dc}
%         h_\wrap{\b\db\d\dd}
%     \,=\,
%         0
%     \,,\quad
%     \e^{\a\b} \e^{\c\d} 
%     \partial_{\a\da} \partial_{\c\dc}
%         h_\wrap{\b\db\d\dd}
%     \,=\,
%         0
%     \,.
%     \nonumber
% \end{align}

% \newpage

% \subsection{Dual Twistor Space Construction}
\subsection{Self-Dual Black Holes from Dual Twistor Space}
\label{AAS}

In \Secsto{EH}{SDPD},
we explicated
how SD black holes are described in the twistor space
via deformed twistor lines.
Recently,
Adamo, Araneda, and Seet \cite{AAS}
have established that
SD black holes can be also derived from the \textit{dual} twistor space---%
a ``wrong'' twistor space,
so to speak.
% Said in another way,
Equivalently,
this means that
ASD black holes can be described in the twistor space.
By building upon this fact,
a more recent work \cite{AASS}
by 
Adamo, Araneda, Seet, and Sharma
have provided a derivation
of the Schwarzschild metric---% 
as a real black hole solution with both SD and ASD parts---%
from the twistor space.
Notably,
their construction could be seen as
offering a nonperturbative solution to the googly problem,
although limited to a certain solution.

A well-known advantage of KS metrics
is its effective linearization property
\cite{xanthopoulos1978exact,Harte:2016vwo}
that allows one to view the KS metric perturbation
as a spin-$2$ field given on top of the flat Minkowski background.
% 
% Similarly,
Somewhat relatedly,
the elegant construction by 
Adamo, Araneda, and Seet \cite{AAS}
derives ASD black hole metrics in the KS form
while employing the ``flat,'' undeformed
incidence relation
in the twistor space.
% 
% Specifically,
In particular,
ASD null Maxwell fields are
derived from
a Penrose transform \cite{Penrose:1969ae}
that performs integral over contours
in each undeformed twistor line $\CP^1 \cong L_x {\:\subset\:} \pt$,
which define
ASD KS metric perturbations
on Minkowski background
through Tod's theorem \cite{Tod:1982mmp}.
The twistor data that encodes 
the KS metric of
each ASD black hole
is a quadric
$\QQ {\:\subset\:} \pt$.

The idea of
Adamo, Araneda, Seet, and Sharma \cite{AASS}
is to examine the so-called \textit{coincidence locus},
which is the locus of points on 
the ``undeformed'' twistor quadric
for an ASD black hole
that form a holomorphic subvariety
with respect to
the deformed complex structure
for a SD black hole.
Normally,
this could be a one-dimensional line
on 
the quadric $\QQ {\:\subset\:} \pt$
for the ASD black hole solution
that is also a deformed twistor line $\LL_x {\:\subset\:} \pt$
for the SD black hole solution.
However,
\rcite{AASS} observes that
for SDTN and ASDTN black holes,
centered around the same point for simplicity,
the coincidence locus
happens to exhibit complex dimension two in $\pt$.
\rcite{AASS} then observes that
% (in \App{SINGLECOPY}'s language)
% the type D single copy Maxwell field
% associated with the ASDTN black hole
the ASD dyon Maxwell field---%
as a single copy of the ASDTN black hole
% ---%
as explained on in \App{SINGLECOPY}---%
defines
a symplectic structure that is,
notably,
K\"ahler compatible with the
deformed complex structure of the SDTN twistor space
when restricting to the coincidence locus,
provided that
both black holes are
centered around the same point.
% As a result,
% they together defines a K\"ahler metric
% that turns out to be
% conformally equivalent to the Schwarzschild metric.
% 
% Consequently,
As a result,
the coincidence locus is
equipped with a K\"ahler triple,
and
the
% resulting
K\"ahler metric
turns out to be
conformally equivalent to the Schwarzschild metric.

% discontinuity Sch/Kerr disturbing - may hinder implementing NJA from twistor space
% (not as SDTN -> SDKTN, but Sch -> Kerr)

% which is the intersection between
% the deformed twistor line $\LL_x {\:\subset\:} \pt$
% for a SD black hole solution
% and
% the quadric $\QQ {\:\subset\:} \pt$
% for an ASD black hole solution.

In this appendix,
we wish to briefly review \rcite{AAS}'s construction.
In \App{AASS},
we will sketch how \rcite{AASS}'s discussion
could be approached
from our perspective.

\subsubsection{Overview}

% In the upcoming \Secs{AAS}{AASS},
% we wish to briefly review
% \refsand{AAS}{AASS},
% respectively.
% In \Sec{AASS},
% we reproduce \rcite{AASS}'s construction of the Schwarzschild metric
% from SDTN and ASDTN data.
% In \Sec{CMET},
% we consider the analogous twistor-space construction of the
% C-metric \cite{Weyl:1917gp,Kinnersley:1970zw}
% from SDPD and ASDPD data,
% adding acceleration.

% \subsection{Self-Dual Black Holes from Dual Twistor Space}
% \label{AAS}

\rcite{note-sdtn} showed that the SDTN solution admits a KS metric.
\rcite{AAS} provided
a twistor theorist's top-down explanation for this fact,
which not only rederived \rcite{note-sdtn}'s metric
but also established a complete classification of
all SD black holes
from dual twistor space.
Remarkably, this discovered a previously unknown KS metric
for the SDPD solution.
% \newpage

\rcite{AAS}'s construction
describes a two-step algorithm.
First,
dual twistor quadrics
yield
SD null Maxwell fields
via Penrose transform \cite{Penrose:1969ae}.
Second,
SD null Maxwell fields
yield
SD KS metrics
via Tod's theorem \cite{Tod:1982mmp}.
The exhaustive classification of dual twistor quadrics
(modulo equivalence by Poincar\'e action)
precisely identifies the three distinct classes of SD black hole solutions:
EH, SDTN, and SDPD.

To clarify, 
\rcite{AAS}'s dual twistor space
refers to the following open subset
of the ``flat'' $\CP^3$,
\textit{without} deformation of the complex structure:
\begin{align}
    \label{pt*}
    \pt^*
    \,=\,
        \Big\{\mem{
            [(\pi_\da , \omega^\a)] \in \CP^3
        \,\Big|\,
            \pi_\da \neq 0
        }\mem\Big\}
    \,.
\end{align}
The projective coordinates of the dual twistor space $\pt^*$
are denoted as
$W^\rmA = (\pi_\da , \omega^\a)$.
The (dual) incidence relation is simply
\begin{align}
    \label{inc0*}
    \omega^\a
    \,=\,
        \pi_\da\mem x^{\da\a}
    \,.
\end{align}

\subsubsection{Dual Twistor Quadrics and Killing Spinor}
% \subsubsection{Dual Twistor Quadrics and Derived Spacetime Objects}
% % 
% The point of departure is a description of
% dual twistor quadrics and
% derived spacetime objects.
A generic quadratic polynomial on the dual twistor space
$\pt^*$
takes the form
\begin{align}
    \label{gen-Q}
    Q(W)
    \,=\,
        \frac{1}{2}\mem
        W^\rmA W^\rmB\hem Q_{\rmA\rmB}
    \,=\,
        \frac{1}{2}\mem
        \omega^\a \omega^\b\hem \sfa_{\a\b}
        % - \omega^\a\mem \sfb_{\a\da}\hem \pi^\da
        + \omega^\a\mem \sfb_\a{}^\da\hem \pi_\da
        + \frac{1}{2}\mem
        \pi_\da \pi_\db\mem \sfc^{\da\db}
    \,,
\end{align}
where $\sfa_{\a\b}$, 
% $\sfb_{\a\da}$, 
$\sfb_\a{}^\da$,
and $\sfc^{\da\db}$
encode the components of
the constant symmetric matrix $Q_{\rmA\rmB}$.
\rcite{AAS} defines that a dual twistor quadric is the zero locus of this $Q(W)$
in $\pt^*$
such that
% the symmetric tensor 
$Q_\wrap{\rmA\rmB}$
is nondegenerate,
i.e.,
is of the form $A_\wrap{(\rmA} B_\wrap{\rmB)}$
with $A_\rmA \neq B_\rmA$.

On dual twistor line,
\begin{align}
    \label{rhoQ=K}
    \rho_x Q
    \,=\,
        \pi_\da \pi_\db\mem K^{\da\db}\hnem(x)
    \transit{where}
    K^{\da\db}\hnem(x)
    \,=\,
        \frac{1}{2}\mem
        x^{\da\a} x^{\db\b} \sfa_{\a\b}
        +
            x^{(\da|\a} \sfb_\a{}^{|\db)}
        + 
        \frac{1}{2}\mem
            \sfc^{\da\db}
    \,.
\end{align}
As noted by Penrose \cite{penrose1967twistoralgebra},
the spinor field $K^{\da\db}\hnem(x)$ extracted in this fashion
% i.e., when extracted from a dual twistor quadric,
defines a $\text{valence-}2$ Killing spinor
in the flat background $\mflat$:
since $\pi_\da \pi_\db\mem
K^{\da\db}\hnem(x)$
descends to $\pt^*$,
\begin{align}
    \label{killing2}
    0 \,=\,
    \pi^\dc \partial_{\c\dc}\mem 
    \BB{
        \pi_\da \pi_\db\mem
        K^{\da\db}(x)
    }
    \qiq
    \partial_\wrap{\a(\da} K_\wrap{\db\dc)}(x)
        \,=\, 0
    \,.
\end{align}
The nondegeneracy of $Q_{\rmA\rmB}$ implies that
$K^{\da\db}$ is non-null,
so it can be represented as
\begin{align}
    \label{K-repab}
    K^{\da\db}
    \,=\,
        \trho\: \frac{
            2\hem \ta^{(\da} \tb^{\db)}
        }{\rp{\tb\ta}}
\end{align}
for some scalar field $\trho$ and
two spinor fields $\ta^\da \neq \tb^\da$
on $\mflat$.
Note that
$K^\da{}_\dc\hem K^\dc{}_\db = \trho^2\mem \delta^\da{}_\db$
and
$K^{\da\db} K_\wrap{\da\db} = -2\trho^2$.
It can be also seen that
$\ta^\da$ and $\tb^\da$
will define shear-free null geodesic congruences
in $\mflat$,
due to the Kerr theorem \cite{penrose1967twistoralgebra}.

Said in another way,
$K_\wrap{\da\db}(x)$
defines an SD conformal Killing-Yano tensor
in the flat background $\mflat$.
As is well-known,
the divergence of a conformal Killing-Yano tensor
gives a Killing vector
in a Ricci-flat manifold
\cite{Semmelmann:2002fra,Jezierski:2005cg,Frolov:2017kze}.
% 
% \begin{align}
%     \xi_a \,=\,
%         \frac{1}{3}\mem
%         \partial_b K^b{}_a
%     \qiq
%     \partial_{(a} K_{b)c}
%     \,=\,
%         \eta_{ab}\mem \xi_c
%         - \eta_{c(a} \xi_{b)}
%     \,,
% \end{align}
% where we assume the four-dimensional flat background $\mflat$.
% 
Concretely,
\begin{align}
    \label{xi=dK}
    \xi_{\a\da}
    \,=\,
        \frac{2}{3}\mem \partial_\wrap{\a\db} K^\db{}_\da
    \qiq
    \xi^{\da\a}
    \,=\,
        x^{\da\b} \sfa_\b{}^\a
        +
            \sfb^{\da\a}
    \,,\quad
    \partial_{\c\dc} K_\wrap{\da\db}
    \,=\,
        \xi_\wrap{\c(\da}\mem \te_\wrap{\db)\dc}
    \,.
\end{align}
Clearly, this $\xi^{\da\a}$
describes generators of
ASD rotations
and translations
in $\mflat$.

Certainly,
we have already encountered
these facts in \App{PTF>F}.

\newpage

\subsubsection{The Two-Step Algorithm}

Given a dual twistor quadric $Q(W)$,
\rcite{AAS} considers the contour integral formula
\begin{align}
    \label{penrose-meth}
    \Phi_\wrap{\da\db}(x)
    \,=\,
        4
        \oint 
            \frac{\rdr}{2\pi i}\,
            \rho_x\bbsq{
                \frac{
                    \pi_\da \pi_\db
                }{
                    \bigbig{
                        \o^\a Q_{,\a}
                    }^{\nem2}
                    \hhem
                    Q
                }
            }
    \,,
\end{align}
while presuming that
$\o^\a Q_{,\a} \neq 0$.
Here,
$Q_{,\a}$ denotes $\partial Q/\partial \omega^\a$,
and $\o^\a$ is the reference.
Since \eqref{penrose-meth}
describes a Penrose transform,
$\Phi_\wrap{\da\db}$
defines a SD Maxwell field on $\mflat$.
% Computation shows that 
In fact,
it is also null
when evaluated around simple poles.
Noting that
$\rho_x\act{ Q_{,\a} } = - \xi_{\a\da}\hhem \pi^\da$,
residue calculation gives
% \begin{align}
%     \label{penrose-meth.comp}
%     \Phi_\wrap{\da\db}
%     % \,=\,
%     \,\,\,=\,\,\,
%         \frac{2}{\trho}\hem
%         \frac{
%             \ta_\da \ta_\db
%         }{
%             % \bigbig{
%             %     % \sqrt{2}\hem
%             %     \o_\c \ta_\dc\mem \xi^{\dc\c}
%             % }^{\nem2}
%             \lsq \ta | \xi | \o \rangle^2
%             \vphantom{\big|}
%         }
%     \transit{or}
%         - \frac{2}{\trho}\hem
%         \frac{
%             \tb_\da \tb_\db
%         }{
%             % \bigbig{
%             %     \o_\c \tb_\dc\mem \xi^{\dc\c}
%             % }^{\nem2}
%             \lsq \tb | \xi | \o \rangle^2
%             \vphantom{\big|}
%         }
%     \,.
% \end{align}
% 
\begin{align}
    \label{penrose-meth.comp}
    \Phi_\wrap{\da\db}
    \,\,\,=\,\,\,
        \frac{1}{\trho}\hem
        \frac{
            \sqrt{2}\hem
            \ta_\wrap{\da}
        }{
            \lsq \ta | \xi | \o \rangle
            \vphantom{\big|}
        }
        \frac{
            \sqrt{2}\hem
            \ta_\wrap{\db}
        }{
            \lsq \ta | \xi | \o \rangle
            \vphantom{\big|}
        }
    \transit{or}
        \frac{1}{\trho}\hem
        \frac{
            \sqrt{2}\hem
            \tb_\wrap{\da}
        }{
            \lsq \tb | \xi | \o \rangle
            \vphantom{\big|}
        }
        \frac{
            \sqrt{2}\hem
            \tb_\wrap{\db}
        }{
            \lsq \tb | \xi | \o \rangle
            \vphantom{\big|}
        }
    \,.
\end{align}
The $\Z_2$ ambiguity of swapping $\ta^\da$ and $\tb^\da$
can be resolved arbitrarily;
for instance, one can make
$\ta^\da \o^\a$ and $\tb^\da \o^\a$
describe
``outgoing'' and ``ingoing''
null congruences,
respectively.

The theorem of Tod \cite{Tod:1982mmp}
then states that
a SD null Maxwell field 
$\Phi_\wrap{\da\db}$
on flat spacetime
defines
a SD KS metric
by
$g = \eta + 2\k\, \o_\a \o_\b\hem \Phi_\wrap{\da\db}\mem dx^{\da\a} \modot dx^{\db\b}$,
through the existence of a scalar field $\Phi$
such that
$\Phi_\wrap{\da\db} = \o^\a \o^\b \partial_\wrap{\a\da} \partial_\wrap{\b\db} \Phi$
and
$\Box\hem\Phi = 0$.
Note that this describes the case when
the {\Pleb} second heavenly equation is
``effectively linearized,''
meaning its left-hand and right-hand sides,
$\Box\hem\Phi$ and $\k\mem \Phi_\wrap{\da\db}\hhem \Phi^{\da\db}$,
are separately vanishing.
Surely,
Tod's formula
amounts to
the metric perturbation stated earlier in
\eqref{hF},
$h_\wrap{\a\da\b\db} = \o_\a \o_\b\hem \Phi_\wrap{\da\db}$.

\subsubsection{Complete Classification and Derivation of Self-Dual Black Holes}
\label{AAS>DER}

Eventually,
\rcite{AAS}
identifies the three distinct classes of SD black holes.
Excluding the trivial case when $K^{\da\db}\hnem(x)$ lacks dependence in spacetime coordinates,
\rcite{AAS} spells out
the exhaustive classification of all dual twistor quadrics,
modulo Poincar\'e action,
as
\begin{subequations}
\label{Qclass}
\begin{align}
\label{Qclass.a}
    Q(W)
    \,=\,
        \tfrac{1}{2}\mem
        \omega^\a \omega^\b\hem \sfa_{\a\b}
    &\qfq
    K^{\da\db}\hnem(x)
    \,=\,
        \tfrac{1}{2}\mem
        x^{\da\a} x^{\db\b} \sfa_{\a\b}
    \,,\\
\label{Qclass.b}
    Q(W)
    \,=\,
        -
            \omega^\a\mem \sfb_{\a\da}\hem \pi^\da
    &\qfq
    K^\da{}_\db(x)
    \,=\,
        x^{\da\a} \sfb_\wrap{\a\db}
        - \tfrac{1}{2}\,
            \delta^\da{}_\db\mem
            (x^{\dc\c} \sfb_{\c\dc})
        \vphantom{\frac{1}{2}}
    \,,
    \kern-0.2em\\
\label{Qclass.c}
    \kern-0.4em
    Q(W)
    \,=\,
        \tfrac{1}{2}\mem
        \omega^\a \omega^\b\hem \sfa_{\a\b}
        + 
        \tfrac{1}{2}\mem
        \pi_\da \pi_\db\mem \sfc^{\da\db}
    &\qfq
    K^{\da\db}\hnem(x)
    \,=\,
        \tfrac{1}{2}\mem
        x^{\da\a} x^{\db\b} \sfa_{\a\b}
        + 
        \tfrac{1}{2}\mem
            \sfc^{\da\db}
    \,.
\end{align}
\end{subequations}
Note that
the associated Killing vector $\xi^{\da\a}$ in \eqref{xi=dK}
is always nonvanishing.
% 
% Consequently,
\rcite{AAS} shows that
the three cases
shown in \eqrefss{Qclass.a}{Qclass.b}{Qclass.c}
respectively
give rise to
the KS metric
% the KS metric perturbation
% $h_\wrap{\a\da\b\db} = \o_\a \o_\b\hem \Phi_\wrap{\da\db}$
of the
EH, SDTN, and SDPD solutions,
via \eqrefs{penrose-meth}{hF}.

To be pedagogical,
let us reproduce this calculation.
For the EH solution,
choosing
the invariant structure
$\sfa_\wrap{\a\b}$ such that
% \begin{align}
%     \sfa_\wrap{\a\b}
%     \,=\,
%         % \frac{1}{b}\mem
%         % M\mem
%         |\sfa|\mem
%     % \,\propto\,
%         \bb{
%             \begin{array}{cc}
%             \displaystyle
%                 \,0\,
%             &
%                 \,1\,
%             \\
%             \displaystyle
%                 \,1\,
%             &
%                 \,0\,
%             \end{array}
%         }{\vphantom{\bb{}}}_{\kern-3.5pt\a\b}
%     \,,
% \end{align}
\begin{align}
    \label{EH:strchoice}
    \sfa_\wrap{\a\b}
    \,=\,
        -\TwoByTwoMatrixCustom{0.46em}{0}{1}{1}{0}
    \qfq
    \sfa_\a{}^\b
    \,=\,
        -\TwoByTwoMatrixCustom{0.46em}{1}{0}{0}{-1}
    \qiq
    \sfa_\a{}^\c\hem \sfa_\c{}^\b
    \,=\,
        \delta_\a{}^\b
\end{align}
realizes the Killing spinor in \eqref{K-repab}
by
\begin{align}
    \label{EH:rab}
    \trho
    \,=\,
        \tfrac{1}{2}\mem \rp{pq}
    \,,\quad
    \ta^\da
    \,\propto\,
    p^\da
    \,,\quad
    \tb^\da
    \,\propto\,
    q^\da
    \,.
\end{align}
On the outgoing branch,
one finds
\begin{align}
    \frac{
        \sqrt{2}\hem
        \ta^\da
    }{
        \lsq \ta | \xi | \o \rangle
        \vphantom{\big|}
    }
    \,=\,
        \frac{
            \sqrt{2}\hem
            p^\da
        }{
            \rp{pq}
        }
    \,,
\end{align}
so \eqref{penrose-meth.comp} gives
$\Phi_\wrap{\da\db} = 4\hhem p_\da p_\db /\rp{pq}^3$,
reproducing \eqref{EH:derivs}.

For the SDTN solution,
choosing
\begin{align}
    \sfb_\wrap{\a\da}
    \,=\,
        \sqrt{2}\hem \delta_{\a\da}
    \,=\,
        \sqrt{2}\hem
        \TwoByTwoMatrixCustom{0.46em}{1}{0}{0}{1}
    % \,,
\end{align}
realizes the Killing spinor in \eqref{K-repab}
by
\begin{align}
    \label{SDTN:rab}
    \trho
    \,=\,
        R
    \,,\quad
    \ta^\da
    \,\propto\,\nem
        \lrp{
            \begin{array}{c}
                1
            \\[2pt]
            \displaystyle
                \frac{x +\nem iy}{R +\nem z}
            \end{array}
        }
    \,,\quad
    \tb^\da
    \,\propto\,\nem
        \lrp{
            \begin{array}{c}
            \displaystyle
                - \frac{x -\nem iy}{R +\nem z}
            \\[2pt]
                1
            \end{array}
        }
    \,,
\end{align}
consistently with \eqref{SDTN:alpha}.
Straightforward computation shows that
\begin{align}
    \frac{
        \sqrt{2}\hem
        \ta^\da
    }{
        \lsq \ta | \xi | \o \rangle
        \vphantom{\big|}
    }
    \,=\,
        \lrp{
            \begin{array}{c}
                1
            \\[2pt]
            \displaystyle
                \frac{x +\nem iy}{R +\nem z}
            \end{array}
        }
    \,,
\end{align}
thus reproducing
the KS metric 
written in \eqref{SDTN:ks}
(descried via \eqrefs{SDTN:g1}{SDTN:alpha}).
Similarly, the KS metric in the ``ingoing'' convention
can be also obtained.

For the SDPD solution,
choosing
\begin{align}
    \sfa_\wrap{\a\b}
    \,=\,
        \frac{2}{b}\hhhem
        \TwoByTwoMatrixCustom{0.46em}{0}{1}{1}{0}
    \,,\quad
    \sfc^{\da\db}
    \,=\,
        b\hem
        \TwoByTwoMatrixCustom{0.46em}{0}{1}{1}{0}
    % \,,\quad
    % \,,
\end{align}
gives \eqref{SDPD:K}.
It
realizes the Killing spinor in \eqref{K-repab}
by
\begin{align}
    \label{SDPD:rab}
    \trho
    \,=\,
        S
    \,,\quad
    \ta^\da
    \,\propto\,\nem
        \lrp{
            \begin{array}{c}
            \displaystyle
                \frac{z+ t}{U \nem+\hnem S}
            \\[12pt]
            \displaystyle
                \frac{x+\nem iy}{S+\nem V}
            \end{array}
        }
    \,,\quad
    \tb^\da
    \,\propto\,\nem
        \lrp{
            \begin{array}{c}
            \displaystyle
                -\frac{x-\nem iy}{S+\nem V}
            \\[12pt]
            \displaystyle
                \frac{z-t}{U \nem+\hnem S}
            \end{array}
        }
    \,,
\end{align}
consistently with \eqref{SDPD:alpha}.
Straightforward computation shows that
\begin{align}
    \frac{
        \sqrt{2}\hem
        \ta^\da
    }{
        \lsq \ta | \xi | \o \rangle
        \vphantom{\big|}
    }
    \,=\,
        \lrp{
            \begin{array}{c}
            \displaystyle
                \frac{z+ t}{U \nem+\hnem S}
            \\[12pt]
            \displaystyle
                \frac{x+\nem iy}{S+\nem V}
            \end{array}
        }
    \,,
\end{align}
thus reproducing
the KS metric 
written in \eqref{SDPD:ks}
(descried via \eqrefs{SDPD:g1}{SDPD:alpha}).
Similarly, the KS metric in the ``ingoing'' convention
can be also obtained.

Note that here
we have worked in a fashion that is
agnostic about
the scale of $\ta^\da$ and $\tb^\db$.
Our convention is such that
in all of these three cases,
it holds that
\begin{align}
    \Phi^{\da\db\dc\dd}
    \,=\,
        \frac{3}{2\trho^5}\,
            K^{(\da\db} K^{\dc\dd)}
    \,=\,
        \frac{1}{\trho^3}\,
            \frac{
                6\hem
                \ta^{(\da}
                \ta^\db
                \tb^\dc
                \tb^{\dd)}
            }{
                \rp{\ta\tb}{}^2
            }
    \,.
\end{align}

\newpage

% \subsection{Real Black Holes from Twistor Space}
\subsection{Coincidence Loci and Real Black Holes}
\label{AASS}

% \subsection{Schwarzschild Black Hole from Twistor Space}
% \label{AASS}

It has long been known
that
real black hole solutions may be 
viewed as
% systems of
combinations of
SD and ASD parts.
As \rcite{AASS} has reminded us,
this insight traces back to
Hawking \cite{hawking1977gravitational}:
the Schwarzschild black hole
can be regarded as a combination of SD and ASD Taub-NUT black holes,
overlapped at the same point.
In the meantime,
the idea of ``patching'' SD and ASD spacetimes together
to construct real spacetimes
has been a longstanding aspiration
of the {\Pleb} school
\cite{plebanski1998linear,robinson-TN44-03,robinson1987some}
as well.\footnote{
    The author thanks Maciej Dunajski for this anecdote.
}

Gross and Perry \cite{Gross:1983hb}
had qualitatively argued for the possibility of
interpreting the Kerr solution as 
a dipolar configuration of Taub-NUT instantons.
An explicit realization of this idea
has been given in \rcite{nja}:
it is shown that
the Kerr metric
represents the nonlinear superposition
of SD and ASD Taub-NUT solutions,
based on a factorization of the ring singularity
in the holomorphic category.
The Newman-Janis algorithm \cite{Newman:1965tw-janis}
is reproduced as
the unique transformation
that splits the Taub-NUT centers apart
to generate Kerr from Schwarzschild.
See also
\cite{Ghezelbash:2007kw,Adamo:2023fbj,gabriel1,Chacon:2021wbr}
for related discussions
from various angles.

Overall,
we see that
an exciting program is being revived:
comprehending and constructing
% real black hole solutions in four dimensions
% four-dimensional real black holes
real black holes in four dimensions
as systems of SD and ASD building blocks.

\rcite{nja}
% established 
% provided
gave
the construction of the Schwarzschild-Kerr family
as
the nonlinear superpositions
% ``superpositions''
of SDTN and ASDTN black holes,
at the level of metrics on \textit{spacetime}.
\rcite{AASS} 
% pursued 
provided
a twistor theorist's implementation for this construction.
Namely,
the question is whether
the Schwarzschild-Kerr family,
which describes real solutions with both SD and ASD curvatures,
can be derived from \textit{twistor space}.

To this end,
\rcite{AASS} considered describing
both SD and ASD black holes in the twistor space.
The SD black hole is implemented in the twistor space
by the original Penrose \cite{Penrose:1976js} framework.
At the same time,
the ASD black hole is implemented also in the twistor space
by the Adamo-Araneda-Seet \cite{AAS} framework.
In the former,
the SD black hole 
arises from 
deformed twistor lines
due to the
deformed complex structure.
In the latter,
the ASD black hole
arises from
a dual twistor quadric.
% Concretely, \rcite{AASS} 
% applies this approach
% to obtain the Schwarzschild solution,
% in which case
% the SD and ASD solutions are
% the SDTN and ASDTN black holes
% centered around the same point.
Concretely, \rcite{AASS} 
specialized to the case of the Schwarzschild solution,
i.e., overlapping centers.

What seems crucial 
in this construction
is that
both descriptions,
of the SD solution and the ASD solution,
are based on KS coordinates.
Also, a subtle yet important aspect is that
\rcite{AASS}
works in the Euclidean signature and
presumes the
Atiyah-Hitchin-Singer \cite{Atiyah:1978wi} form of the nonlinear graviton.
Below, we wish to give
a minimal review of
some key elements in \rcite{AASS}'s construction
from our perspectives.

\subsubsection{The Coincidence Locus}

Let us approach \rcite{AASS}'s construction
% from our perspective.
in a correspondence space point of view.
Let $\PT$ be the curved twistor space of the SDTN solution.
Working in the correspondence space $\F \to \PT$
realized with KS coordinates,
consider
\begin{align}
    \label{TN:Q}
    \QQ
    \,=\,
        \Big\{\mem\mem{
            (x,[\l]) \nem\inn \F
        \mem\,\Big|\,\hem\mem
            Q(x,\l) = 0
        }\mem\,\Big\}
    \,,
\end{align}
where
\begin{align}
    \label{ASDQ}
    Q(x,\l)
    % \,=\,
    %     - \l^\a\hhem \xi_{\a\da}\hem x^{\da\b} \l_\b
    \,=\,
        -\sqrt{2}\hem
            \l^\a\hhem \delta_{\a\da}\hem 
            x^{\da\b}\hhem
            \l_\b
    % \,=\,
    %     2R\mem 
    %         \frac{
    %             \lp{\a\l}
    %             \lp{\b\l}
    %         }{
    %             \lp{\a\b}
    %         }
\end{align}
amounts to
% the ASDTN twistor quadric.
the twistor quadric for the ASDTN solution.
% whose center is
% shifted by a certain constant displacement
% ${-2a^{\da\a}}$
% for full generality.
It should be emphasized that
\eqref{ASDQ} plugs in
the undeformed incidence relation
(unlike as in \Sec{STATES}).
Since $x^{\da\a} \l_\a$ is not holomorphic,
one does not expect that 
it
descends to the curved twistor space $\PT$.

Explicitly,
the SDTN Lax vector field
$L_\dc = \l^\c E_\wrap{\c\dc}$
is given by
\begin{align}
    \label{SDTN:Lax}
    L_\dc
    \,=\,
    %     \l^\c \partial_\wrap{\c\dc}
    %     + \k\mem \l_0\,
    %         \Phi^\dd{}_\dc\mem 
    %         \o^\d
    %         \partial_\wrap{\d\dd}
    % \,=\,
        \l^\c \partial_\wrap{\c\dc}
        + \frac{\k\hem\lp{\o\l}}{R}\mem
            % \frac{
            %     \ta_\dc\mem \ta^\dd 
            % }{
            %     \langle \o | \delta | \ta \rsq^2
            % }\mem
            \ta_\dc\mem \ta^\dd
            \o^\d
            \partial_\wrap{\d\dd}
    \,,
\end{align}
in the fixed-scale convention
in \eqref{SDTN:alpha}.
The second term does not annihilate $x^{\da\a}\l_\a$.
Simple calculation shows that
acting \eqref{SDTN:Lax} on \eqref{ASDQ}
gives
\begin{align}
    \label{TN:LQ}
    L_\dc\act{
        Q(x,\l)
    }
    \,=\,
        - \frac{
            \sqrt{2}\hhem\k\hem\lp{\o\l}^2
            \ta_\dc
        }{
            R
            % \hem
            % \langle \o | \delta | \ta \rsq^2
        }\,
        \langle \l | \delta | \ta \rsq
    \,,
\end{align}
which is generically nonzero.
That is,
\begin{align}
    \label{TN:CIcond}
    L_\dc\act{
        Q(x,\l)
    }
    \,=\,
        0
    \qfq
    \l_\a
    \,\propto\,
        \delta_{\a\da}\mem \ta^\da
    \,.
\end{align}

We may comment on how
this result can be also approached from
our explicit deformed incidence relation
in \eqref{SDTN:incidence}.
Using \eqref{SDTN:incidence},
\eqref{ASDQ} is equivalently written as
\begin{align}
    \label{ASDQ.Fform}
    Q(x,\l)
    \,=\,
        - \sqrt{2}\hem
            \l^\a\hhem \delta_{\a\da}\hem
            F^\da(x,\l)
        + 2\k\hem
            \lp{\o\l}^2\hhem 
            \bb{\nem\hnem{
                \frac{x +\nem iy}{R +\nem z}
            }\hnem\nem}
    \,,
\end{align}
where the $\log$ term disappears 
by virtue of $\l^\a \l_\a = 0$.
The first term in \eqref{ASDQ.Fform}
evidently gets annihilated by the Lax vector field in \eqref{SDTN:Lax}.
Hence it suffices to compute
\begin{align}
\begin{split}
    L_\dc\act{
        \zeta
    }
    \,&=\,
        \l_1\mem E_{0\dc}\act{\zeta}
        - \l_0\mem E_{1\dc}\act{\zeta}
    \,,\\
    \,&=\,
        \l_1\mem \frac{1}{\sqrt{2}\hem R}\mem
            \ta_\da
        -\l_0\mem \frac{\zeta}{\sqrt{2}\hem R}\mem
            \ta_\da
    \,=\,
        \frac{\ta_\da}{\sqrt{2}\hem R}\mem
        \BB{
            \l_1 - \zeta\mem \l_0
        }
    \,,
\end{split}
\end{align}
where we have denoted
$\zeta := (x \mplus iy)/(R \mplus\hnem z)$ as in \eqref{SDTN:zeta}
and used 
our previously computed results in \eqref{SDTN:zetaEs}.
This shows that
$Q(x,\l)$ descends to $\PT$
precisely when $\zeta$ does,
which means
\begin{align}
    \frac{\l_1}{\l_0}
    \,=\,
        \zeta
    \qfq
    1 - \frac{x +\nem iy}{R +\nem z}\mem \frac{\l_0}{\l_1}
    \,=\,
        0
    \,,
\end{align}
reproducing the condition in \eqref{TN:CIcond}.
Note that this is also precisely when
the argument of the logarithm in
$F^\da(x,\l)$ vanishes,
i.e., the locus around which the SDTN $\m$-coordinates
exhibit a multivaluedness.

% \newpage

With this observation,
\rcite{AASS} restricts their attention
to a holomorphic subvariety,
which we understand as
\begin{align}
    \label{TN:chi}
    \chi
    \,=\,
        \Big\{\mem\mem{
            (x,[\l]) \nem\inn \F
        \mem\,\Big|\,\hem\mem
            Q(x,\l) = 0
            \,\,\,\text{and}\,\,\,
            L_\dc\act{
                Q(x,\l)
            } = 0
        }\mem\,\Big\}
    \,.
\end{align}
\eqref{TN:chi} is the definition of the coincidence locus.

\newpage

\subsubsection{A Relation Between Principal Spinors}
\label{AASS>REL}

It is easily seen that $Q(x,\l)$ factorizes as
\begin{align}
    Q(x,\l)
    \,=\,
        2\r\,
        \frac{
            \lp{\a\l} \lp{\b\l}
        }{
            \lp{\b\a}
        }
    \,.
\end{align}
Explicitly,
\begin{align}
    \label{ASDTN:rab}
    \rho
    \,=\,
        R
    \,,\quad
    \a^\a
    \,\propto\,\nem
        \lrp{
            \begin{array}{c}
                1
            \\[2pt]
            \displaystyle
                \frac{x -\nem iy}{R +\nem z}
            \end{array}
        }
    \,,\quad
    \b^\a
    \,\propto\,\nem
        \lrp{
            \begin{array}{c}
            \displaystyle
                - \frac{x +\nem iy}{R +\nem z}
            \\[2pt]
                1
            \end{array}
        }
    \,.
\end{align}
It follows that
$Q(x,\l) = 0$
amounts to either
$\l_\a \propto \a_\a$
or
$\l_\a \propto \b_\a$.

In the meantime,
the calculations in the previous page
have shown that
$L_\dc\act{Q(x,\l)} = 0$
amounts to
$\l_\a \propto \delta_{\a\da}\mem \ta^\da$:
recall \eqref{TN:CIcond}.
Remarkably
(though not surprisingly
in light of the relation between Euclidean and Lorentzian conjugations),
the principal spinors $\ta^\da$, $\tb^\da$ of the SDTN solution
in \eqref{SDTN:rab}
are related to
the principal spinors $\a^\a$, $\b^\a$ of the ASDTN solution
in \eqref{ASDTN:rab}
as
\begin{align}
    \label{TN:barelation}
    \b_\a
    \,\propto\,
        \delta_{\a\da}\mem \ta^\da
    \,,\quad
    \a_\a
    \,\propto\,
        \delta_{\a\da}\mem \tb^\da
    \,,
\end{align}
provided that their centers coincide.
Therefore,
$\l_\a \propto \delta_{\a\da}\mem \ta^\da$
simply collapses to
$\l_\a \propto \b_\a$.

Consequently,
the coincidence locus in \eqref{TN:chi} 
simply selects the ``ingoing branch''
of $\QQ$ in \eqref{TN:Q}:
\begin{align}
    \label{TN:chi.exl}
    \chi
    \,=\,
        \Big\{\mem\mem{
            (x,[\l]) \nem\inn \F
        \mem\,\Big|\,\hem\mem
            \lp{\b\l}
                = 0
        }\mem\,\Big\}
    \,.
\end{align}
In this way, the coincidence locus exhibits complex dimension two, not one.

\subsubsection{Holomorphic and Anti-Holomorphic Forms}

When working in the correspondence space,
the one-form $e^{\da\a} \l_\a$
exhibits zero contraction with
the Lax vector field $L_\da = \l^\a E_{\a\da}$.
Hence it will be regarded as spanning the $(1,0)$-space.
When specializing on the coincidence locus, one replaces $\l_\a$ with $\beta_\a$.

To identify the basis for the $(0,1)$-space,
\rcite{AASS} carefully examines
the Dolbeaut operator
in their Euclidean signature and
Atiyah-Hitchin-Singer \cite{Atiyah:1978wi} setup.
The result amounts to
declaring
$dx^{\da\a} \a_\a$ as $(0,1)$
on the coincidence locus.
Let us simply import this fact,
although we are technically working in the holomorphic category
(hence $(1,0)$ and $(0,1)$ should be understood in a complexified sense).

In sum, $e^{\da\a} \b_\a$ spans the $(1,0)$-space whereas
$dx^{\da\a} \a_\a$ spans the $(0,1)$-space.
Since $\ta_\da\hem dx^{\da\a} = \ta_\da\hem e^{\da\a}$,
a useful identity arises that
\begin{subequations}
\begin{align}
\begin{split}
    \label{id-adxo}
    \lsq \ta | dx | \o \rangle
    % \,&=\,
    %     \frac{1}{\lp{\a\b}}\mem\BB{
    %         \lsq \ta | dx | \b \rangle
    %         \hem\lp{\a\o}
    %         -
    %         \lsq \ta | dx | \a \rangle
    %         \hem\lp{\b\o}
    %     }
    % \,,\\
    \,&=\,
        \frac{1}{\lp{\a\b}}\mem\BB{
            \lsq \ta | e | \b \rangle
            \hem\lp{\a\o}
            -
            \lsq \ta | dx | \a \rangle
            \hem\lp{\b\o}
        }
    \,,
\end{split}
\end{align}
which implies
\begin{align}
    \label{id-dxb}
    dx\hem |\b\rangle
    \,=\,
        e\hem |\b\rangle
        + \frac{1}{R}\,
            |\ta\rsq\hem
        \frac{\lp{\o\b}}{\lp{\a\b}}\mem\BB{
            \lp{\o\a}
            \lsq \ta | e | \b \rangle
            -
            \lp{\o\b}
            \lsq \ta | dx | \a \rangle
        }
    \,.
\end{align}
\end{subequations}

\newpage

\subsubsection{Derivation of the Schwarzschild Metric}

Finally,
the field strength of the ASD dyon
in KS coordinates
(see \eqrefss{Is:D}{nd.1}{WeylDC})
is given by
\begin{align}
    \label{fasdd}
    F
    \,=\,
        \frac{1}{R^2}\,
            \frac{2\hem \te_\wrap{\da\db}}{\lp{\a\b}}\,
                dx^{\da\a} \a_\a
                \wedge
                dx^{\db\b} \b_\b
    \,,
\end{align}
which is a closed two-form.
In fact,
\rcite{AASS} views it as a symplectic form
on the coincidence locus.
By using the identity in \eqref{id-dxb},
\eqref{fasdd} translates to
\begin{align}
    \label{fasdd2}
    F
    \,=\,
        \frac{2}{R^2}\,
        \frac{1}{\lp{\a\b}}\mem
        \bb{
            \te_\wrap{\da\db}
            + 
            \frac{\lp{\o\a}\lp{\o\b}}{\lp{\a\b}}\mem
            \frac{\k}{R}\,
                \ta_\da 
                \ta_\db
        }\mem
            \bigbig{
                dx^{\da\a} \a_\a
            }
                {\,\wedge\,}
            \bigbig{
                e^{\db\b} \b_\b
            }
    \,,
\end{align}
which is evidently $(1,1)$.
Hence \eqref{fasdd2} is K\"ahler compatible with the complex structure
as a symplectic form.
The K\"ahler metric is
\begin{align}
    \label{gkahler1}
    g
    \,=\,
        - \frac{4i}{R^2}\,
        \frac{1}{\lp{\a\b}}\mem
        \bb{
            \te_\wrap{\da\db}
            + 
            \frac{\lp{\o\a}\lp{\o\b}}{\lp{\a\b}}\mem
            \frac{\k}{R}\,
                \ta_\da 
                \ta_\db
        }\mem
            \bigbig{
                dx^{\da\a} \a_\a
            }
                {\,\odot\,}
            \bigbig{
                e^{\db\b} \b_\b
            }
    \,.
\end{align}
% Noting that
% \begin{align}
%     \te_\wrap{\da\db}\mem
%         dx^{\da\a} \a_\a
%         \odot
%         dx^{\db\b} \b_\b
%     \,=\,
%         \frac{\lp{\a\b}}{2}\,
%         \e_\wrap{\a\b}\mem
%         \te_\wrap{\da\db}\mem
%             dx^{\da\a}
%             \odot
%             dx^{\db\b}
%     \,,
% \end{align}

When converting to the ``flat'' basis,
the two terms in the big bracket in \eqref{gkahler1}
cooperate to
establish a cancellation of a 
$\lsq\ta|dx|\a\rangle \motimes \lsq\tb|dx|\b\rangle$
component,
yielding
\begin{align}
    \label{gkahler2}
    g
    \,=\,
        - \frac{4i}{R^2}\,
        \frac{1}{\lp{\a\b}}\mem
        \bb{
            \frac{\lp{\a\b}}{2}\,
                \e_\wrap{\a\b}
                \te_\wrap{\da\db}
                \mem
                dx^{\da\a} \modot dx^{\db\b}
            + 
            \frac{\lp{\o\b}^2}{\lp{\a\b}}\mem
            \frac{\k}{R}\,
                \lsq\ta|dx|\a\rangle^2
        }
    \,.
\end{align}
Reviving the rescaling freedom of $\ta^\da$ gives
\begin{align}
    \label{gkahler3}
    g
    \,=\,
        - \frac{i}{R^2}\hnem
        \lrp{
            2\hem
                \e_\wrap{\a\b}
                \te_\wrap{\da\db}
                \mem
                dx^{\da\a} \modot dx^{\db\b}
            + 
            \frac{4\k}{R}\,
                \frac{
                    {\lsq\ta|dx|\a\rangle}^{\nem2}
                    \vphantom{\b}
                }{
                    % \langle\o|\delta|\ta\rsq^2
                    {\lsq\ta|\delta|\o\rangle}^{\nem2}
                    \vphantom{\big|}
                }
                \frac{\lp{\o\b}^{\nem2}}{\lp{\a\b}^{\nem2}}
        }
    \,.
\end{align}
The bracketed combination in \eqref{gkahler3}
gives the Schwarzschild metric
in the outgoing KS form,
of mass $\k$.
% \begin{align}
%     g_\text{Sch}
%     \,=\,
%         dt^2 - dx^2 - dy^2 - dz^2
%         \mem+\mem \frac{2\k}{R}\mem
%             \bb{
%                 dt - \frac{
%                     x\mem dx + y\mem dy + z\mem dz
%                 }{R}
%             }^{\nem\nem\hnem2}
%     \,.
% \end{align}
Namely,
the K\"ahler metric
of the coincidence locus
is conformal to the KS metric of the Schwarzschild solution.

% We may
% comprehend \eqref{gkahler3}
% by taking 
% the SD ``type N'' Maxwell field
% $\Phi_\wrap{\da\db} = R^{-1}\mem \ta_\da \ta_\db / {\lsq\ta|\delta|\o\rangle}^{\nem2}$
% and
% the ASD ``type D'' Maxwell field
% $F_\wrap{\a\b} = R^{-2}\mem 2\hem \a_\wrap{(\a} \b_\wrap{\b)} / \lp{\a\b}$
% as building blocks:
% \begin{align}
%     g
%     \,=\,
%         -\frac{i}{R^2}\mem\bb{
%             \eta
%             + \frac{4\k}{R}
%         }
% \end{align}

A few remarks are in order.
First of all,
suppose one desires to derive the Kerr solution.
In this case,
$Q(x,\l)$ in \eqref{ASDQ} can be replaced with
\begin{align}
    \label{ASDQa}
    Q(x,\l)
    % \,=\,
    %     - \l^\a\hhem \xi_{\a\da}\hem x^{\da\b} \l_\b
    \,=\,
        -\sqrt{2}\hem
            \l^\a\hhem \delta_{\a\da}\hem 
            \bigbig{
                x^{\da\b} 
                \mplus 2a^{\da\b}
            }
            \l_\b
    % \,=\,
    %     2R\mem 
    %         \frac{
    %             \lp{\a\l}
    %             \lp{\b\l}
    %         }{
    %             \lp{\a\b}
    %         }
    \,,
\end{align}
while the SDTN center needs not be shifted necessarily.
With a nonzero separation, $2a^{\da\a} \neq 0$,
the relation in \eqref{TN:barelation}
is lost.
Hence the entire construction falls apart.
Regarding this point,
\rcite{AASS} seems to envision
an approach that
utilizes the deformed quadric framework of \rcite{Araneda:2022xii}.

In spirit of the Newman-Janis shift \cite{Newman:1965tw-janis,Crawley:2021auj,nja},
it will be ideal to have
a derivation of the Schwarzschild-Kerr family
(in the holomorphic category)
in a way that the deformation adding spin
is smooth and continuous.
Some other ideas include
a twistor setup 
with holomorphic quadric
as briefly pursued in \Sec{STATES>BH}
and
an ambitwistor setup
that treats SD and ASD black hole modes
on an equal footing
around the flat background.
The former may have an advantage
that the physical semantics is clear.
The latter could
essentially boil down
\rcite{nja}'s nonlinear superposition theorem
to a certain consistency equation in ambitwistor space.
% These directions may not preclude connections with
% the coincidence locus.

\newpage

\subsection{Deformed Incidence Relations Revisited}
\label{APP>IF}

A notable elegance of \rcite{AAS}'s formalism,
reviewed in \App{AAS},
is that
it covariantly manifests the invariant structures
that characterize each SD black hole solution
in KS coordinates.
For instance, the EH solution
is characterized by
a sole ASD plane $\sfa_{\a\b}$,
while the SDTN solution
is characterized by
a sole isomorphism $\sfb_{\a\da}$
between the ASD and SD spinor spaces,
i.e., a fixed non-null direction in spacetime.
% The SDPD solution is characterized by
% not only an ASD plane $\sfa_{\a\b}$
% but also a SD plane $\sfc^{\da\db}$,
% which are independent in principle.

On a related note,
it is natural to anticipate that
a unique advantage of 
constructing the deformed incidence relations
in KS coordinates,
as is done in this paper,
would be that
the results may be organized in terms of the invariant structures
$\sfa_{\a\b}$, $\sfb_{\a\da}$, $\sfc^{\da\db}$.

Indeed, the SDTN answer in \eqref{SDTN:incidence}
could be written in terms of the invariant structure $\sfb^{\da\a} \propto \delta^{\da\a}$
and the reference structure $\o_\a$.
% We will revisit this point.
To reproduce,
\begin{align}
    \label{SDTN:IF}
    F^\da(x,\l)
    \,&=\,
        x^{\da\a} \l_\a
        + \k\mem
            \xi^{\da\a}\mem
            \bbsq{
                \l_\a
                \hem
                \log\nem\bb{\nem{
                    1
                    -  
                    \frac{
                        \lsq \ta | \xi | \i \rangle
                        \vphantom{\big|}
                    }{
                        \lsq \ta | \xi | \o \rangle
                        \vphantom{\big|}
                    }
                    \frac{\lp{\o\l}}{\lp{\i\l}}
                }\nem}
                + \o_\a\mem
                    \frac{
                        \lsq \ta | \xi | \i \rangle
                        \vphantom{\big|}
                    }{
                        \lsq \ta | \xi | \o \rangle
                        \vphantom{\big|}
                    }
                    \frac{\lp{\o\l}}{\lp{\i\l}}
            }
    \,,
\end{align}
where $\xi^{\da\a} = \sfb^{\da\a} = \sqrt{2}\hem \delta^{\da\a}$.

We may revisit our results about the EH and SDPD solutions
from this perspective.
For the EH solution,
the quadratic holomorphic coordinates $X^{\da\db}(x,\l)$ in \eqref{EH:quadX}
describes
\begin{align}
\begin{split}
    \label{EH:IF}
    X^{\da\db}
    \,&=\,
        f^\da f^\db
        + \frac{2\k}{
            K^{\da\db} K_\wrap{\da\db}
        }\mem
            f_-^\da f_-^\db
    \,,
\end{split}
\end{align}
if $f^\da(x,\l) := x^{\da\a} \l_\a$ and
\begin{align}
    f_\pm^\da(x,\l)
    \,:=\,
        x^{\da\a}\hem
        (\Pi_\pm)_\a{}^\b\hem
        \l_\b
    \,,\quad
    (\Pi_\pm)_\a{}^\b
    \,=\,
        \frac{1}{2}\mem
        \bb{
            \delta_\a{}^\b
            \pm 
            \frac{\sfa_\a{}^\b}{|\sfa|}
        }\mem
    \,.
\end{align}
% Here, $\sfa_\a{}^\c \sfa_\c{}^\b = |\sfa|^2\mem \delta_\a{}^\b$;
% for our choice in \eqref{EH:strchoice},
% $|\sfa| = 1$.
Our choice in \eqref{EH:strchoice} describes
$\sfa_\wrap{\a\b} = |\sfa|\hem \bigbig{
    \o_\a \i_\b + \i_\a \o_\b
}$
with $|\sfa| = 1$,
so
$f^\da_- = x^{\da0} \l_0$.
% \footnote{
%     Relevant or not, we see that
%     % Parenthetically, note also that
%     $F^\doz F^\diz$ in \eqref{summary:EH}
%     describes
%     \begin{align}
%         \bsfa_\wrap{\da\db}\hem
%             F^\da F^\db
%         \mem=\mem
%             \bsfa_\wrap{\da\db}\mem
%             x^{\da\a}\hem x^{\db\b}\mem
%             \bb{
%                 \l_\a\hhem \l_\b
%                 - \frac{16\k}{(x^2)^2}\mem
%                     \i_\a\hhem \i_\b
%                     \mem \lp{\o\l}^2
%             }
%         \,,
%     \end{align}
%     if we introduce
%     $\bsfa_\wrap{\da\db} \propto (\s_1)_\wrap{\da\db}$.
%     % \begin{align}
%     %     \bsfa_\wrap{\da\db}
%     %     \,=\,
%     %         \TwoByTwoMatrixCustom{0.46em}{0}{1}{1}{0}
%     %     \,.
%     % \end{align}
% }
% 
For the SDPD solution,
the expressions seem generically unwieldy,
but it could be pointed out that
$F^\doz F^\diz$ in \eqref{SDPD:F0F1}
describes%
% \footnote{
%     Relevant or not,
%     a similar rewriting is possible
%     for the EH solution's $F^\doz F^\diz$ in \eqref{summary:EH}
%     if one introduces
%     $(\s_1)_\wrap{\da\db}$.
% }
\begin{align}
\label{SDPD:IF}
\begin{split}
    \sfc_\wrap{\da\db}\hem
        F^\da F^\db
    \mem=\mem
    \sfc_\wrap{\da\db}\hem
    \bb{
        f^\da f^\db
        + 2\k\hem b\,
            \ta^\da \ta^\db
        \frac{\lp{\o\l}^2}{
            \lsq \ta | \xi | \i \rangle
            \vphantom{\big|}^2
        }
    }
    \,.
\end{split}
\end{align}
Again, here we are completely agnostic about the scale choice for $\ta^\da$.

\subsection{Single Copy Images}
\label{SINGLECOPY}

In this appendix,
we quickly point out
the coexistence of two single copy images for
all SD black holes.
This peculiarity was first pointed out by \rcite{note-sdtn}.

We begin by a short overview of the classical double copy program.
The KS double copy correspondence 
\cite{monteiro2014black}
associates 
KS gauge potentials in Maxwell theory
with
KS metrics in gravity.
The Weyl double copy correspondence
\cite{Luna:2018dpt}
associates
type $[1,1]$ Maxwell field strengths
with
type D Weyl tensors in gravity,
whose foundation
can be traced back to
some facts about valence-$2$ Killing spinors
\cite{Walker:1970un,Hughston:1972qf},
in fact.
The SD double copy correspondence
\cite{monteiro2011kinematic,monteiro2014black}
considers the case in which
the {\Pleb} second heavenly equation
and its Yang-Mills counterpart \cite{Parkes:1992rz}
are effectively linearized,
and then identifies the {\Pleb} scalars
after color stripping.

To elaborate,
the Weyl double copy states the relationship
\cite{Luna:2018dpt}
\begin{align}
    \label{WeylDC}
    \tphi^{(2,0)}_{\da\db\dc\dd}
    \,\,\propto\,\,
        \phi^{(1,0)}_{(\da\db}
        \mem
        \frac{1}{
            \phi^{(0,0)}
        }\,
        \phi^{(1,0)}_{\dc\dd)}
    \,.
\end{align}
In fact,
it has been observed
\cite{Chacon:2021wbr,white2021twistorial,Guevara:2021yud}
that
their twistor representatives
(recall \App{PTF>F})
also satisfy the relation
\begin{align}
    \label{Mrel}
    \M^{(s)}
    \,=\,
        \frac{1}{Q^{1+s}}
    \qiq
    \M^{(2)}
    \,=\,
        \M^{(1)}\hem
        \frac{1}{\M^{(0)}}\mem
        \M^{(1)}
    \,.
\end{align}
Notably,
Guevara \cite{Guevara:2021yud}
showed that these twistor representatives
can be derived as
the half-Fourier transforms of 
three-point scattering amplitudes
with respect to the massless leg,
thus justifying the identification of
the relation in \eqref{Mrel}
as double copy
(in a Kawai-Lewellen-Tye \cite{KLT} fashion).
Guevara \cite{Guevara:2021yud}'s
result is yet limited to
the SDTN quadric,
while
we see that the relation in \eqref{Mrel}
can be stated for any SD black hole, in fact.

We see that
the proposition in \eqref{nd}
(up to chirality inversion)
establishes
the KS double copy for all SD black hole solutions
such that the Weyl double copy correspondence in \eqref{WeylDC}
holds at the same time.
In this way,
the KS and Weyl double copy correspondences agree.

Note that the KS double copy correspondence
between the SD dyon and the SDTN solution
was discovered and established by \rcite{note-sdtn}.
For the SDPD solution,
the KS double copy for the SDPD solution
has not been identified in the literature
to our knowledge.
Our discussion here asserts that 
the single copy gauge potential is
\begin{align}
    \label{ASDPD}
    A 
    \,=\,
        \frac{1}{S}\mem \bb{
            \frac{z+t}{S+U}\, d(z-t)
            + \frac{x+iy}{S+V}\, d(x-iy)
        \hnem}
    \,.
\end{align}
It can be shown that this represents 
% the gauge potential of 
a causally disjoint pair of accelerating SD dyons.

On the contrary,
we see from 
\eqref{nd.1} and
the formula for the metric perturbation
stated beneath \eqref{nd}
(again up to chirality inversion)
that
the SD double copy 
associates
each SD black hole
with {\Pleb} scalar $\Phi$
to gauge potentials
whose field strength
is the SD null Maxwell field
$\Phi_\wrap{\da\db}$.
Certainly, this is based on Tod \cite{Tod:1982mmp}'s theorem.

This peculiar disagreement between
the two groups of classical double copy proposals---%
KS and Weyl on one side,
SD on the other side---%
was first reported by \rcite{note-sdtn}.
Here, we see that the issue is validated
and persists for all SD black holes.
The former group associates type D solutions with type [1,1] solutions,
whereas the latter group associates type D soultions with type [2] solutions.
In the main article,
we always referred to the KS/Weyl double copy
when using the word ``single copy.''

Note that the SD double copy
had also been framed as ``operator KS double copy''
\cite{monteiro2011kinematic}.
When taken in this way,
\rcite{note-sdtn}
provided a physical resolution
by showing that
a nonlocal operator,
originating from propagators between chiral matters,
establishes
an operator double copy between
SD dyon and SDTN.
The nonlocality in this construction
precisely generates the Dirac/Misner strings.
See also \rcite{Ilderton:2025gug} for a related work.

\newpage

\newpage
\let\c\oldc
\let\i\oldi
\bibliography{references.bib}

\end{document}